\documentclass[rivista]{cimento}

\usepackage{amssymb}
\usepackage{amsfonts}
\usepackage{epsfig}
\usepackage{cite}

\newcommand{\nablav}{\mbox{\boldmath$\nabla$}}
\newcommand{\de}{\mbox{d}}

\newcommand{\msunyr}{M_{\odot}/\mbox{yr}}

\newcommand{\dln}{\left|\frac{\de\ln\Omega}{\de\ln R}\right|}

\title{Self-gravitating accretion discs}

\author{Giuseppe Lodato\from{lei}}

\instlist{
  \inst{lei} Department of Physics and Astronomy, University of Leicester, University Road, LE1 7RH, Leicester, UK}

\PACSes{
\PACSit{95.30.Lz}{Hydrodynamics}
\PACSit{97.10.Gz}{Accretion and accretion disks}
}

\begin{document}

\maketitle

\begin{abstract}

I review recent progresses in the dynamics and the evolution of self-gravitating accretion discs. Accretion discs are a fundamental component of several astrophysical systems on very diverse scales, and can be found around supermassive black holes in Active Galactic Nuclei (AGN), and also in our Galaxy around stellar mass compact objects and around young stars. Notwithstanding the specific differences arising from such diversity in physical extent, all these systems share a common feature where a central object is fed from the accretion disc, due to the effect of turbulence and disc instabilities, which are able to remove the angular momentum from the gas and allow its accretion. In recent years, it has become increasingly apparent that the gravitational field produced by the disc itself (the disc's self-gravity) is an important ingredient in the models, especially in the context of protostellar discs and of AGN discs. Indeed, it appears that in many cases (and especially in the colder outer parts of the disc) the development of gravitational instabilities can be one of the main agents in the redistribution of angular momentum. In some cases, the instability can be strong enough to lead to the formation of gravitationally bound clumps within the disc, and thus to determine the disc fragmentation. As a result, progress in our understanding of the dynamics of self-gravitating discs is essential to understand the processes that lead to the feeding of both young stars and of supermassive black holes in AGN. At the same time, understanding the fragmentation conditions is important to determine under which conditions AGN discs would fragment and form stars and whether protostellar discs might form giant gaseous planets through disc fragmentation. 

\end{abstract}

\section{INTRODUCTION}

Disc-like or flattened geometries are very common in astrophysics, from the large scale of spiral galaxies down to the small scales of Saturn's rings. In both these examples, the discs are characterized by a prominent structure either in the form of a spiral structure (in the case of galaxy discs) or in the form of gaps and spirals on small scales (in the case of Saturn's rings). The dynamics underlying the development of such structures is determined by the propagation of density waves, and the important role of the disc's self-gravity in their development has been clearly recognized (see, for example, \cite{bertinbook}). In the above examples the system is either{\it  collisionless}, such as in the case of spiral galaxies, where the dominant component is constituted of stars, or {\it particulate}, such as in the case of Saturn's rings, where the dominant components (mainly rocks and pebbles) do undergo inelastic collisions, but the system cannot be simply described in terms of hydrodynamics. In the last thirty years increasing attention has been given to {\it fluid} discs, where the dynamically active component is gaseous. Here, dissipative effects, associated with friction or viscosity, can significantly affect the dynamics of the disc. Through dissipation the fluid elements in the disc can lose their energy or, more fundamentally, their angular momentum, as I describe in more detail below, and can fall towards the bottom of the potential well, hence accreting on to a central gravitating body. Such a system, where a disc feeds a central object through accretion, under the effect of viscous forces, is called an {\bf accretion disc}. 

Accretion discs are found in very different contexts and over a wide range of physical scales. On the largest scale, they are one of the main physical ingredient to power the central engine of Active Galactic Nuclei (AGN), through accretion on to a supermassive black hole. The mass of the central black hole ranges from $10^6M_{\odot}$ to $10^9M_{\odot}$ and the accretion discs can extend out to large distances, of the order of about 1 pc, where rotating gaseous discs have often been detected through water maser emission \cite{miyoshi95,greenhill97,kondratko06}. On the galactic scale, accretion discs around compact objects, such as white dwarfs, neutron stars and stellar mass black holes, are often found in binary systems, where the compact object (with a mass of the order of $1M_{\odot}$) is fed by material outflowing from the companion. Historically, this has been one of the first contexts where accretion disc theory has found widespread application (a detailed review with an emphasis on accretion in galactic binary system and AGN can be found in the textbook by Frank, King and Raine \cite{franck}). Another class of objects where accretion discs play an important role are Young Stellar Objects (YSO). Here, the central accreting object is a young protostar, which receives most of its mass from a surrounding protostellar discs. In this case, protostellar discs play also another key role, as the site where planet formation takes place. Understanding the dynamics of the disc and the development of the various instabilities that might occur in it, and possibly lead to turbulence, is therefore essential if one wants to understand some properties of our own Solar system, as well as those of extra-solar planets. A thorough description of the current observational state of protostellar discs can be found, for example, in the recent Protostars and Planets V book \cite{PPV}.

As for the case of spiral galaxies, also for accretion discs the effects of
the disc self-gravity can be very important. Indeed, also in this case the
development of gravitational instabilities can lead to the formation of grand
design spiral structures, which deeply affect the structure and the dynamics
of the disc. The analysis of such instabilities naturally reveals several
similarities between the case of spiral galaxies and that of accretion discs,
but also some important differences. First of all, as mentioned above,
accretion discs are fluid and, as such, they are intrinsically
dissipative. Indeed, the evolution of gravitationally unstable accretion
discs, as I will discuss below, is strongly dependent on the gas cooling
rate. On the other hand, gravitationally unstable discs of stars, in the absence of any gas, have a natural tendency to dynamically heat up. Thermal energy, or disordered kinetic energy, plays a key role in stabilizing self-gravitating discs. However, while for stellar discs this energy is mostly in the form of disordered stellar motions, with the possibility of an anisotropic velocity dispersion, for gaseous discs the disordered kinetic energy content is mostly in the form of thermal energy. Furthermore, for gaseous discs it is expected that the role of resonances is much reduced with respect to a collisionless system. Finally, while for galaxy discs the disc mass can be a substantial fraction of the total mass of the system and therefore give a large contribution to the gravitational potential, in the case of accretion discs the disc mass is often much smaller than the mass of the central object (with some important exceptions, described below). 

Historically, accretion disc theory has concentrated on the non self-gravitating case, the effects of self-gravity having only been discussed occasionally \cite{pacinski78,kolikalov79,linpringle87,linpringle90}. In the last ten years, on the other hand, the important role of the disc self-gravity has been clearly recognized, partly due to improved observations, which have shown that in several observed systems (on all scales, from AGN to protostars) the disc mass can be high enough to have a dynamical role, and partly due to the increased computational resources, which have allowed a detailed numerical investigation of the development of gravitational instabilities in the non-linear regime. 
However, the early papers mentioned above already reveal the main directions around which research in this field would have later developed. Currently, most of the interest revolves around the three issues of self-regulation \cite{pacinski78}, disc fragmentation \cite{kolikalov79} and angular momentum transport induced by gravitational instabilities \cite{linpringle87,linpringle90}. This again reveals a difference with respect to the galaxy discs case, where most of the attention has been traditionally given to the morphology induced by self-gravity. 

In this paper, I present an overview of the dynamics of self-gravitating
accretion discs. Accretion disc theory is a very vast topic and has been
treated extensively elsewhere \cite{franck,pringle81}, so I will not repeat it here. I will rather summarize the most salient general features, with particular emphasis on those features which are most affected by the disc self-gravity. Similarly, a detailed description of all the different astrophysical systems where self-gravitating accretion discs play a role (protostellar discs, AGN,...) is obviously beyond the scope of this review. The structure of the paper is as follows. In section \ref{sec:dynamics} I describe the basic equations that determine the evolution of accretion discs and the main effects of the disc self-gravity. In section \ref{sec:GI} I discuss the development of gravitational instabilities, and present some models which incorporate in a simple way the salient physics behind their development. In section \ref{sec:simulations} I review the current state of numerical simulations of gravitationally unstable gaseous discs. In section \ref{sec:transport} I focus on the transport properties induced by gravitational instabilities, while in section \ref{sec:fragmentation} I describe the related issue of disc fragmentation. Finally, in the last two sections I will present some examples of observed systems where the concepts described in this paper find a natural application, and in particular, in section \ref{sec:examples} I will discuss the impact of self-gravity in the dynamics of protostellar discs with some application related to planet formation. In section \ref{sec:AGN} I focus on the impact of self-gravity in AGN discs also showing how self-gravitating accretion discs might have played an important role at high red-shifts, by allowing the formation of the seeds of supermassive black holes.

\section{SELF-GRAVITATING ACCRETION DISC DYNAMICS}
\label{sec:dynamics}

\subsection{The thin disc approximation} 
One of the main properties of accretion discs is that they are often thin. This means that the typical lengthscale in the vertical direction, the disc thickness $H$, is much smaller that the radial distance from the central object $R$.  For example, AGN discs are generally very thin, with aspect ratios $H/R\approx 0.001 - 0.01 $, while protostellar discs are comparably a bit thicker, with $H/R\approx 0.1$. As we will discuss below, this has a significant impact on the conditions under which the disc is self-gravitating, and hence this intrinsic difference between the AGN case and the protostellar case will be reflected in a significantly different behaviour of these two kind of systems in the self-gravitating regime. 

The thin disc condition allows us to treat the disc, to a first approximation, as infinitesimally thin, and to introduce a {\it small quantity}, the aspect ratio $H/R\ll 1$. This implies that most of the equations we are going to use can be integrated in the vertical direction, and rather than dealing with quantities {\it per unit volume} (such as the density $\rho$), we will deal instead with quantities {\it per unit surface} (such as the surface density $\Sigma=\int_{-\infty}^{\infty}\rho\de z$). When ``volume'' quantities are needed (for example the viscosity $\nu$), these will generally be understood as vertically averaged. 

The fundamental ordering of lengthscales $H/R\ll 1$ is also related to a similar ordering in terms of velocities. Indeed, one can show (see section \ref{sec:vertical}) that this ordering is equivalent to requiring that the sound speed $c_{\rm s}$ is much smaller than the rotational velocity $v_{\phi}$. It is useful to anticipate another important relation between the relevant velocities, derived by requiring that accretion takes place on a long timescale. This condition implies that the radial velocity $v_{R}$ should be smaller than both the sound speed and the rotational speed. We can therefore summarize these relations as:
\begin{equation}
v_{R}\ll c_{\rm s}\ll v_{\phi}.
\end{equation}

Finally, it is worth adding a word of caution on the applicability of the thin disc condition. Indeed, in some cases the disc might not be very thin. This happens for example, in the hotter inner regions of AGN discs. Protostellar discs, as mentioned above, are generally relatively thick. One should therefore be cautious when using the thin disc approximation, especially when considering quantities averaged in the vertical direction. 

\subsection{Basic equations}
The evolution of accretion discs is determined by the basic equations of viscous fluid dynamics: the continuity equation and Navier-Stokes equations. Consider an accretion disc rotating around a central object of mass $M$. Given the geometry of the problem, we adopt cylindrical polar coordinates centred on the central object and we assume that to first order the disc is axisymmetric, so that no quantity depends on the azimuthal angle $\phi$. Consider then a disc with surface density $\Sigma(R,t)$. In cylindrical coordinates, the continuity equation (integrated in the vertical direction, following the thin disc approximation) reads
\begin{equation}
\frac{\partial\Sigma}{\partial t}+\frac{1}{R}\frac{\partial}{\partial R}\left(R\Sigma v_{R}\right)=0.
\label{eq:continuity}
\end{equation}

To determine the velocity ${\bf v}$, we use the momentum equation. For an inviscid fluid this takes the form of Euler's equation. For the moment, we consider the full three-dimensional (i.e. not vertically integrated) equation
\begin{equation}
\rho\left[\frac{\partial {\bf v}}{\partial t}+({\bf v\cdot \nablav}){\bf v}\right]= -\nablav P-\rho\nablav{\Phi}.
\label{eq:euler}
\end{equation}
On the right-hand side there are the various forces acting on the fluid. First of all, we have the term describing pressure forces, where $P$ is the pressure and $\rho$ is the density. We then have gravity, represented by the last term on the right-hand side, where $\Phi$ is the gravitational potential. This in general includes both the contribution of the central point mass and that of the disc. A detailed description of these different contributions is provided in the next section. For an infinitesimally thin disc, we can obtain the potential associated with the gas surface density $\Sigma$ (the `self-gravity') from the solution to Poisson's equation:
\begin{equation}
\nabla^2\Phi_{\rm sg}=4\pi G\Sigma\delta(z),
\label{eq:poisson}
\end{equation}
where $\delta(z)$ is the Dirac $\delta$-function, indicating that the disc density is confined to the midplane. 

For an accretion disc, however, Euler equation is not sufficient, because it does not include the important terms due to viscosity. Including viscous forces, the relevant momentum equation is the Navier-Stokes equation. This reads
\begin{equation}
\rho\left[\frac{\partial {\bf v}}{\partial t}+({\bf v\cdot \nablav}){\bf v}\right]= -(\nablav P - \nablav\cdot\sigma)-\rho\nablav{\Phi}.
\label{eq:navier}
\end{equation}
The second term on the right hand side in equation (\ref{eq:navier}) contains the stress tensor $\sigma$ and describes the effect of viscous forces. This term plays a very important role in accretion disc dynamics. The nature of the disc viscosity and of the stress tensor $\sigma$ has been traditionally one of the main unsolved theoretical issues in accretion disc dynamics. In its simplest form, it can be assumed to be given by classical shear viscosity, so that the only non-vanishing component of $\sigma$ in a circular shearing flow is the $R\phi$ component, proportional to the rate of strain $R\Omega'$ (where $\Omega'=\de\Omega/\de R$):
 \begin{equation}
 \sigma_{R\phi} = \eta R\frac{\de\Omega}{\de R},
 \label{eq:stress}
 \end{equation}
 where $\Omega=v_{\phi}/R$ is the angular velocity and we have introduced the shear viscosity coefficient $\eta$. Equation (\ref{eq:navier}) has three components in the radial, vertical and azimuthal directions, respectively, and each one defines some important properties of accretion discs. In the following three sections we will examine in turn each of these three equations. 

\subsection{Radial equilibrium: centrifugal balance}
\label{sec:radial}

Let us first consider the radial component of equation (\ref{eq:navier}). Here, the ordering of velocities described above turns out to be particularly useful. In particular, on the left-hand side, the first term, that contains the radial velocity $v_{R}$ is negligible with respect to the second term, which (when using the appropriate expression for the differential operators in cylindrical coordinates) gives rise to the centrifugal term $v_{\phi}^2/R$. On the right-hand side, the viscous term vanishes and the pressure term can be obtained using the equation of state. If the gas is barotropic (i.e. if pressure only depends on density), the sound speed is simply defined as:
\begin{equation}
c_{\rm s}^2=\frac{\de P}{\de\rho}.
\label{eq:state}
\end{equation}
We then have:
\begin{equation}
\frac{1}{\rho}\frac{\partial P}{\partial R}=\frac{c_{\rm s}^2}{\rho}\frac{\partial \rho}{\partial R}\sim \frac{c_{\rm s}^2}{R}.
\end{equation}
Since $c_{\rm s}\ll v_{\phi}$, also this term is second order with respect to the leading term. The only term on the right hand side able to balance the centrifugal force is therefore gravity. We then have, to first order:
\begin{equation}
\frac{v_{\phi}^2}{R}\simeq \frac{\partial\Phi}{\partial R}.
\end{equation}
For accretion discs the gravitational field is generally dominated by the central compact object, but in some cases the disc self-gravity can also produce a sizable effect. The central object contribution to the field  is simply given by
\begin{equation}
 \frac{\partial\Phi}{\partial R}=\frac{GM}{R^2}
 \label{eq:keplerian}
\end{equation}
Thus, in the limit where we neglect self-gravity, the disc rotation is Keplerian, with $v_{\phi}=v_{\rm K}\equiv \sqrt{GM/R}$ and where the angular velocity is given by $\Omega=\Omega_{\rm K}\equiv\sqrt{GM/R^3}$. 

The disc contribution to the gravitational field depends on the matter distribution through Poisson's equation. For an infinitesimally thin disc, the relation between $\Sigma$ and the disc gravitational field can be written, for example, using the complete elliptic integrals of the first kind, $K$ and $E$
\begin{equation}
 \frac{\partial\Phi_{\rm disc}}{\partial R}(R,z)=\frac{G}{R}\int_0^{\infty}\left[K(k)-\frac{1}{4}\left(\frac{k^2}{1-k^2}\right)\left(\frac{R'}{R}-\frac{R}{R'}+\frac{z}{RR'}\right)E(k)\right]\sqrt{\frac{R'}{R}}k\Sigma(R')\de R',
\end{equation}
where $k^2=4RR'/[(R+R')^2+z^2]$ \cite{binney,BL99}. The above equation in general describes a complicated relation between $\Sigma$ and the gravitational field, which can be obtained numerically. However, there are a few cases where the relation is particularly simple and can be derived analytically. One interesting case is the one described by Mestel \cite{mestel63}, where $\Sigma=\Sigma_0R_0/R$ and the disc extends out to infinity. In this case the gravitational field at the disc midplane takes a particularly simple form:
\begin{equation}
 \frac{\partial\Phi_{\rm disc}}{\partial R}(R)=2\pi G\Sigma(R).
 \label{eq:mestel}
\end{equation}
In the extreme limit, where the central object contribution in negligible and a Mestel disc dominates the gravitational field \cite{bertin97}, the rotation curve is thus flat:
\begin{equation}
v_{\phi}^2=R \frac{\partial\Phi_{\rm disc}}{\partial R}(R)=2\pi G\Sigma(R)R=2\pi G \Sigma_0R_0,
\label{eq:mestel2}
\end{equation}
and the angular velocity $\Omega\propto 1/R$. Interestingly, as described in more detail in section \ref{sec:selfregulation} below, it turns out that a Mestel disc is the natural solution to the accretion disc equations in a steady state, in the limit of strongly self-gravitating discs \cite{bertin97}. Now, while in actual accretion discs the central object mass is not expected to be negligible, it is interesting to note that in many cases observations of protostellar discs imply surface densities $\Sigma\propto 1/R$ \cite{beckwith90}). 

Having obtained the rotation curve in the two limiting cases of negligible disc self-gravity and negligible central object gravity, we can now ask under which conditions do we expect the different contributions to dominate. Comparing Eq. (\ref{eq:keplerian}) to Eq. (\ref{eq:mestel}), we see that in order for the disc contribution to be comparable to the central object we require
\begin{equation}
M_{\rm disc}(R)\approx 2\pi \Sigma(R)R^2\approx M,
\end{equation}
that is, we require disc masses of the order of the central object mass. This is confirmed by detailed models of self-regulated discs \cite{BL99}. In most cases, the accretion disc mass is much lower than the central object and so the Keplerian approximation is generally appropriate. However, there are a few important exceptions where non-Keplerian rotation has been observed and can be interpreted in terms of relatively massive discs. This occurs for example for some protostellar discs around massive stars \cite{cesaroni94,cesaroni05} and also in some AGN discs, such as in the nucleus of the active galaxy NGC 1068 \cite{greenhill97,LB03a}. We defer a more detailed description of such cases to section \ref{sec:examples} and \ref{sec:AGN} below. 

The above derivation is only valid to first approximation, since we have neglected the pressure forces in the radial direction. This is generally a good approximation, but in some cases the corrections due to pressure effects might play an important role. In particular, this is the case when one considers the dynamics of small solid bodies within the disc (a very important component involved in the process of planet formation, as discussed below in section \ref{sec:planet}). The solids are not subject to pressure and would thus move on exactly Keplerian orbits, therefore generating a small velocity difference with respect to the gas (in a process somewhat analogous to the so called asymmetric drift in stellar dynamics), which in turn determines a very fast migration of the solids \cite{weiden77,rice04,rice06}. Let us then calculate this correction. The radial equation of motion, including pressure terms is:
\begin{equation}
\frac{v_{\phi}^2}{R}=\frac{1}{\rho}\frac{\partial P}{\partial R}+\frac{v_{\rm circ}^2}{R},
\label{eq:gasdrift1}
\end{equation}
where we have indicated with $v_{\rm circ}^2=R(\de\Phi/\de R)$ the circular velocity in the absence of pressure. If we introduce the sound speed $c_{\rm s}$ and make the simple assumption that the density $\rho$ has a power law dependence on radius, with power law index $-\beta$ (so that $\rho\propto R^{-\beta}$), we obtain:
\begin{equation}
v_{\phi}=v_{\rm circ}\left[1-\beta \left(\frac{c_{\rm s}}{v_{\rm{circ}}}\right)^2\right]^{1/2}.
\label{eq:gasdrift2}
\end{equation}
In cases where the disc is hot, in the sense that the sound speed is non negligible with respect to $v_{\rm circ}$, then pressure effects will give a sizable contribution to the rotation curve.

\subsection{Vertical structure: hydrostatic equilibrium}
\label{sec:vertical}

We now consider the vertical component of equation (\ref{eq:navier}). Here, since the velocity in the vertical direction is very small (the disc being confined to the equatorial plane), we can neglect the left-hand side of the equation altogether. The viscous force vanishes as well, since the only non-zero component of the stress is in the $R\phi$ direction. We are therefore left with just two terms to balance: gravitational force and pressure force in the vertical direction. Such a situation, where gravity is balanced by pressure, is called ``hydrostatic balance''. The equation reads:
\begin{equation}
\frac{1}{\rho}\frac{\partial P}{\partial z}=-\frac{\partial \Phi}{\partial z}.
\label{eq:hydro}
\end{equation}
We again first consider the case of non self-gravitating discs. In this case, the vertical component of gravity is simply:
\begin{equation}
-\frac{GM}{r^2}\frac{z}{r}\sim -\frac{GM}{R^2}\frac{z}{R},
\label{eq:vertgrav}
\end{equation}
where $r$ is the spherical radius and the approximation is valid for small $z$. We can then rewrite Equation (\ref{eq:hydro}) as:
\begin{equation}
\frac{c_{\rm s}^2}{\rho}\frac{\partial \rho}{\partial z}=-\frac{GMz}{R^3}.
\label{eq:vertical}
\end{equation}
The solution to this equation is straightforward if the sound speed is independent of $z$. The vertical density profile in this case turns out to be a Gaussian:
\begin{equation}
\rho(z)= \rho_0\exp\left[-\frac{GMz^2}{2R^3c_{\rm s}^2}\right]=\rho_0\exp\left[-\frac{z^2}{2H_{\rm nsg}^2}\right],
\end{equation}
where $\rho_0$ is the midplane density and we have introduced a typical vertical scale-length in the non-self-gravitating case $H_{\rm nsg}$:
\begin{equation}
H_{\rm nsg}=\frac{c_{\rm s}}{\Omega_{\rm K}}.
\end{equation}
It is also worth introducing a slightly modified scale-length\footnote{The small difference between $H_{\rm nsg}$ and $H_{\star}$ is generally neglected in many papers, and the two expressions are both commonly used to indicate the disc thickness in the non-self-gravitating case.} $H_{\star}=\sqrt{\pi/2}H_{\rm nsg}$, so that the surface density $\Sigma=\int_{-\infty}^{\infty}\rho\de z$ is related to the midplane density by $\Sigma=2\rho_0H_{\star}$. Note the simple relation between the thickness, the sound speed and the angular velocity. The disc aspect ratio is then:
\begin{equation}
\frac{H_{\rm nsg}}{R}=\frac{c_{\rm s}}{v_{\rm K}},
\end{equation}
which then demonstrates, as anticipated above, that the requirement that the disc be thin is equivalent to requiring that the disc rotation is highly supersonic, i.e. that $v_{\rm K}\gg c_{\rm s}$. 

Let us now consider the case of a disc dominated by self-gravity. In this case, we simply have to replace the gravitational force on the right hand side of equation (\ref{eq:vertical}) with the force produced by a slab of gas with surface density $\Sigma$. If the slab is radially homogeneous we have
\begin{equation}
\frac{c_{\rm s}^2}{\rho}\frac{\partial \rho}{\partial z}=-2\pi G\Sigma(z),
\end{equation}
where $\Sigma(z)=\int_{-z}^{z}\rho(z')\de z'$.
The solution of the hydrostatic balance in this case is more difficult but can be done analytically in the case of constant sound speed (such case is generally referred to as the ``self-gravitating isothermal slab'') \cite{spitzer42}. The density profile in this case is not Gaussian, but is given by:
\begin{equation}
\rho(z)=\rho_0\frac{1}{\cosh^2(z/H_{\rm sg})},
\end{equation}
where the thickness in the self-gravitating case is:
\begin{equation}
H_{\rm sg}=\frac{c_{\rm s}^2}{\pi G \Sigma},
\end{equation}
where in this case the surface density and the midplane density are related by $\Sigma=2\rho_0H_{\rm sg}$.

Now, we can ask the same question that we asked above for the radial contribution to the gravitational field. How massive has to be the disc, in order for self-gravity to significantly affect its vertical structure? Note that  the vertical component of the gravitational field produced by the disc is of the order of $2\pi G\Sigma\sim GM_{\rm disc}/R^2$. On the other hand, the vertical component of the star's gravitational field is smaller than the radial ($GM/R^2$) by a factor $H_{\rm nsg}/R$ (cf. Eq. (\ref{eq:vertgrav})) and thus \cite{pacinski78} the vertical structure of the disc is affected by self-gravity already when the disc mass is of the order of:
\begin{equation}
\frac{M_{\rm disc}}{M}\approx \frac{H_{\rm nsg}}{R}\ll 1.
\label{eq:massratio}
\end{equation}
Thus, for thin discs the vertical structure of the disc is affected by the disc self-gravity, even when the disc mass is much smaller than the central object mass. Another way of looking at the same question is to compare the two expressions for the thickness obtained above:
\begin{equation}
\frac{H_{\rm sg}}{H_{\rm nsg}}=\frac{c_{\rm s}\Omega_{\rm K}}{\pi G\Sigma}.
\label{eq:ratio}
\end{equation}
The two expressions become comparable when the dimensionless parameter on the right hand side is of order unity. It is then easy to see that this occurs when $M_{\rm disc}/M\approx H_{\rm nsg}/R$. The parameter on the right (to within factors of order unity) is nothing else than the well known $Q$ parameter that controls the development of gravitational instabilities in the disc \cite{toomre64,bertinbook2}. These will be discussed at length starting from section \ref{sec:GI} below. 

\begin{figure}
\centerline{\psfig{figure=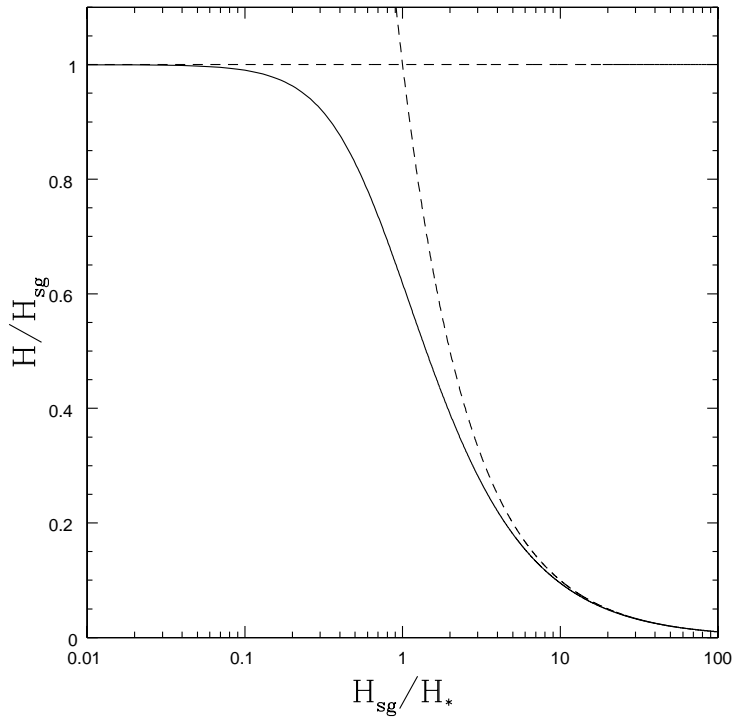,width=10cm}}
\caption{\small Disc thickness from on Eq. (\ref{eq:thickness}) as a function of $H_{\rm sg}/H_{\star}$. The two dashed lines indicate the limiting cases $H=H_{\rm sg}$ and $H=H_{\star}$.} \label{fig:thickness}
\end{figure}

It is also possible to solve the vertical hydrostatic balance equation in the combined case where both the central object and the disc contribute to the gravitational field. In this case, we have:
\begin{equation}
\frac{1}{\rho}\frac{\partial \rho}{\partial z}=-\frac{2\pi G\Sigma(z)+\Omega_{\rm K}^2z}{c_{\rm s}^2}.
\end{equation}
This equation does not have an analytical solution and has to be solved numerically. However, an approximate but accurate solution can be obtained also in this case \cite{BL99}, with the thickness given by
\begin{equation}
H=\frac{H_{\rm sg}}{2}\left(\frac{H_{\star}}{H_{\rm sg}}\right)^2\left[\sqrt{1+4\left(\frac{H_{\rm sg}}{H_{\star}}\right)^2}-1\right],
\label{eq:thickness}
\end{equation} 
which in this form holds for Keplerian rotation. Further correction terms for non-Keplerian cases can be easily incorporated in the above expression \cite{BL99}. Note that Eq. (\ref{eq:thickness}) correctly reproduces the limiting cases of fully self-gravitating discs, $H_{\rm sg}\ll H_{\star}$, where $H\approx H_{\rm sg}$ and that of non-self-gravitating disc, $H_{\star}\ll H_{\rm sg}$, where $H\approx H_{\star}$. In the following we will simply use $H$ to indicate the disc thickness. Figure \ref{fig:thickness} shows the thickness calculated from Eq. (\ref{eq:thickness}) as a function of $H_{\rm sg}/H_{\star}$, along with the two limiting cases $H=H_{\rm sg}$ and $H=H_{\star}$. Note that when $H_{\rm sg}\approx H_{\star}$ (which, as I discuss below, is a common case for self-gravitating discs), we have $H\approx 0.62 H_{\rm sg}=0.62H_{\star}$. 

\subsection{Angular momentum conservation: the issue of `anomalous' viscosity}
\label{sec:alpha}

Let us finally discuss the last component of Navier-Stokes equation, in which the viscous term plays an important role. First, we integrate Eq. (\ref{eq:navier}) in the vertical direction:
 \begin{equation}
\Sigma\left(\frac{\partial {\bf v}}{\partial t}+({\bf v\cdot \nablav}){\bf v}\right)= -(\nablav P - \nablav\cdot T)-\rho\nablav{\Phi},
\label{eq:navier2}
\end{equation}
where $T$ is the vertical integral of the stress tensor (Eq. (\ref{eq:stress})), whose only non-vanishing component is
\begin{equation}
T_{R\phi}=\nu\Sigma R\frac{\de\Omega}{\de R}.
\label{eq:stress2}
\end{equation}
In Eq. (\ref{eq:stress2}) the vertically averaged kinematic viscosity $\nu$ is defined through
\begin{equation}
\nu\Sigma={\int_{-\infty}^{\infty}\eta\de z}.
\end{equation}

Combining the $\phi$ component of Eq. (\ref{eq:navier2}) with continuity equation (Eq. (\ref{eq:continuity})) and after a little algebra, we obtain the expression for angular momentum conservation for a viscous disc:
\begin{equation}
\frac{\partial}{\partial t}\left(\Sigma Rv_{\phi}\right)+\frac{1}{R}\frac{\partial}{\partial R}(Rv_R \Sigma Rv_{\phi}) = \frac{1}{R}\frac{\partial}{\partial R}(\nu\Sigma R^3\Omega'),
\label{eq:angmom}
\end{equation}
where $\Omega'=\de\Omega/\de R$. The physical interpretation of the above expression is readily apparent. The left hand side is the Lagrangian derivative of the angular momentum per unit mass $\Sigma Rv_{\phi}$, while the right hand side is the torque exerted by viscous forces. 

With the help of the continuity equation (Eq. (\ref{eq:continuity})), we can obtain the radial velocity from equation (\ref{eq:angmom}):
\begin{equation}
v_R=\frac{1}{R\Sigma (R^2\Omega)'}\frac{\partial}{\partial R}(\nu\Sigma R^3\Omega'),
 \label{eq:angmom2}
\end{equation}
which can be inserted back in equation (\ref{eq:continuity}) to finally give:
\begin{equation}
\frac{\partial\Sigma}{\partial t}=-\frac{1}{R}\frac{\partial}{\partial R}\left[\frac{1}{(R^2\Omega)'}\frac{\partial}{\partial R}\left(\nu\Sigma R^3\Omega'\right)\right],
\label{eq:diffusion}
\end{equation}
which is the master equation that describes the evolution of an accretion disc. In general, in Eq. (\ref{eq:diffusion}), $\nu$ can depend on both radius and $\Sigma$, and even $\Omega$, for self-gravitating discs, may be a function of $\Sigma$, thus making Eq. (\ref{eq:diffusion}) a non-linear evolutionary equation. However, if $\Omega$ and $\nu$ are independent of $\Sigma$, then Eq. (\ref{eq:diffusion}) is a diffusion equation and a simple analysis of its structure shows that the disc surface density evolves on a timescale of the order of
\begin{equation}
t_{\rm visc}=\frac{R^2}{\nu}.
\end{equation}
In particular, in the important case of Keplerian rotation, the diffusion equation has the following form:
\begin{equation}
\frac{\partial\Sigma}{\partial t}=\frac{3}{R}\frac{\partial}{\partial R}\left[R^{1/2}\frac{\partial}{\partial R}\left(\nu\Sigma R^{1/2}\right)\right].
\label{eq:diffusionkeplerian}
\end{equation}

The above discussion shows the extremely important role of viscosity in the evolution of accretion discs. However, unfortunately, the ultimate physical origin of viscosity is also the major unsolved problem in accretion disc theory. As it was soon realized, standard kinetic viscosity due to collisions between gas molecules is far too low to account for the transport of angular momentum needed in observed accretion discs. In order to see this, let us consider the viscous timescale introduced above, $t_{\nu}=R^2/\nu$. It is convenient to express this quantity in units of the dynamical timescale $t_{\rm dyn}=\Omega^{-1}$:
\begin{equation}
\frac{t_{\nu}}{t_{\rm dyn}}=\frac{R^2\Omega}{\nu}.
\label{eq:reynolds}
\end{equation}
Note that the ratio above is nothing else than the Reynolds number $Re$ of the flow. Standard, collisional viscosity can be expressed as the product between the typical random velocity of molecules (that will be of order of the sound speed $c_{\rm s}$) times the collisional mean free path $\lambda=1/(n\sigma_{\rm coll})$, where $n$ is the number density of the gas and $\sigma_{\rm coll}$ is the collisional cross-section. We then have:
\begin{equation}
\lambda = \frac{1}{n\sigma_{\rm coll}}=\frac{\mu m_{\rm p}}{\rho \sigma_{\rm coll}} = \left(\frac{2\mu m_{\rm p}}{\Sigma \sigma_{\rm coll}}\right) H,
\end{equation}
where $m_{\rm p}\approx 10^{-24}$g is the proton mass, $\mu$ is the average molecular weight (we can take it to be $\approx $ 2, for simplicity) and where $\rho$ and $\Sigma$ are the usual volume and surface density of the disc, respectively. Now, substituting $\nu=\lambda c_{\rm s}$ in equation (\ref{eq:reynolds}), we get:
\begin{equation}
\frac{t_{\nu}}{t_{\rm dyn}}=\left(\frac{\Sigma\sigma_{\rm coll}}{2\mu m_{\rm p}}\right)\left(\frac{H}{R}\right)^{-2}.
\label{reynolds2}
\end{equation}
To give some ideas of the numbers involved, let us assume that the collisional cross section is simply of the order of the size of an hydrogen molecule $\sigma_{\rm coll}\approx 10^{-16}$ cm$^{2}$. In order to give a rough estimate of the disc surface density let us consider a typical protostellar disc, and assume (to be conservative) that the disc mass is $\approx 0.005 M_{\odot}$ and that the disc size is $\approx 50$ AU, which gives us $\Sigma\approx 0.005 M_{\odot}/(50 \mbox{AU})^2\approx  10 \mbox{g/cm}^2$.  Inserting this numbers in Eq. (\ref{reynolds2}), and also assuming that $H/R\approx 0.1$, we get $t_{\nu}/t_{\rm dyn}\approx 10^{11}$. Since the dynamical timescale for protostellar discs is of the order of a few years, the above estimates would lead to the conclusion that in order to accrete a disc of mass $0.005M_{\odot}$ from a distance of 50 AU, it would take much longer than the Hubble time! Clearly, the magnitude of viscosity must be much larger than the simple collisional one estimated above. Similar estimates can also be done for discs in other contexts, such as AGN and galactic binaries, where the gas is expected to be ionized and the transport coefficient will be mostly determined by plasma processes \cite{franck} 

However, the very small magnitude of collisional viscosity offers a possible way out from the apparent paradox. Indeed, as remarked above, the ratio of the viscous timescale to the dynamical one is also equal to the Reynolds number of the flow. The fact that collisional viscosity gives such a large estimate of this ratio also implies that the Reynolds number of the accretion flow is correspondingly large. Now, it is well known that for high Reynolds numbers a flow is subject to the development of turbulence and we should therefore expect that the flow in an accretion disc should be highly turbulent. In these conditions, viscosity can be much higher because in this case angular momentum is exchanged not through collisions of individual gas molecules, but by the mixing of fluid elements moving around in the disc due to turbulence. The typical length-scale of such motion can be several orders of magnitude larger than the collisional mean free path and consequently transport becomes much more effective. The question is then how to characterize such an `anomalous'  viscosity. For decades, modelers have relied on a very simple and successful parameterization of viscosity in terms of an unknown dimensionless parameter, called $\alpha$ \cite{shakura73}. The argument is very simple and draws from the fact that the stress tensor has the physical dimension of a pressure, that is a density times the square of a velocity. The simplest assumption is then to take the stress tensor to be just proportional to the (vertically integrated) pressure $\Sigma c_{\rm s}^2$:
\begin{equation}
T_{R\phi}=\frac{\de\ln\Omega}{\de\ln R}\alpha \Sigma c_{\rm s}^2,
\label{eq:alpha}
\end{equation}
where $\de\ln\Omega/\de\ln R$ ($\sim$ -3/2 for a quasi Keplerian rotation curve) is just a number (to remind us that the viscous stress is proportional to the rate of strain), and where $\alpha$ is the proportionality factor between the stress tensor and the pressure. 

Another way of expressing the $\alpha$-prescription is by considering the kinematical viscosity $\nu$. Indeed, Eq. (\ref{eq:alpha}) is equivalent to:
\begin{equation}
\nu=\alpha c_{\rm s} H.
\label{eq:alpha2}
\end{equation}
This form of the $\alpha$-prescription offers a simple way to put some constraints on $\alpha$. The magnitude of turbulent viscosity is given roughly by $\nu\sim \hat{v} l$, where $l$ is the typical size of the largest eddies in the turbulent pattern ad where $\hat{v}$ is the typical turbulent velocity. Now, it is unlikely that the turbulence will be highly supersonic, since otherwise it would be easily dissipated through shocks and we thus have $\hat{v}\lesssim c_{\rm s}$. An upper limit to the size of the largest eddies $l$ is obviously given by the disc thickness $H$ (if we consider an isotropic turbulence). These two upper limits, taken together, clearly imply that $\alpha<1$. Typically quoted values for $\alpha$, derived from comparing theoretical evolutionary models to observations range from $\alpha\approx 0.01$ \cite{hartmann98} for protostellar discs to $\alpha\approx 0.1$ for galactic binaries \cite{lasota01}, although in some more uncertain cases (such as for the outbursting protostellar systems called FU Orionis objects), the models tend to indicate a somewhat smaller value $\alpha\approx 10^{-3}$ \cite{bellin94,LC04}. 

Note that the arguments above do not necessarily imply that the stress tensor is strictly proportional to the gas pressure. Rather they merely indicate that we may expect the ratio of the two to be smaller than unity. While in many cases the parameter $\alpha$ is taken to be a constant, it obviously does not need to be so. In this respect, the $\alpha$-prescription is not a theory of viscosity in accretion discs, since it only replaces an unknown parameter, $\nu$, with another unknown parameter, $\alpha$. By using the $\alpha$-prescription we simply measure the stress in units of the local pressure, and we realize that in these units the stress should not be larger than unity, if it is driven by local turbulence. 

Modern accretion disc theory is thus based on the assumption that `anomalous' transport of angular momentum is provided by turbulent viscosity. Clearly, in order to maintain the turbulence for long enough, energy needs to be continuously injected at the largest scales. This is usually associated with the development of instabilities in the disc. Hydrodynamic, magneto-hydrodynamic and gravitational instabilities have all been considered in this respect and in recent years several numerical simulations have shown that many of the instabilities considered can indeed lead to sustained turbulence. MHD instabilities are often considered to be the most likely cause of `anomalous' viscosity. A detailed discussion of these processes can be found elsewhere \cite{balbusreview}. In the context of this paper, I will put more emphasis on the role of gravitational instabilities, and the transport of angular momentum and energy associated with gravitational instabilities will be discussed in detail below in section \ref{sec:transport}. Gravitational instabilities are the most likely alternative to MHD instabilities in providing a source of transport in accretion discs, and their role is now well established, especially in the context of protostellar discs \cite{hartmann06,clarke07,kratter07}.

Before moving on, it is worth pointing out that the term ``turbulence'' is sometime misused in astronomy, often indicating a simply fluctuating velocity field, which does not necessarily display the typical turbulent cascade from large to small scales. While in some cases simulations of hydrodynamic instabilities do produce turbulent vortices, often the structure consists of larger scale coherent disturbances. This is particularly true for the case of gravitational instabilities, that develop in the form of spiral structures and for which the term ``gravo-turbulent'', sometimes used, is probably inappropriate. 

\subsection{Simple Keplerian disc models}

\begin{figure}
\centerline{\psfig{figure=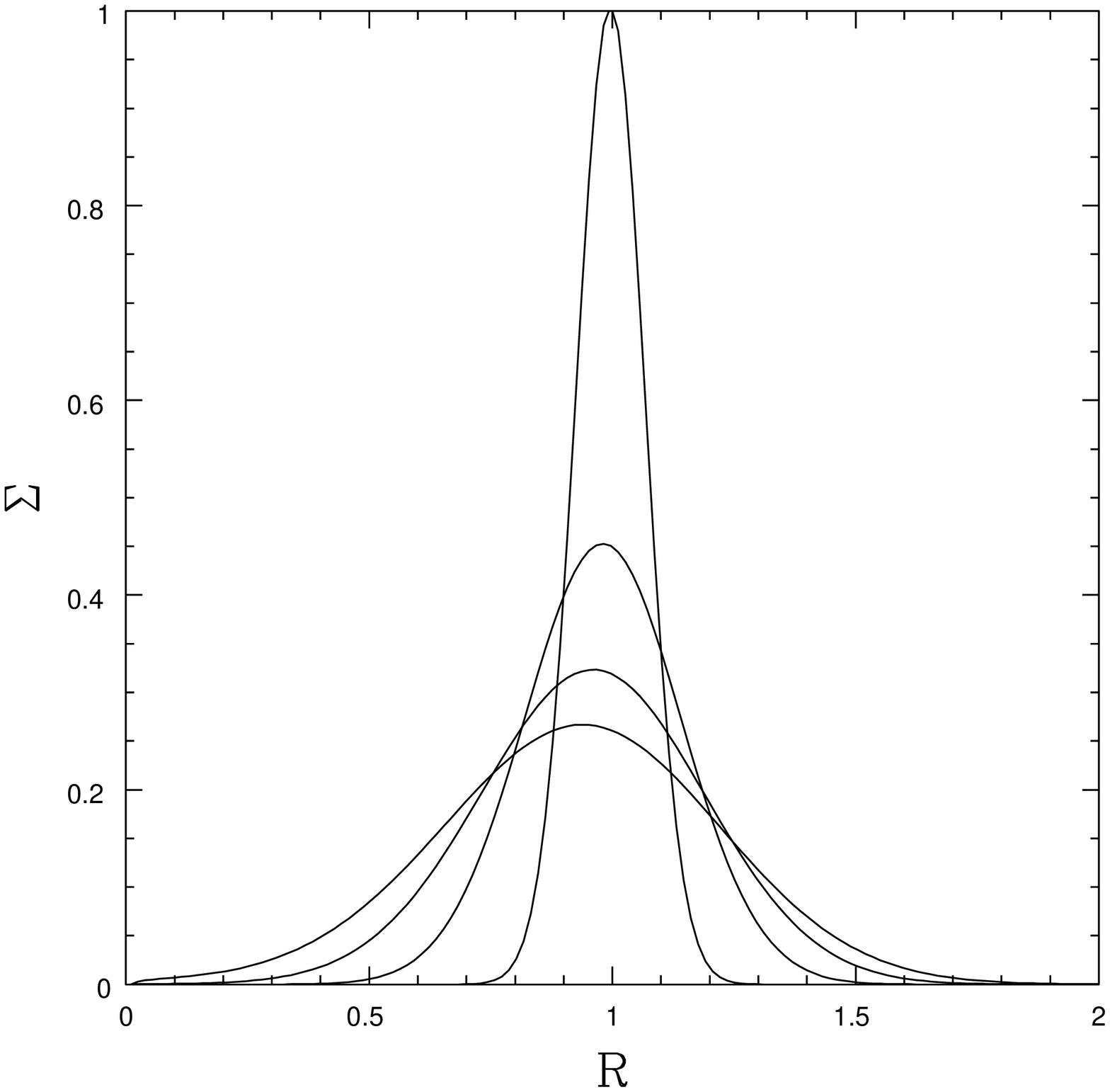,width=7cm}\psfig{figure=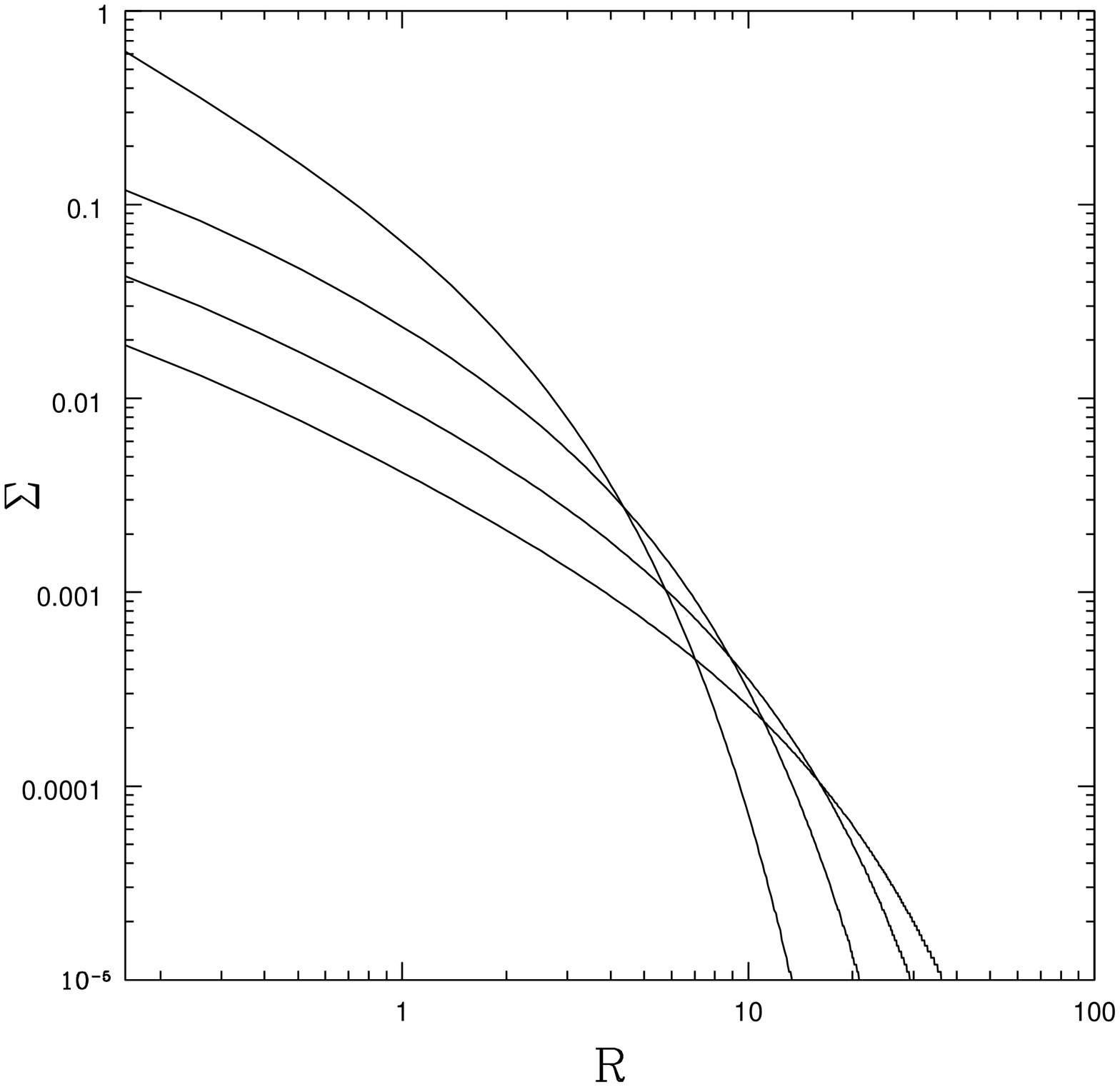,width=7cm}}
\caption{\small Left panel: evolution of the surface density of a Keplerian spreading ring, under the effect of a constant viscosity $\nu$. The lines refer (from top to bottom) to $\tau$=0.01, 0.05, 0.1 and 0.15 (see text for the definition of $\tau$). Right panel: evolution of a truncated power-law surface density, in the case where $\nu\propto R$. Units are arbitrary.} 
\label{fig:spreading}
\end{figure}

Let us see what kind of evolution is associated with the equation of motion
described above in a few simple cases. Let us first consider the case of
Keplerian rotation, described by Eq. (\ref{eq:diffusionkeplerian}). A
particularly simple but instructive configuration is when the viscosity $\nu$
is a constant, independent of radius. In this case Eq.
(\ref{eq:diffusionkeplerian}) can be solved analytically \cite{lyndenbell74}.
We take as initial condition for the surface density an infinitesimally thin
ring of mass $m$, whose shape is a $\delta$ function centered at some radius
$R_0$:
\begin{equation}
\Sigma(R,t=0) = \frac{m}{2\pi R_0}\delta(R-R_0)
\end{equation}
The solution to equation (\ref{eq:diffusionkeplerian}) in this case is:
\begin{equation}
\Sigma(x,\tau) = \frac{m}{\pi R_0^2}\frac{x^{-1/4}}{\tau}e^{-(1+x^2)/\tau}I_{1/4}\left(\frac{2x}{\tau}\right),
\end{equation}
where $I_{1/4}$ is a modified Bessel function of the first kind, $x=R/R_0$ and
$\tau=12\nu t/R_0^2$. The evolution of the surface density is shown in the
left panel of Fig. \ref{fig:spreading}, where the different lines refer (from
top to bottom) to different times $\tau$ = 0.01, 0.05, 0.1 and 0.15. This
figure illustrates quite clearly why this example is called ``the spreading
ring''. We see here that, indeed, under the action of viscous forces the disc,
rather than merely accreting, spreads both inwards ad outwards. This outward
spreading is needed in order to conserve angular momentum. Viscosity simply
acts on the fluid to redistribute angular momentum between different annuli
transporting it from the inside out, the total angular momentum being
conserved. In this way, as the inner parts of the disc lose their angular
momentum and accrete, the outer parts take up the excess angular momentum and
move outward. A detailed analysis of the analytic solution
\cite{franck,pringle81} shows that the transition between inward and outward
motion occurs at a radius of order $R_{\rm tr} \sim t\nu/R_0\sim
R_0(t/t_{\nu})$ and is therefore an increasing function of time. Therefore, at
late times, only the outermost parts of the disc move outwards, and most of
the mass is eventually accreted. For $t\rightarrow\infty$, all the mass in the
ring is accreted and the angular momentum is transported to infinitely large
radii by a negligibly small amount of mass.

Another important class of solution \cite{lyndenbell74,hartmann} is found in cases where the viscosity has a simple power-law dependence on radius, $\nu\propto R^{b}$. In this case, it is possible to find a self-similar analytical solution, where the initial density profile is such that accretion proceeds almost steadily out to a typical radius $R_1$ at which the density is exponentially truncated. An example of the evolution in this case is shown in the right-hand panel of Fig. \ref{fig:spreading}, for the case $b=1$. Also in this case, as for the spreading ring, the disc spreads significantly outwards, even if most of the mass is eventually accreted. The transition between inward and outward moving portion of the disc for the self-similar solutions occurs at a transition radius $R_{\rm tr}\sim R_1(t/t_{\nu}(R_1))^{1/(2-b)}$ and therefore increases with time (for the particular case where $b=1$ described above, we have that $R_{\rm tr}$ increases linearly with time). As the disc empties out, the accretion rate drops steadily.  

\subsection{Steady-state solutions}
\label{sec:steady}
In a steady state, the continuity equation takes the form:
\begin{equation}
\dot{M}=-2\pi Rv_R\Sigma,
\end{equation}
where the constant $\dot{M}$ is the mass accretion rate and the signs have been chosen in such a way that when $v_R$ is negative (i.e. directed inwards) the accretion rate is positive. Analogously, angular momentum conservation becomes:
\begin{equation}
\dot{M}\Omega R^2-2\pi\nu\Sigma |\Omega'| R^3 = \dot{J}
\label{eq:steadyang}
\end{equation}
where $\dot{J}$ is the constant net flux of angular momentum, and is determined by two contributions: the first term on the left hand side, which indicates the angular momentum advected with the accretion process, and the second term, which indicates the outward flux produced by viscous torques. 

$\dot{J}$ is sometimes determined by applying the so-called ``no torque'' condition, according to which at the disc inner radius $R_{\rm in}$ the angular velocity profile flattens (due to the presence of a boundary layer where the disc connects to the central object) so that its gradient and therefore the viscous torque vanish. For Keplerian rotation, this implies $\dot{J}=\dot{M}(\Omega R^2)_{\rm in} = \dot{M}\sqrt{GMR_{\rm in}}$. Inserting this in equation (\ref{eq:steadyang}), we obtain:
\begin{equation}
3\pi\nu\Sigma = \dot{M}\left(1-\sqrt{\frac{R_{\rm in}}{R}}\right).
\label{eq:nusigma}
\end{equation}

At large radii, $R\gg R_{\rm in}$, where the effects of a finite $\dot{J}$ are negligible, the surface density and the viscosity satisfy the following simple relation:
\begin{equation}
\dot{M}=\left|\frac{\de\ln\Omega}{\de\ln R}\right|2\pi\nu\Sigma,
\label{eq:nusigma2}
\end{equation}
that is, surface density and viscosity are inversely proportional to each other. An interesting consequence of the above relation arises when we use the $\alpha$-prescription in the self-gravitating case, where $H=c_{\rm s}^2/\pi G\Sigma$. Indeed, in this case Eq. (\ref{eq:nusigma2}) becomes:
\begin{equation}
\dot{M}=2\alpha\frac{c_{\rm s}^3}{G}\dln,
\end{equation}
which shows that for a self-gravitating disc the accretion rate $\dot{M}$ is only dependent on $\alpha$ and on the sound speed $c_{\rm s}$. In particular, a self-gravitating disc with a constant $\alpha$ in a steady state is approximately isothermal.

\subsection{Temperature profile and spectral energy distribution} 
\label{sec:sed}
Let us conclude the analysis of the main properties of accretion discs by considering the energetics associated with the accretion process. One of the most important consequences of accretion is the release of gravitational potential energy as the accreting matter falls into the potential well. The power per unit surface $D(R)$ dissipated by viscous stresses is given by:
\begin{equation}
D(R)=\nu\Sigma(R\Omega')^2=\dln^2\nu\Sigma\Omega^2\approx\dln\frac{\dot{M}}{2\pi}\Omega^2,
\label{eq:power}
\end{equation}
where the last equality holds approximately at large radii, where we can use Eq. (\ref{eq:nusigma2}). The dissipation rate is therefore proportional to $\Omega^2$, which can be a relatively steep decreasing function of radius. For example, if the potential is dominated by the central point mass and the rotation curve is Keplerian, we have $D(R)\propto R^{-3}$. In the other extreme condition of a purely self-gravitating disc, with a flat rotation curve, the decrease is slightly more gentle, with $D(R)\propto R^{-2}$. In fact, whenever the disc contributes to the rotation curve, it results in a gentler decrease of $\Omega$ at large radii, therefore producing a relatively larger energy dissipation, especially in the outer disc. 

\begin{figure}
\centerline{\psfig{figure=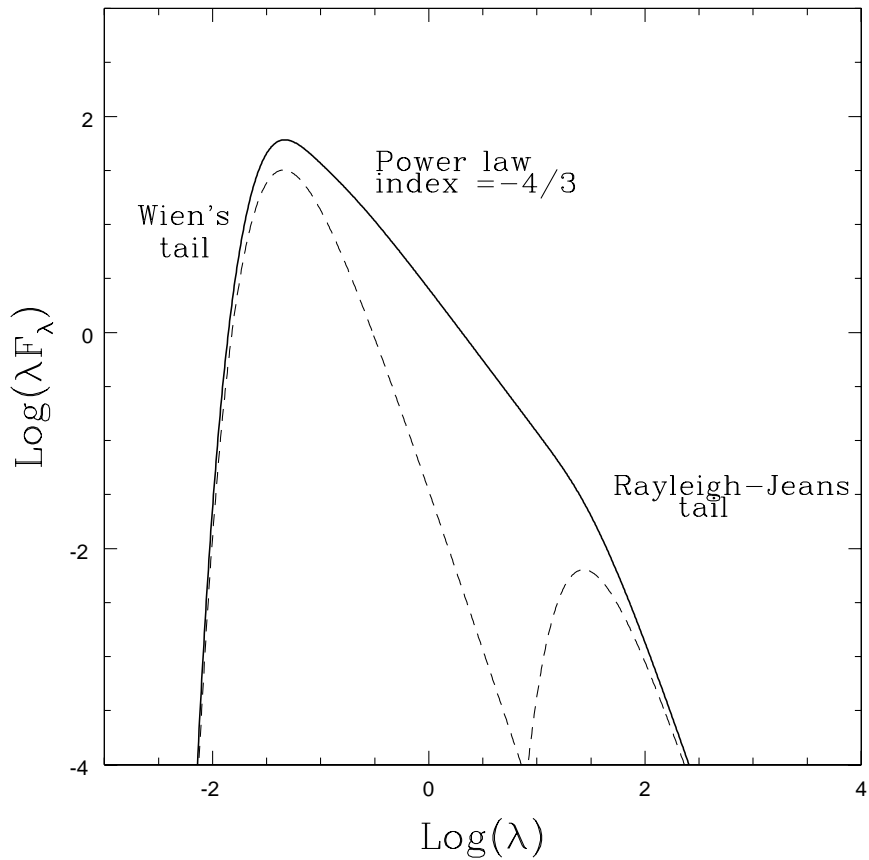,width=12cm}}
\caption{\small SED of a Keplerian optically thick disc. At short wavelength the SED approaches the Wien's tail corresponding to the temperature in the inner disc $T_{\rm in}$. At large wavelengths the SED approaches the Rayleigh-Jeans tail corresponding to the outer disc temperature $T_{\rm out}$. At intermediate wavelengths the SED is a power law with index depending on the temperature scaling with radius. For the Keplerian disc shown here the power law index is $n=-4/3$. The two dashed lines show for comparison two pure blackbody spectra corresponding to $T_{\rm out}$ and $T_{\rm in}$. Units are arbitrary.} 
\label{fig:SED}
\end{figure}

The above discussion provides a way to estimate to first approximation what is the spectral energy distribution (SED) produced by an actively accreting disc. Let us assume that the energy dissipated through viscosity is released as blackbody emission at the disc surface (which would occur if the disc is sufficiently opaque). Then the effective temperature $T_{\rm eff}$ of the emitted spectrum is simply given by $2\sigma_{\rm B}T_{\rm eff}^4$, where $\sigma_{\rm B}$ is Stefan-Boltzmann constant and where the factor 2 is introduced because the energy is emitted by the two sides of the disc. This then leads to 
\begin{equation}
T_{\rm eff}(R) = \left(\frac{\dot{M}}{4\pi\sigma_{\rm SB}}\dln\Omega^2\right)^{1/4} = \left(\frac{3GM\dot{M}}{8\pi\sigma_{\rm SB} R^3}\right)^{1/4},
\end{equation}
where the last equality holds for a Keplerian rotation curve. Each annulus of the disc thus emits as a blackbody at a different temperature. The full spectrum of the disc is then obtained by superimposing the different blackbody spectra from different annuli. If $F_{\lambda}$ is the flux emitted at wavelength $\lambda$, we thus have
\begin{equation}
F_{\lambda}= \frac{\cos i}{d^2}\int_{R_{\rm in}}^{R_{\rm out}} 2\pi R B_{\lambda}[T_{\rm s}(R)]\de R,
\end{equation}
where $d$ is the distance between the disc and the observer, $i$ is the inclination and $B_{\lambda}(T)$ is the Planck function. The shape of the SED is a `stretched' blackbody spectrum, extending over a much wider wavelength range than a simple Planck spectrum, from the shortest wavelength $\lambda_{\rm min}\approx hc/kT_{\rm in}$, associated to the temperature in the inner disc, to the longest $\lambda_{\rm max}\approx hc/kT_{\rm out}$, corresponding to the outer disc temperature. Fig. \ref{fig:SED} shows a typical SED produced by a Keplerian disc, with the two dashed lines corresponding to blackbody spectra at $T=T_{\rm min}$ and $T=T_{\rm max}$. The short wavelength part of the SED approaches the Wien's tail appropriate to the inner disc temperature $T_{\rm in}$. The long wavelength SED approaches the Rayleigh-Jeans tail appropriate to the outer disc temperature $T_{\rm out}$. At intermediate wavelengths, it can be shown \cite{lyndenbell69,adams88} that, if the temperature profile is a power law with respect to radius, with power law index $-q$, then the SED $\lambda F_{\lambda}$ is also a power law with respect to wavelength, with power law index $n=2/q-4$. We thus get the typical Keplerian accretion disc spectrum, where $q=3/4$ and thus $n=-4/3$. For a flat rotation curve disc, we have $q=1/2$, which then leads to $n=0$ and the SED is flat at intermediate wavelengths. This reflects the fact that for a flatter rotation curve the emission at large distances is enhanced, thus enhancing the long-wavelength contribution to the SED. 

It is interesting to notice that in many cases the observed SED from accretion discs (for example, in the case of protostellar discs) is indeed relatively flat, indicating a flat temperature profile. However, in most cases, the disc mass is not high enough to produce substantial deviations from Keplerian rotation in order to reproduce such flat SED, except possibly for some interesting cases (and in particular for the so called FU Orionis objects), which will be discussed in more detail in Section \ref{sec:YSO} below. The observational evidence for a relatively hot outer disc is generally interpreted in terms of an additional source of heating, coming for example from irradiation from the central star or from the inner disc. A full treatment of the energy balance in an accretion disc requires solving the radiative transfer equation through the disc, and therefore depends on the detailed optical properties of the gas. This has been done under a variety of conditions in several papers \cite{chiang97,dullemondPPV}.

Finally it is worth pointing out the relationship that is established between the viscosity parameter and the cooling time in thermal equilibrium. Indeed, in equilibrium (and assuming that there is no other source of heating for the disc) the heating rate provided by viscosity has to balance the cooling rate. The cooling rate per unit surface can be simply written as
\begin{equation}
C=\frac{\Sigma c_{\rm s}^2}{\gamma(\gamma-1)\tau_{\rm cool}},
\end{equation}
where $\gamma$ is the ratio of specific heats and where we have introduced a cooling timescale $\tau_{\rm cool}$. Equating this expression to the heating rate due to viscosity (Eq. (\ref{eq:power})) and using the $\alpha$-prescription we get
\begin{equation}
\alpha=\dln^{-2}\frac{1}{\gamma(\gamma-1)\Omega\tau_{\rm cool}}.
\label{eq:alphatcool}
\end{equation}

\section{GRAVITATIONAL INSTABILITIES AND SELF-REGULATION}
\label{sec:GI}

In the sections above, we have described the equilibrium configuration of accretion discs and their secular evolution due to viscosity. We have described what is the effect of the disc self-gravity on such equilibrium state. In particular, we have seen that for small but non negligible total disc masses $M_{\rm disc}/M\approx H/R$, self-gravity affects the disc vertical structure, while the rotation curve is only affected by self-gravity when the disc mass becomes comparable with $M$. However, one of the most important effects associated with self-gravity is the possibility of developing gravitational instabilities. The physical origin of the instability is essentially related to the standard Jeans instability for a homogeneous fluid, where large scale disturbances cannot be stabilized by pressure gradients, thus determining a typical wavelength (the so-called Jeans wavelength) associated with the instability. In a rotating disc, as I will show here below, the additional effect of rotation can also stabilize these large scale perturbations.  

As I have already remarked above, gravitational instabilities might be particularly important in providing a significant source of transport of angular momentum through the disc. Under certain conditions, the non-linear outcome of the instability can be relatively violent, leading to the fragmentation of the disc into gravitationally bound clumps. Most of the recent research on self-gravitating discs has concentrated on exploring the conditions under which the instability takes the form of a long-lived spiral structure, leading to a steady transport of angular momentum, or alternatively determines the disc fragmentation. In this section and in the following ones, I discuss in detail such issues. 

\subsection{Development of gravitational instabilities}

The dynamical processes underlying the development and the maintenance of gravitational instabilities in discs have been discussed in several papers and reviews, mostly in view of an application to the discs of spiral galaxies \cite{bertinbook2,binney}. On the other hand, many processes relevant to the dynamics of stellar discs are also applicable to the case of a fluid disc. The starting point of such analysis is the determination of the propagation properties of density waves in a self-gravitating disc, through a standard linear perturbation analysis of the relevant dynamical equations, i.e. the continuity equation (Eq. (\ref{eq:continuity})), the momentum equation for an inviscid disc (Eq. (\ref{eq:euler})), supplemented by the equation of state (Eq. (\ref{eq:state})), to relate pressure and density, and  by Poisson's equation (Eq. (\ref{eq:poisson})) in order to connect the disc density to the gravitational potential. The difficulty here arises from the long-range nature of gravitational force, so that the gravitational potential at a given location depends on the whole distribution of the gas density $\Sigma$, rather than on its local value. A local analysis can nonetheless be carried out in the WKB approximation, for tightly wound perturbations, for which the pitch angle $m/(kR)\ll 1$, where $k$ is the radial wavenumber and $m$ is the azimuthal wavenumber (corresponding to the number of arms of a spiral disturbance) . In this approximantion, the dispersion relation is:
\begin{equation}
(\omega-m\Omega)^2=c_{\rm s}^2k^2-2\pi G\Sigma|k| +\kappa^2,
\label{eq:wkb}
\end{equation}
where $\omega$ is the frequency of the perturbation and the epicyclic frequency $\kappa$ is given by
\begin{equation}
\kappa^2=\frac{2\Omega}{R}\frac{\de(\Omega R^2)}{\de R}.
\end{equation}
Note that the epicyclic frequency $\kappa$ is generally of the order of the angular velocity $\Omega$, the exact proportionality depending on the rotation curve. For example, for Keplerian rotation, we have $\kappa=\Omega$, while for a flat rotation curve, we have $\kappa=\sqrt{2}\Omega$.

A detailed discussion of the physics behind this dispersion relation can be found elsewhere \cite{bertinbook2,bertinbook}. For our purposes it suffices to describe the most salient features of it. Let us consider axisymmetric disturbances, for which $m=0$. Clearly, whenever $\omega^2$ is positive, a given perturbation simply propagates in a wave-like way. On the other hand, if $\omega^2$ is negative, an exponentially growing instability arises. We thus see from the right-hand side of Eq. (\ref{eq:wkb}) that the second term, which is negative and is associated with the disc self-gravity, is a potentially destabilizing term, while the first, associated with pressure forces, and the third, related to rotation, are stabilizing terms. In particular, while pressure stabilizes the disc at small wavelengths (large $k$), rotation is more effective at stabilizing large scale disturbances (small $k$). Equation (\ref{eq:wkb}) is a simple quadratic expression for the wavelength $k$ and it can be easily seen that for axisymmetric perturbations the right hand side is positive definite (i.e. $\omega^2$ is positive and the perturbation is stable) if:
\begin{equation}
Q=\frac{c_{\rm s}\kappa}{\pi G\Sigma}>1.
\label{eq:toomre}
\end{equation}
The dimensionless parameter $Q$ is therefore essential in determining the local axisymmetric stability of self-gravitating accretion discs (a similar parameter also determines the stability properties of stellar discs, \cite{toomre64}). We have already encountered a parameter very similar to $Q$, when discussing the relative importance of the central object and of the disc in determining the disc thickness (Eq. {\ref{eq:ratio}). We had seen that when $c_{\rm s}\Omega/\pi G\Sigma\approx 1$ then the disc self-gravity contributes to the vertical hydrostatic balance equation as much as the central object. We now see that for the very same range of parameters we also expect the disc to be close to marginal gravitational instability. Note that in the tight winding approximation $m$ does not enter explicitly in the dispersion relation (except in the Doppler shifted frequency) so that both axisymmetric perturbations ($m=0$) and non-axisymmetric ones ($m\neq 0$) share the same instability properties.

For discs that are gravitationally unstable, that is in cases where $Q\lesssim 1$ in the equilibrium state, the most unstable wavelength can be easily obtained by finding the wavenumber $k_0$ for which the right-hand side of Eq. (\ref{eq:wkb}) attains a minimum. This is given by
\begin{equation}
k_0=\frac{\pi G\Sigma}{c_{\rm s}^2}=\frac{1}{H_{\rm sg}}.
\label{eq:wavelength}
\end{equation}
We thus see that the most unstable wavelength is of the order of the disc thickness. The condition for the WKB approximation to hold is then $kR=R/H_{\rm sg}\gg 1$ and it is satisfied for marginally stable discs if the disc is thin, or equivalently if the disc mass is much smaller than the central object mass, $M_{\rm disc}\ll M$ (cf. Eq. (\ref{eq:massratio})). 

It is well known (for example, from early $N$-body numerical simulations \cite{hohl71,ostriker73}) that massive discs are subject to violent large scale (long-wavelength) non-axisymmetric instabilities, even in cases where the parameter $Q$ is above unity, and therefore the disc is predicted to be stable in the WKB approximation. Indeed, a local dispersion relation can also be obtained in the case of relatively long radial wavelengths in the non-axisymmetric case \cite{laubertin78}. This relation is more complicated than the simple quadratic form described above and depends on a new parameter, in addition to $Q$, defined as:
\begin{equation}
J=m\frac{\pi G\Sigma}{R\kappa^2}\frac{4\Omega}{\kappa}\dln^{1/2}.
\end{equation}
Being proportional to $\Sigma$, $J$ is a measure of how massive the disc is. Indeed, to give an idea of the numbers involved, consider a quasi-Keplerian disc, where $\kappa\approx \Omega$ and where $\Sigma\propto R^{-1}$. In this case we have $J=\sqrt{6}mM_{\rm disc}(R)/M$. If $J\ll 1$ (light disc or zero $m$), the dispersion relation reduces to the standard quadratic form. On the other hand, for $J\approx 1$, the perturbations are significantly more unstable. A detailed description of this regime can be found elsewhere \cite{bertin89,bertinbook2}. In particular, when the disc mass is a sizable fraction of that of the central object, low $m$ disturbances can be violently unstable. Interestingly, as I will discuss below, this change of behaviour between light and massive discs has been observed numerically in a recent survey of the development of gravitational instabilities in fluid discs \cite{LR04,LR05}.

To summarize, in the tightly wound approximation we expect a gaseous disc to become gravitationally unstable when $Q<1$. Non-axisymmetric disturbances, with large azimuthal wavenumber $m$ can be unstable even for relatively larger value of $Q$, even though if $Q$ rises above $\approx 2$, then even non-axisymmetric disturbances are stabilized. Finally, it is worth noting that taking into account finite-thickness effects in the tightly wound approximation leads to a stabilization of the disc, so that the marginal stability value of $Q$ is reduced below unity \cite{vandervoort70}.

\subsection{Self-regulation}

Having established that massive discs are linearly unstable, we should now ask what is the non-linear evolution of the instability and whether it can lead to a sustained transport of angular momentum. The stability criterion, Eq. (\ref{eq:toomre}) offers a natural way to describe this. Indeed, we should note that the stability parameter $Q$ is proportional to the sound speed (and hence to temperature), so that colder discs are more unstable. Now, let us consider a disc which is initially hot, so that $Q\gg 1$ and the disc is stable. In the absence of other transport and heating mechanisms, the disc cools down due to radiative cooling until eventually $Q\approx 1$. At this stage, the disc develops a gravitational instability in the form of a spiral structure. Compression and shocks induced by the instability lead to an efficient energy dissipation and as a result the disc will heat up. In turn, the stability condition works like a kind of `thermostat', so that heating turns on only if the disc is colder than a given temperature. If the `thermostat' works, we should then expect the instability to self-regulate in such a way that the disc is always kept close to marginal stability, and thus a self-gravitating disc will evolve to a state where $Q\approx \bar{Q}$, where $\bar{Q}$ is a constant of order unity. Here again, it is reassuring to see that numerical simulations \cite{LR04} reproduce closely this behaviour. 

Note here an important difference between a stellar and a fluid disc. In order for the above self-regulation process to work, the presence of a dissipative component is essential. While this is obviously the case for a gaseous disc, a pure stellar disc lacks this dissipation channel. For a galaxy disc, therefore, self-regulation requires the presence of a gaseous component in addition to the dominant stellar one \cite{bertin88}. 

\subsection{Self-regulated accretion discs models}
\label{sec:selfregulation}

The self-regulation mechanism described above can be easily introduced in steady state models of self-gravitating accretion discs \cite{bertin97}. Indeed, for a self-gravitating disc, as shown in Section \ref{sec:steady} at large radii we have:
\begin{equation}
\dot{M}=\frac{2\alpha c_{\rm s}^3}{G}\dln,
\end{equation}
where we have also used the $\alpha$-prescription. If we supplement this equation by the self-regulation condition:
\begin{equation}
Q=\frac{c_{\rm s}\kappa}{\pi G\Sigma}=\bar{Q}\approx 1,
\end{equation}
we thus easily see that in the case of a disc dominated potential (that is, if the central star mass is negligible) a Mestel disc with $\Sigma\propto R^{-1}$ provides an analytical self-similar solution to the model. In particular, we have:
\begin{equation}
c_{\rm s}=\left(\frac{G\dot{M}}{2\alpha}\right)^{1/3},
\end{equation}
\begin{equation}
v_{\rm c}=\frac{2\sqrt{2}}{\bar{Q}}\left(\frac{G\dot{M}}{2\alpha}\right)^{1/3},
\end{equation}
\begin{equation}
2\pi G\Sigma R = \frac{8}{\bar{Q}^2}\left(\frac{G\dot{M}}{2\alpha}\right)^{2/3}.
\end{equation}
The solution above has remarkably simple properties, with a flat profile of $v_{\rm c}$, $c_{\rm s}$ and $Q$ and where the surface density has a simple power-law dependence on radius. However, it is clearly approximate and only applies to large radii, where the effects of $\dot{J}$ and especially of the presence of a central point mass $M$ can be neglected. In particular, in realistic accretion discs the latter condition is difficult to achieve and we should therefore extend the analysis to include a central mass in the model. In particular, we expect the transition from an inner Keplerian disc to an outer disc, where the rotation curve would tend to flatten, to occur at a distance of the order of
\begin{equation}
R_{\rm s}=2GM\left(\frac{\bar{Q}}{4}\right)^2\left(\frac{G\dot{M}}{2\alpha}\right)^{-2/3}.
\label{eq:rs}
\end{equation}
In this way we can use the simple model above to check the influence of the disc self-gravity on the rotation curve of some observed systems. Let us then estimate the transition radius $R_{\rm s}$ for a number of different cases. We first consider a typical protostellar disc, for which $\dot{M}\approx 10^{-8}\msunyr$, the central star mass is of the order of $1M_{\odot}$ and where we assume $\alpha=0.01$. Inserting these numbers in Eq. (\ref{eq:rs}), we get $R_{\rm s}\approx 7000$AU. We thus see that in order for self-gravity to be important, the disc should extend out to very large distances, much larger than the observed protostellar disc sizes. Typical protostellar discs should therefore not be significantly non-Keplerian. On the other hand, as already mentioned above, in some cases the accretion rate of young protostellar discs can be significantly larger. For example, during the so-called FU Orionis outbursts (see below Section \ref{sec:YSO} for more details) the accretion rate can reach $\dot{M}\approx 10^{-4}\msunyr$. In such cases, $R_{\rm s}$ is reduced to $\approx 10-20$ AU, so that the outer parts of the disc might shown some deviation from Keplerian rotation. Finally, let us consider a typical disc in an Active Galactic Nucleus, for which $\dot{M}\approx 0.1\msunyr$ and $M\approx 10^7M_{\odot}$. In this case, we get $R_{\rm s}\approx 5$pc. Interestingly, as mentioned in the Introduction, water maser emission from AGN discs has been observed exactly at such distances and we therefore expect that at least some of them will display some degree of non-Keplerian rotation.

\begin{figure}
\centerline{\psfig{figure=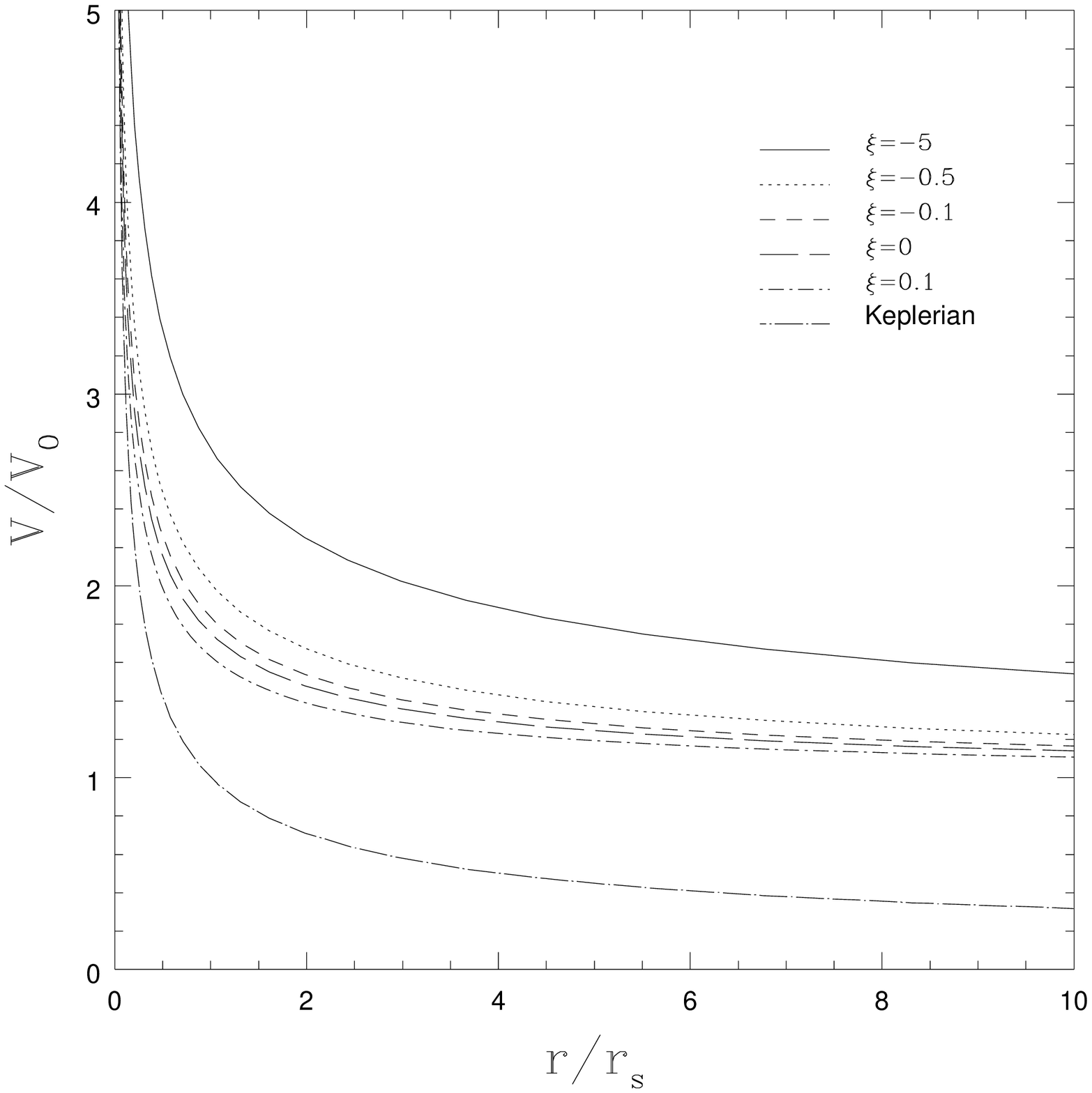,width=7.cm}\psfig{figure=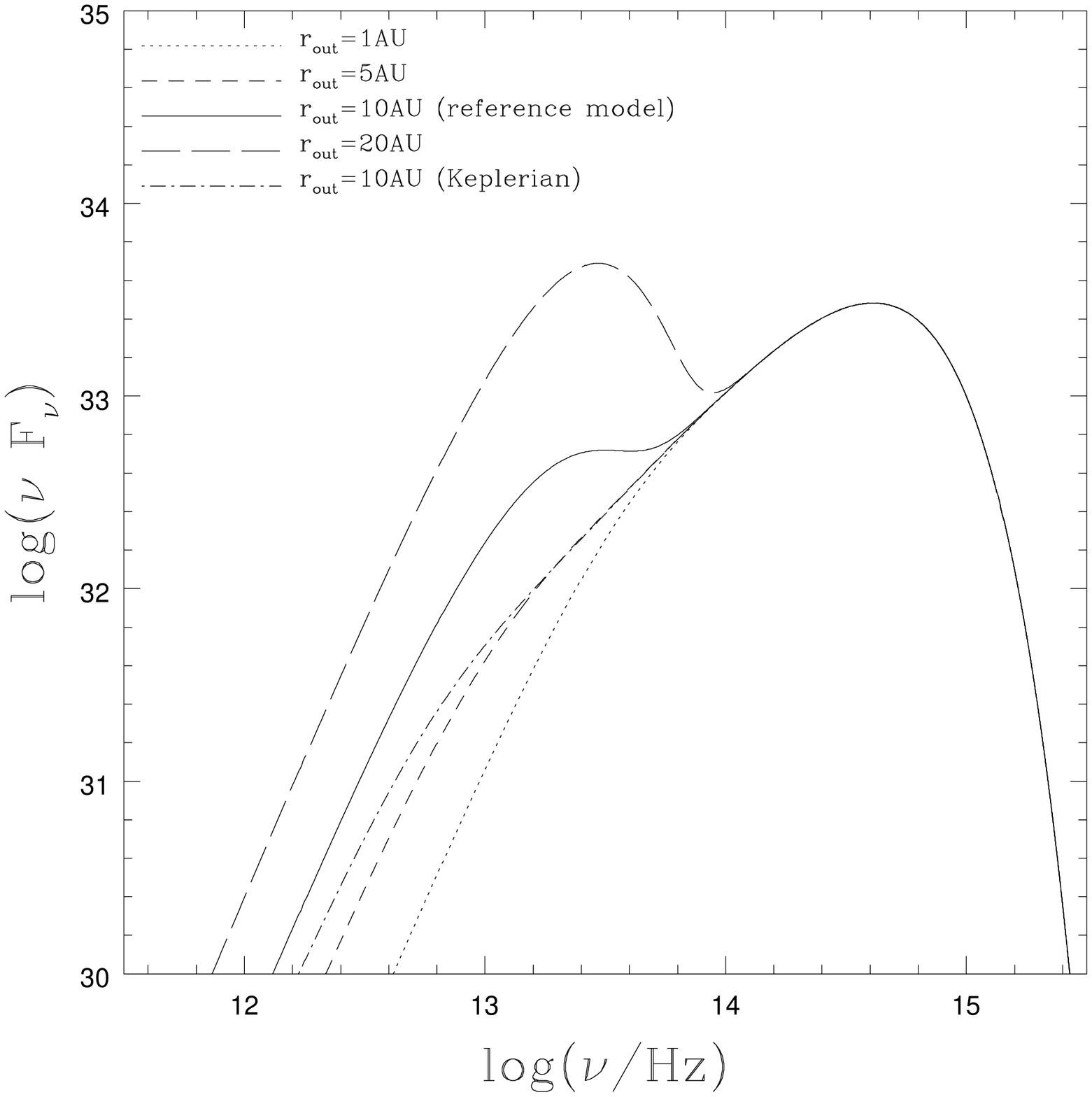,width=7.cm}}
\caption{\small Left: examples of rotation curves for self-regulated disc models in the presence of a central point mass. In the inner region the rotation curve is Keplerian, while it becomes progressively flatter in the outer disc, asymptotically reaching the self-similar configuration. More details can be found in \cite{BL99}. Right: example SEDs for self-regulated discs. At large radii self-regulated discs are hotter than their non-self-gravitating counterparts, resulting in a secondary bump in the SED at long wavelengths \cite{LB2001}, which becomes more pronounced if the disc outer radius is increased.} 
\label{fig:selfregulated}
\end{figure}

In order to compare these self-regulated models to observations, it is then important to extend the analytical self-similar solution to the more general case where it connects to an inner Keplerian disc. Such models have been constructed \cite{BL99} and some examples of rotation curves obtained in this more general case can be seen in Fig. \ref{fig:selfregulated}. Such models can be used for a detailed comparison with observed rotation curves, as will be discussed in Section \ref{sec:AGN} below. Additionally, it is worth mentioning that numerical simulations which try to reproduce the formation of protostellar discs \cite{vorobyov07}, tend to be characterized by fairly massive discs in their earliest stages, with properties that resemble closely the ones described semi-analytically by these models. 

Self-regulated disc models also tend to be relatively hotter in their outer regions, compared to their non-self-gravitating counterparts. This is due both to the self-regulation mechanism, and the effective heating associated with it, and by the fact that the disc has a flatter rotation curve in its outer parts, as discussed in Section \ref{sec:sed}. As a consequence, self-regulated SEDs tend to be flatter than the standard disc SED, shown in Fig. \ref{fig:SED}. and often display a secondary ``bump'' at long wavelengths \cite{LB2001}. An example SED is shown in Fig. \ref{fig:selfregulated}, where the different curves refer to different choices of the disc outer radius (see \cite{LB2001} for details). These models were originally proposed in order to explain some observations of protostellar discs, and will be discussed more in detail in section \ref{sec:YSO} below. Interestingly, very similar SEDs are also produced in numerical simulations of self-gravitating protostellar discs \cite{boley06}, and similar semi-analytic models for self-gravitating disc SEDs have later been proposed also for AGN discs \cite{sirko03}.

\section{NUMERICAL SIMULATIONS OF SELF-GRAVITATING ACCRETION DISCS}
\label{sec:simulations}

In the sections above we have discussed the development of gravitational instabilities essentially in terms of linear perturbation analysis. We have already noted that the non-linear evolution of such instabilities can result in a feedback loop which self-regulates the disc and keeps it close to marginal stability. It is then possible, as we have seen, to develop analytical models where such complex behaviour is introduced in a simple way in the basic steady state disc equations. Clearly, in order to check whether self-regulation is indeed established once the instability becomes non-linear, and to investigate secular effects, like the transport of angular momentum induced by the spiral structure, it is important to perform numerical simulations of the disc well into the non-linear regime. Many simulations of this kind were performed in the early '70s in the context of collisionless stellar systems by means of $N$-body techniques \cite{hohl71,ostriker73,miller70,hohl73}. 

Here, we are interested in the description of a dissipative, gaseous disc and thus these early $N$-body simulations were not be able to catch some aspects of the disc hydrodynamics. In the last fifteen years complex, three-dimensional hydrodynamic simulations have become an essential tool to study accretion discs dynamics and have provided an important `experimental' counterpart to analytical investigations, such as those described in the previous sections. In this respect, a set of simulations should be considered as a `numerical experiment' where our models are checked against measurements of a controlled system, often unavailable from astronomical observations. Simulations have evolved significantly throughout the years, partly due to the increased computational power (which has enable to reach higher resolution and accuracy), and partly to the development of progressively more refined algorithms to include new physical processes in the models. 

A key role in the evolution of self-gravitating accretion discs is played by the cooling process. As discussed above, it is the interplay between cooling and internal heating due to gravitational instability that allows to reach a self-regulated condition. As will be discussed in more detail in section \ref{sec:transport} below, the cooling rate also determines the efficiency with which the gravitational instability redistributes angular momentum and allows accretion. Most of the effort in this area of research is therefore aimed at introducing into the codes progressively more realistic implementations of  the cooling processes in accretion discs. In this section we provide an overview of the development of numerical simulations of accretion discs in the last few years. Specific results concerning the transport induced by gravitational instability and the possibility of disc fragmentation are discussed in Sections \ref{sec:transport} and \ref{sec:fragmentation} below.

Some early attempts at simulating self-gravitating accretion discs were made by using collisionless $N$-body methods \cite{anthony83}. However, the earliest hydrodynamical simulations of self-gravitating accretion discs were performed later \cite{laughlin94} in the context of the formation of protostellar discs. Such simulations employed a very simplified thermodynamical description of the gas, that was considered to be `locally isothermal' in the sense that every parcel of fluid was assumed to keep its internal energy throughout the simulation. Several other early simulations were performed using different equations of state, such as a polytropic equation \cite{laughlin96,laughlin98} and were used to describe the development of gravitational instability, but the effects of the choice of thermal state of the gas were generally overlooked. More emphasis on thermodynamical aspects was placed later \cite{pickett98,pickett2000}, when a comparison between models evolved isentropically and models which enforced an isothermal behaviour revealed that gravitational instabilities developed much more violently in an isothermal configuration, leading to a very fast redistribution of angular momentum and in extreme cases to the disruption of the disc. A posteriori, this behaviour is simply understood based on the arguments described in Section \ref{sec:selfregulation} above. The assumption of isothermality prevents the development of the instability to feed back thermal energy in the equilibrium state, therefore not allowing the non-linear stabilization of the disturbance and the saturation of the gravitationally unstable modes. 

All the simulations described so far start from an initial configuration which is expected to be highly unstable, with $Q$ values of the order of unity or below. It is therefore not surprising that as soon as the simulation is set to evolve, strong disturbances suddenly appear and grow out of control in the case of isothermal simulations. The next question is therefore what would be the response of a disc which is initially stable and is driven to an unstable configuration through cooling. Clearly, in order to consider this case, one has to move away from either an isothermal or a polytropic equation of state and consider instead the full evolution of the energy equation. This was initially done by considering a simple cooling prescription \cite{rice03a,rice03b,gammie01,rice03c,LR04,LR05,RLA05,pickett03,mejia05,mayer04} where the cooling rate is given by:
\begin{equation}
\left.\frac{\de u}{\de t}\right|_{\rm cool}=-\frac{u}{\tau_{\rm cool}},
\end{equation}
where $u$ is the internal energy of the fluid and $\tau_{\rm cool}$ is the cooling timescale, which is set as a free parameter in the simulations. It is important to stress that the above description is not meant to reproduce any specific cooling law, but is just a convenient way (a `toy' model) of exploring the role of the cooling timescale in the outcome of the gravitational instability. Different investigators use different prescriptions for the specific form of the cooling time, which in some cases is taken to be a constant while in other cases is taken to be proportional to the dynamical timescale, so that $\Omega \tau_{\rm cool}$ is kept constant. Despite the different prescriptions for the cooling time and the different type of codes used (local shearing sheet models \cite{gammie01}, global Eulerian grid-based models \cite{pickett03,mejia05}, global Lagrangian particle-based models \cite{LR04,LR05,rice03a,rice03b,rice03c,RLA05,mayer04}) the results appear to converge into an essentially coherent picture, at least for low mass discs. It is now well established that the long term behaviour depends on the relative values of the cooling and dynamical timescale. If the cooling timescale is larger than a few dynamical timescales, an initially stable (large $Q$) disc cools down until $Q$ becomes of the order of unity. At this stage, the disc becomes gravitationally unstable and develops a spiral structure which provides a heating source, through compressional heating and shock dissipation, able to balance the externally imposed cooling. Once in thermal equilibrium, the disc is characterized by an approximately constant value of $Q$ very close to marginal stability. In such a state a spiral structure persists in the disc, to provide the required heating. Therefore the self-regulation mechanism described above determines the disc structure and evolution. Figure \ref{fig:simulation} shows the result of one such simulations, where in this case $\tau_{\rm cool}\Omega=7.5$ and the total disc mass $M_{\rm disc}=0.1M$ \cite{LR04}. In the left panel the colour plot shows the disc surface density, in which a spiral structure is clearly seen. The right panel shows the azimuthally and vertically averaged value of $Q$ as a function of radius. The disc in this case extends from $R=0.25$ to $R=25$ in code units. It is then seen that far from the boundaries (where the density drops and $Q$ correspondingly grows) the disc is self-regulated, with $Q\approx 1$ over a wide radial range. The three curves refer to three different times during the evolution, showing that the disc has reached thermal equilibrium.

\begin{figure}
\centerline{\psfig{figure=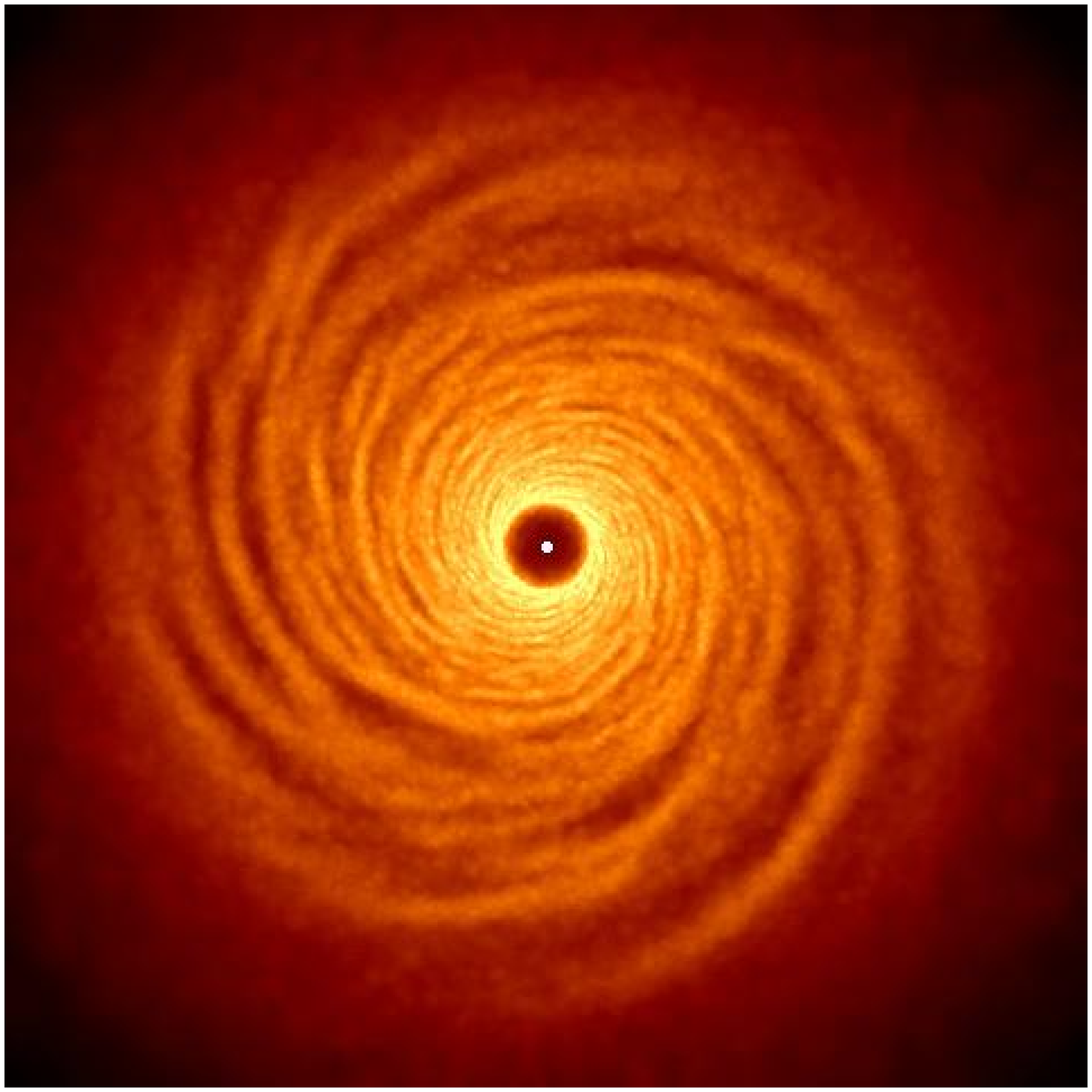,width=7.1cm}\psfig{figure=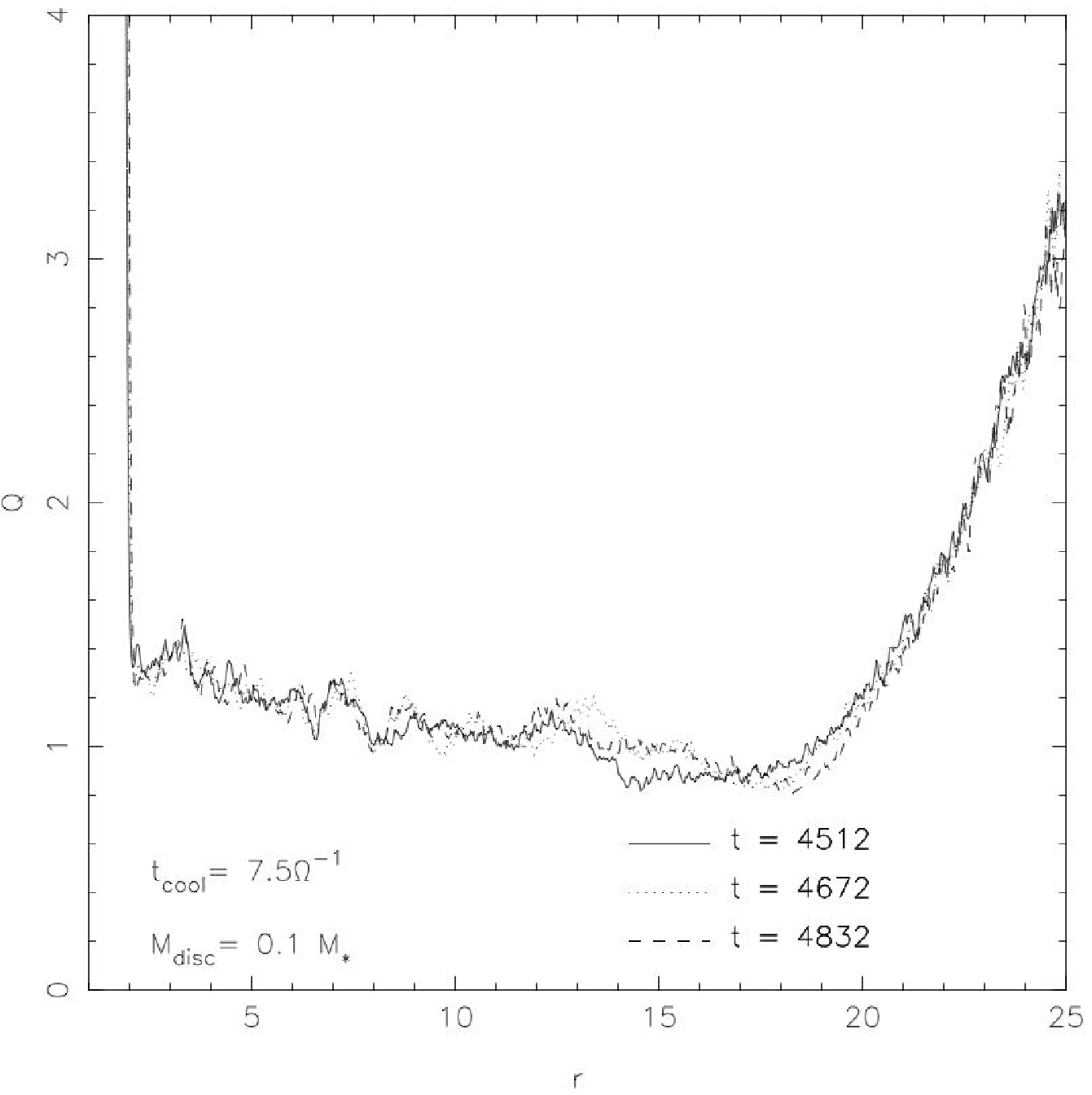,width=7.5cm}}
\caption{\small Numerical simulation of the development of a gravitational instability in a cooled disc with $\tau_{\rm cool}\Omega=7.5$ and where the disc mass $M_{\rm disc}=0.1M$ \cite{LR04}. Left: Surface density of the disc, showing the prominent spiral structure. Right: radial profile of the azimuthal and vertical average of $Q$. The disc is self-regulated over a wide radial range. The three curves refer to three different times during the evolution, showing that the disc has reached thermal equilibrium.} 
\label{fig:simulation}
\end{figure}

\begin{figure}
\centerline{\psfig{figure=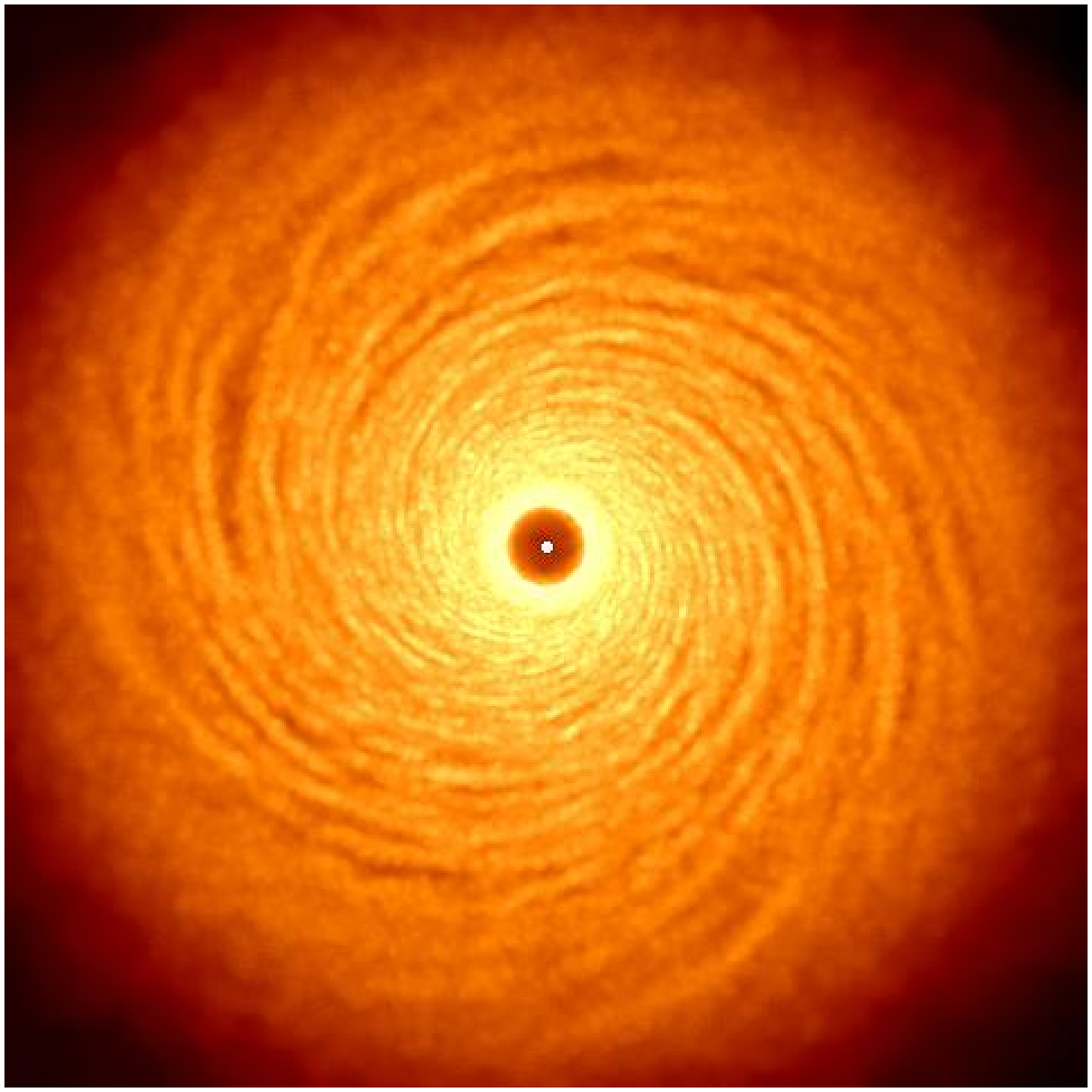,width=4.6cm}\psfig{figure=Md01image_2.ps,width=4.6cm}\psfig{figure=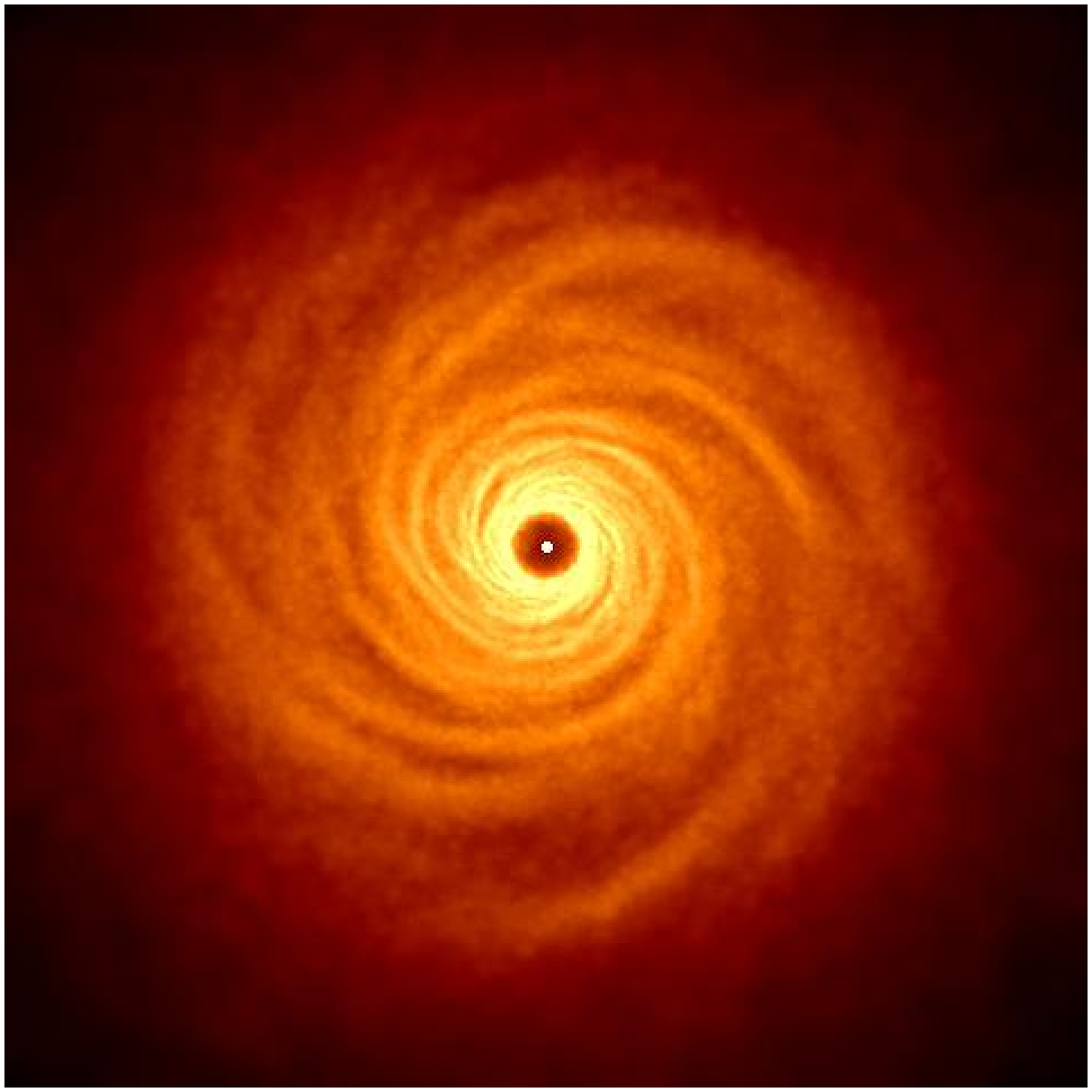,width=4.6cm}}
\centerline{\psfig{figure=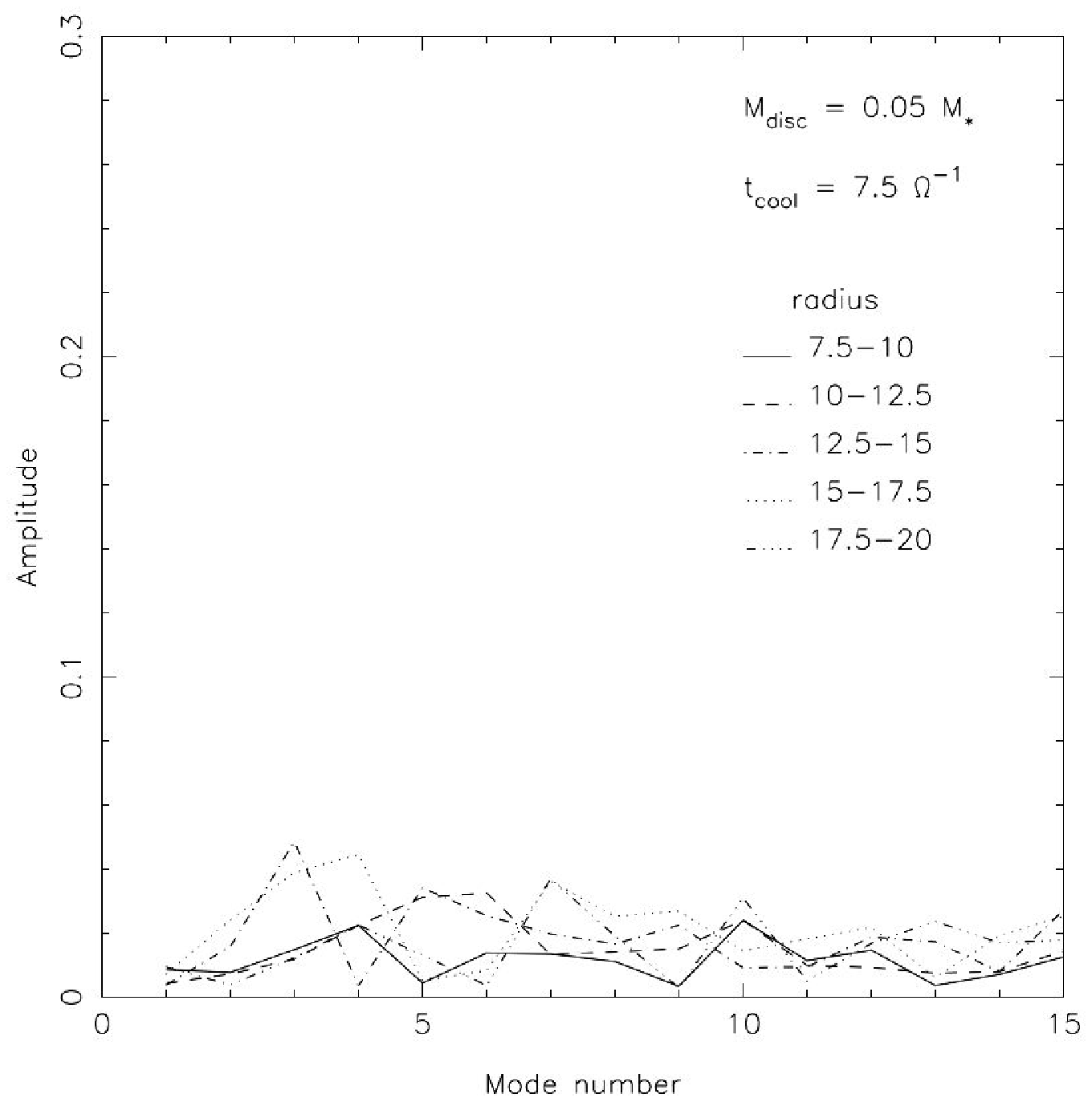,width=5cm}\psfig{figure=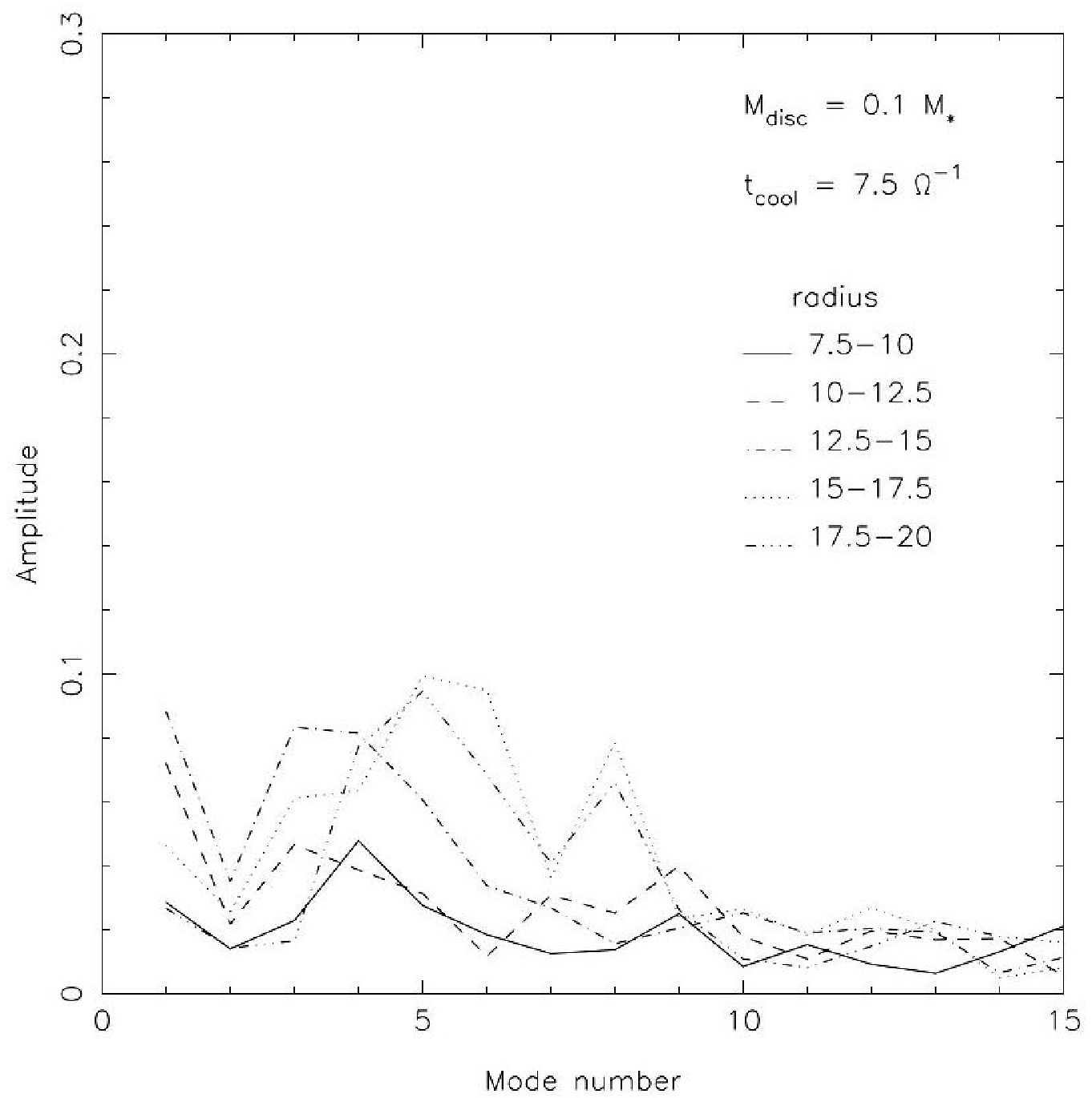,width=5cm}\psfig{figure=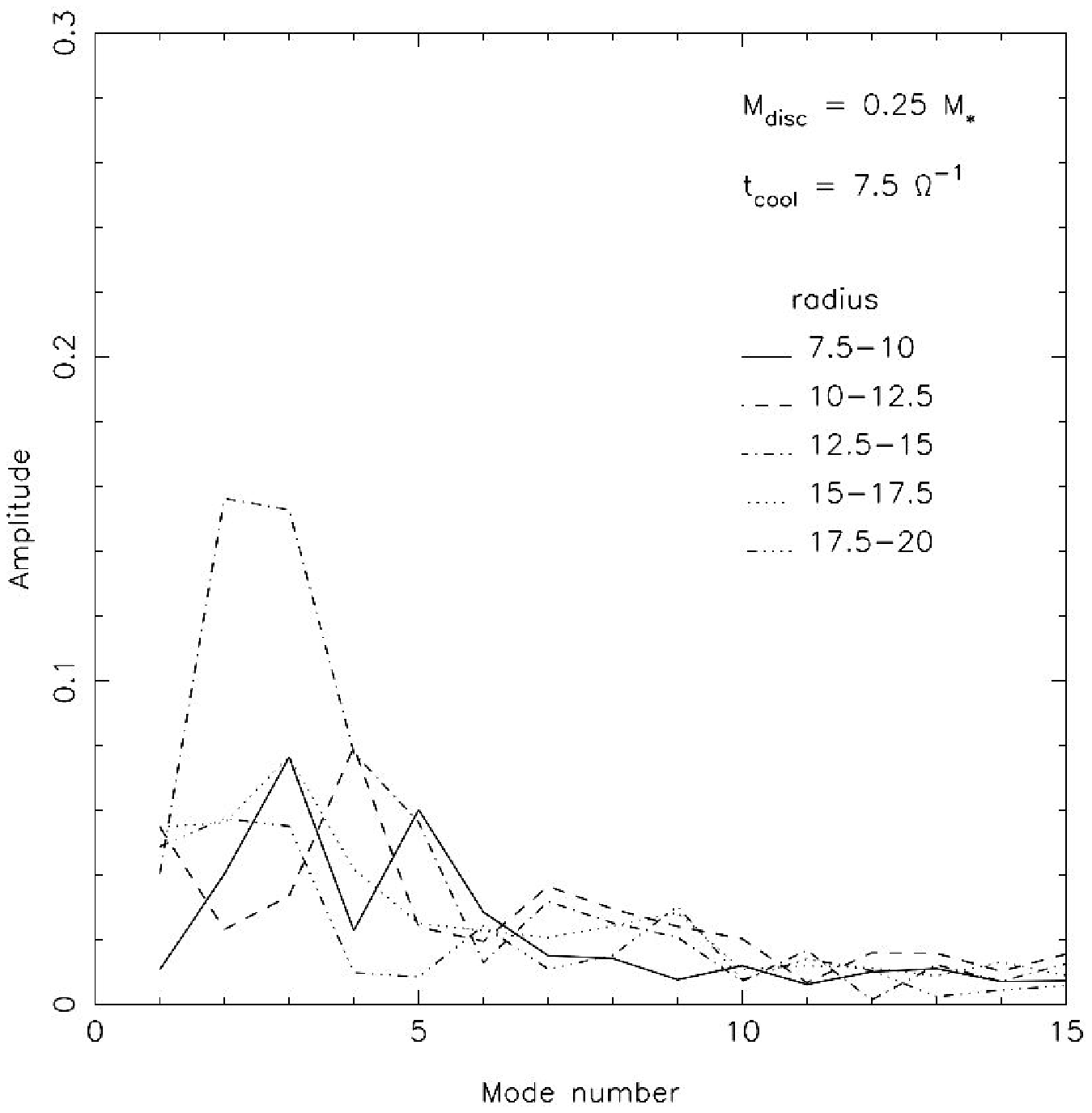,width=5cm}}
\caption{\small Top panels: Surface density structure for a set of self-gravitating disc simulations with $\tau_{\rm cool}\Omega=7.5$. From left to right the mass ratio is $M_{\rm disc}/M=0.05$, 0.1 and 0.25. Lower panels: corresponding mode analysis. The amplitude of various modes with different $m$ is shown.} 
\label{fig:modes}
\end{figure}

\begin{figure}
\centerline{\psfig{figure=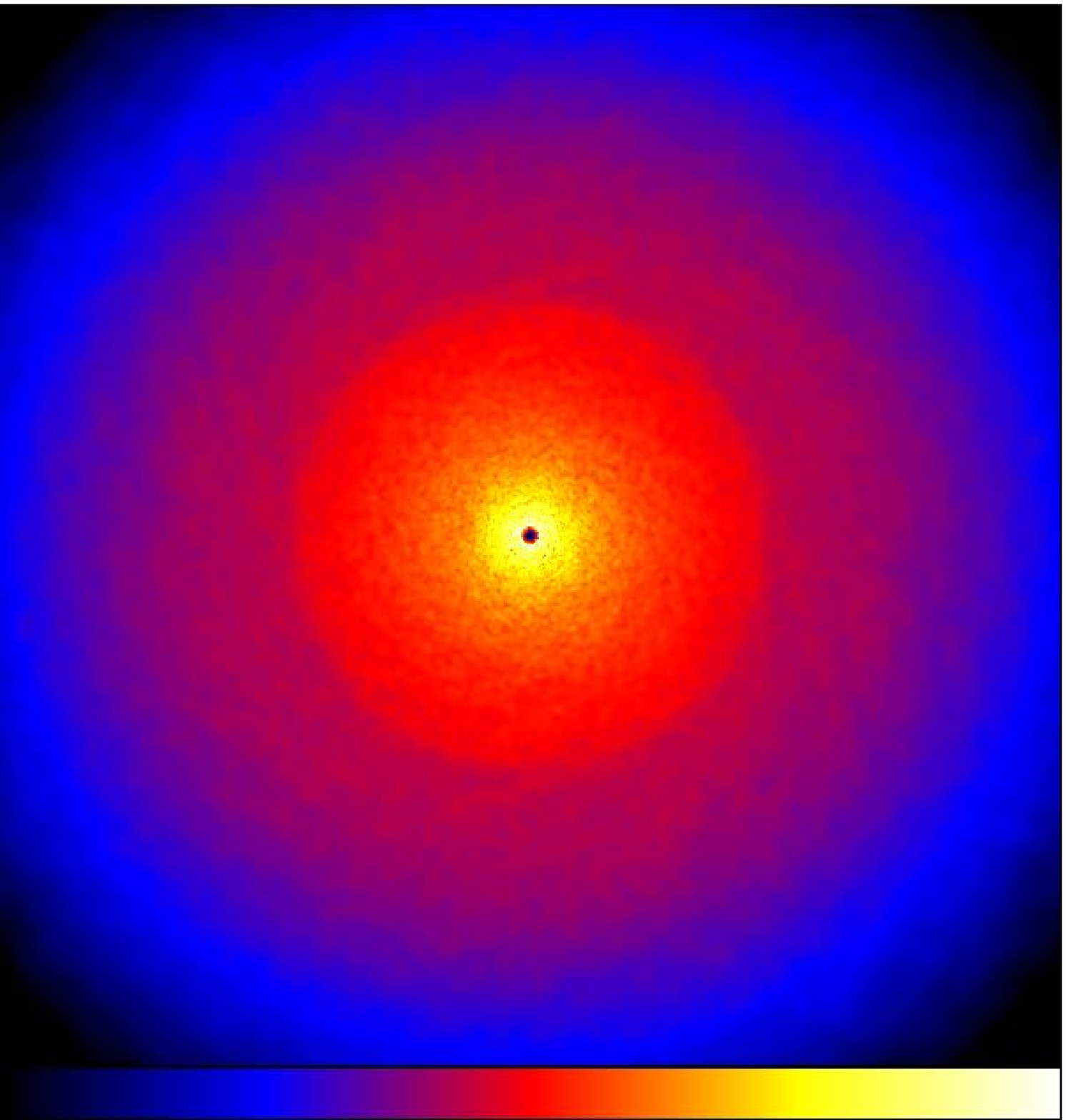,width=6.5cm}\psfig{figure=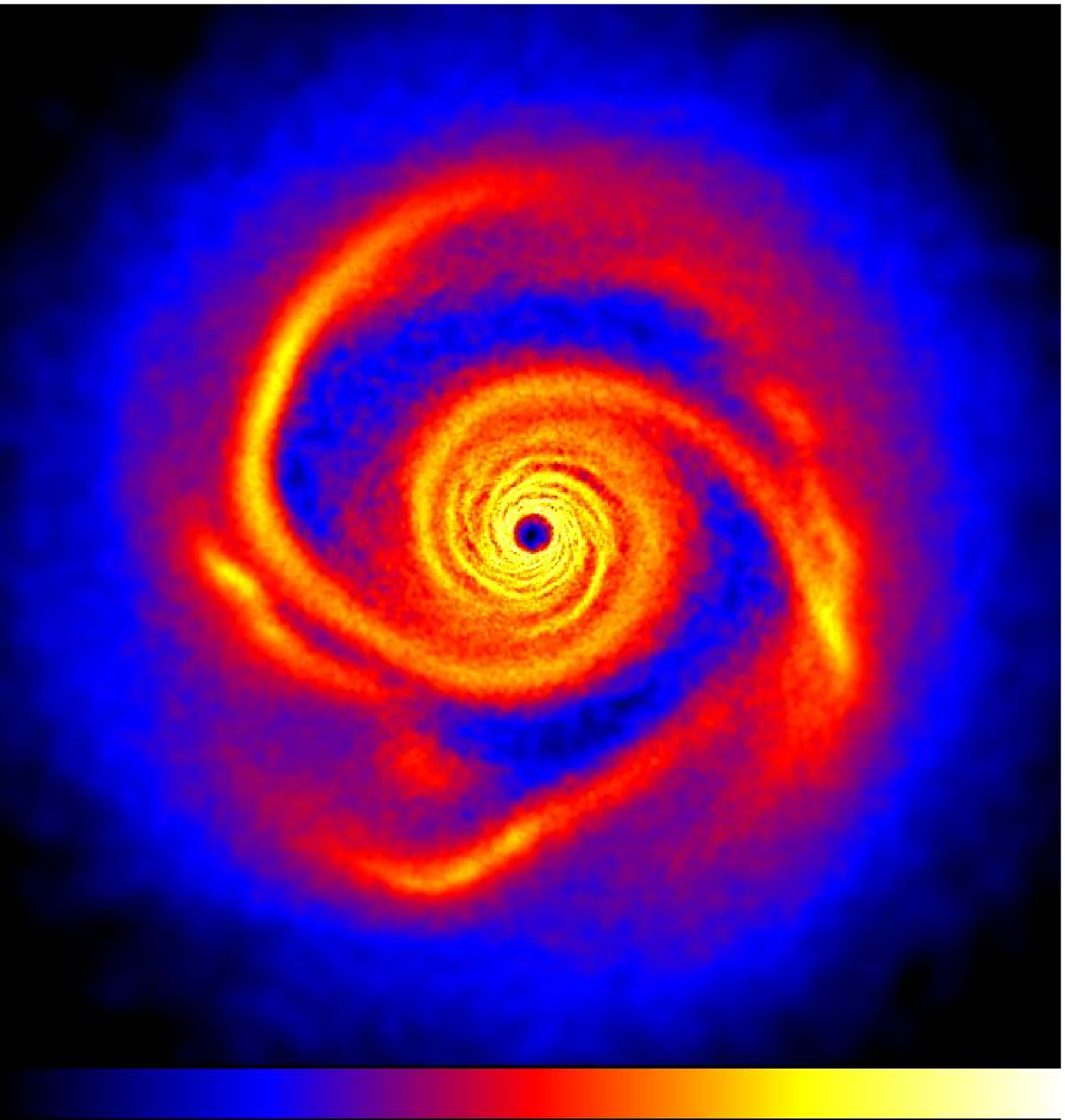,width=6.5cm}}
\centerline{\psfig{figure=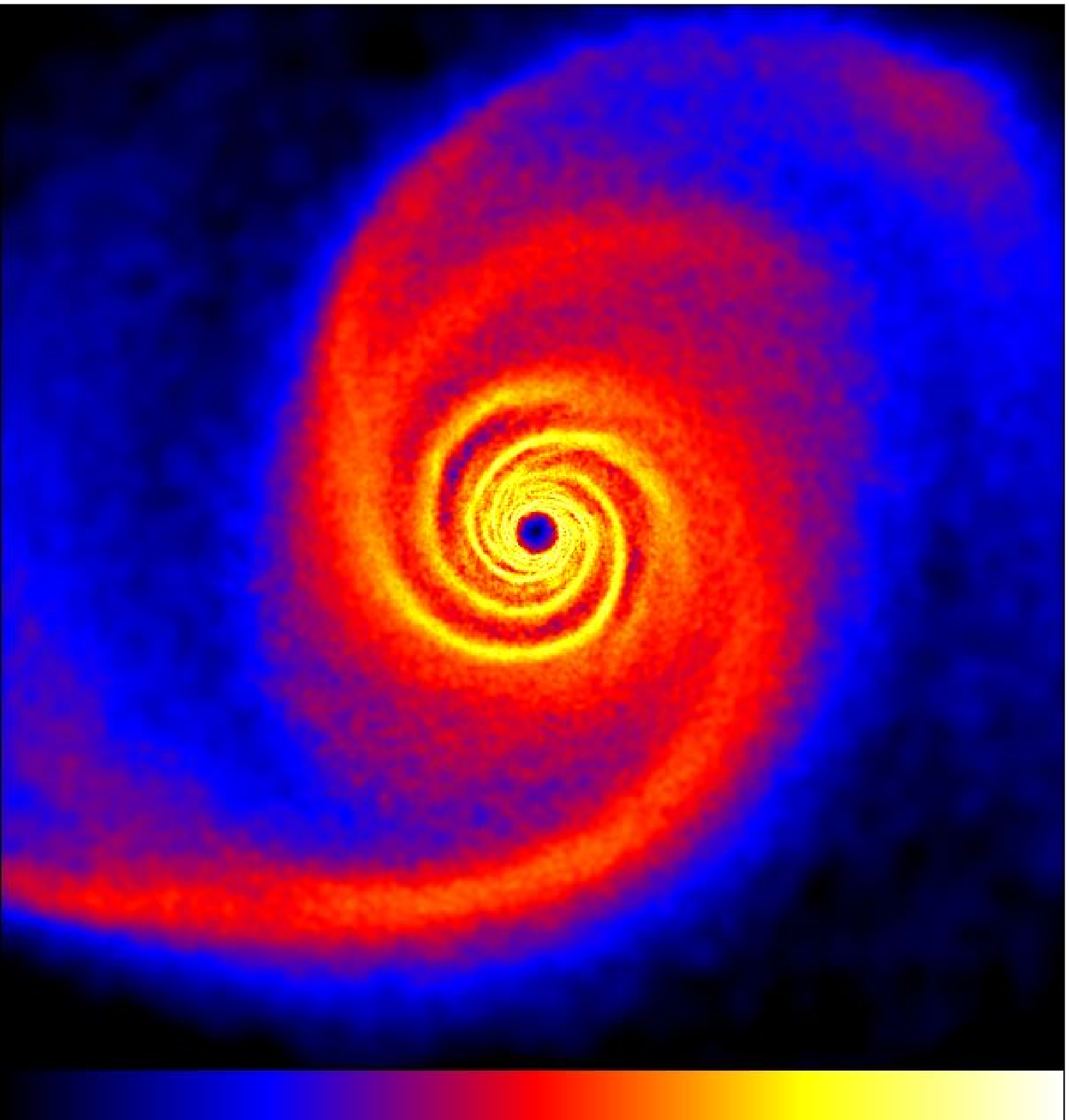,width=6.5cm}\psfig{figure=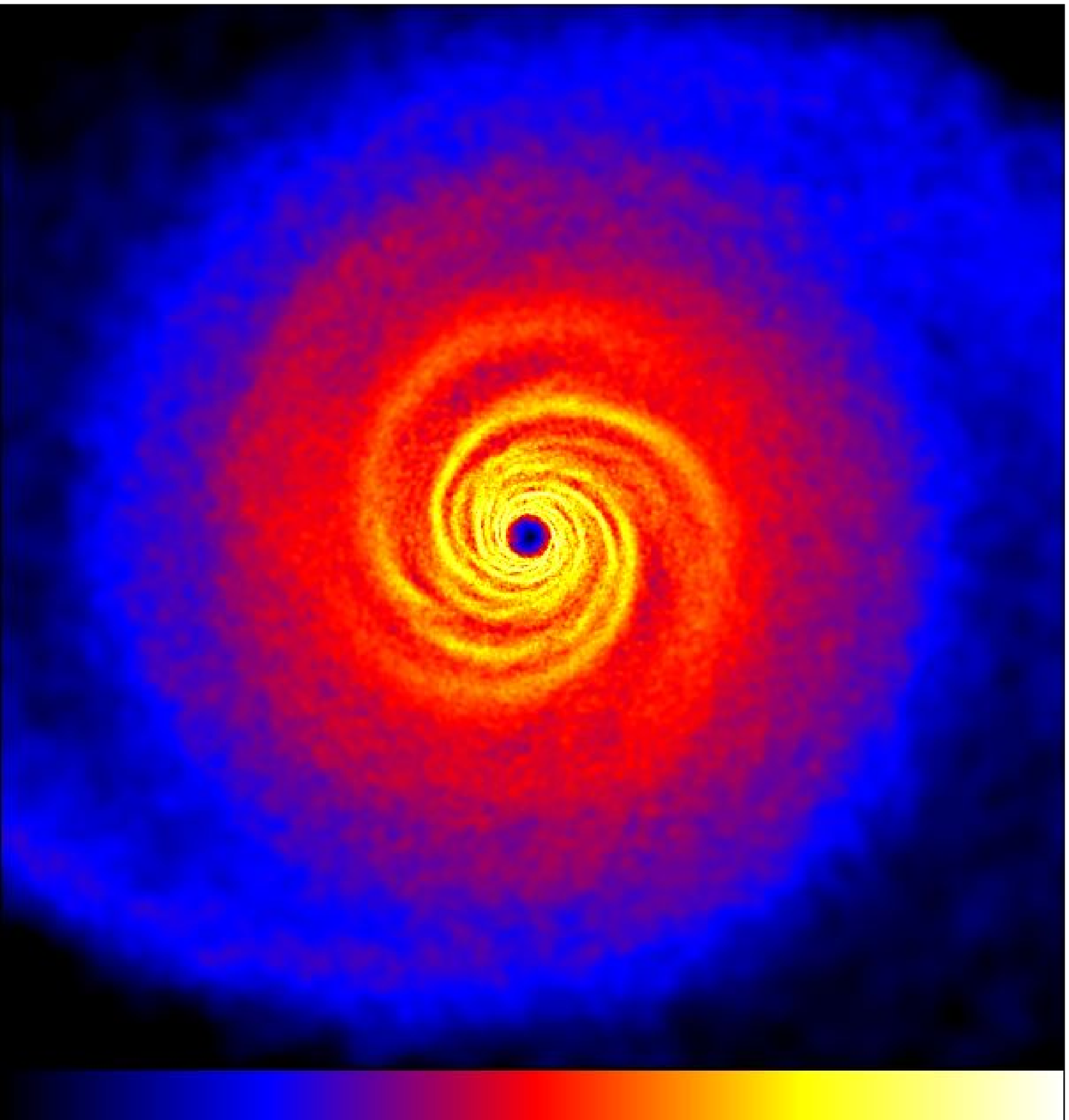,width=6.5cm}}
\caption{\small Evolution of the surface density of the disc during the development of a transient episode of spiral activity in a simulation of a massive $M_{\rm disc}=0.5M$ accretion disc. The four snapshots refer to four different times during the simulation, $t=0$ (upper left), $t=1.9$ (upper right), $t=2.5$ (lower left), $t=3$ (lower right), where the times are given in units of the orbital period at the disc outer edge. From \cite{LR05}.} 
\label{fig:massive}
\end{figure}

 \begin{figure}
 \centerline{\psfig{figure=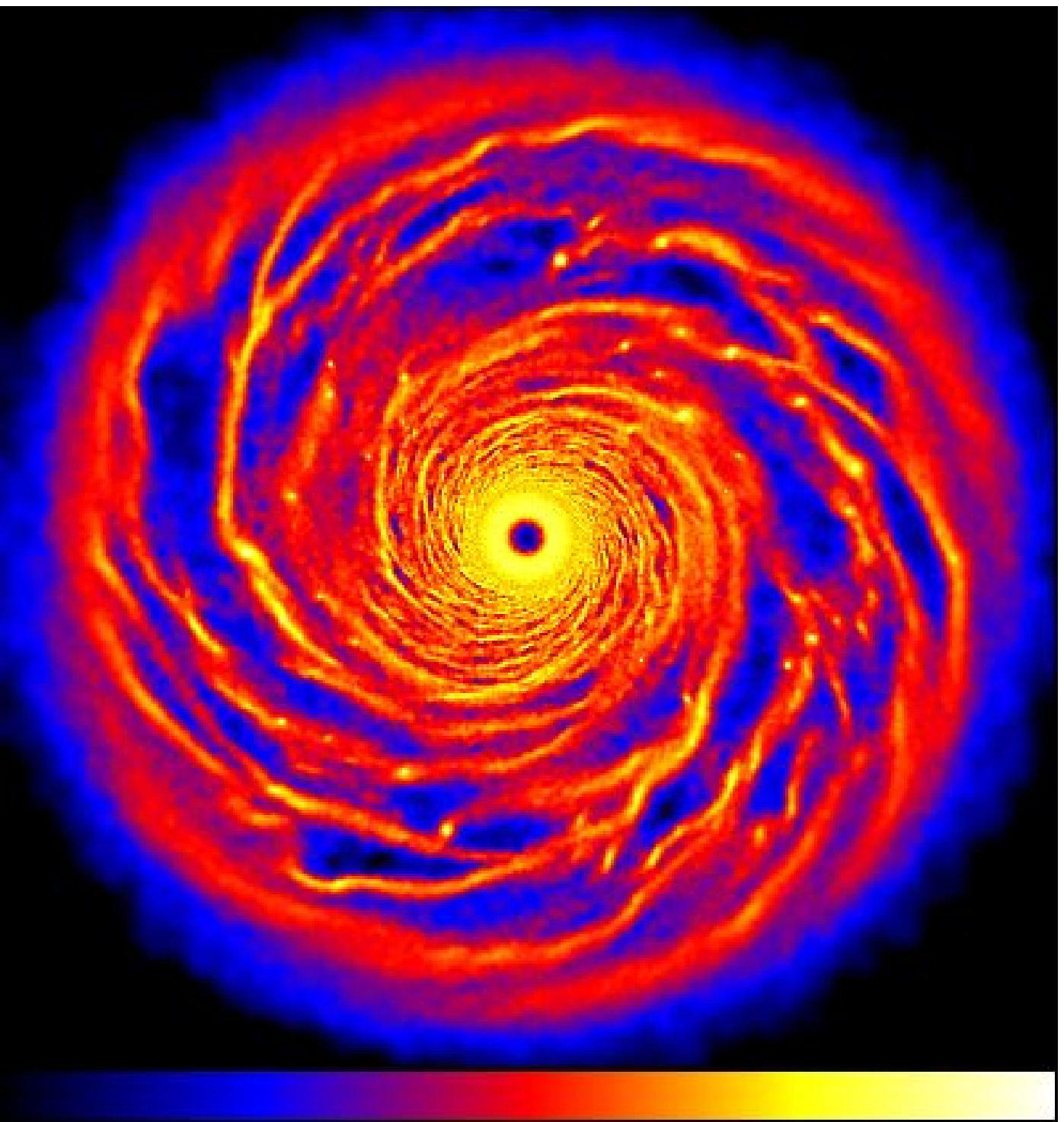,width=11cm}}
\caption{\small Numerical simulation of a self-gravitating disc with $M_{\rm disc}=0.1M$ and $\tau_{\rm cool}=3\Omega^{-1}$. Once unstable, the disc breaks up into numerous gravitationally bound clumps.} 
\label{fig:fragmentation}
\end{figure}

In \cite{LR04} several simulations are performed, with varying mass ratio between the disc and the star, $M_{\rm disc}/M = 0.05$, 0.1 and 0.25. This allows a comparison of the morphology and of the dynamics of the spiral structure developed in the different cases. The upper panel of Fig. \ref{fig:modes} shows a snapshot of the density structure three simulations once they have reached a self-regulated state, while the lower panel shows the power associated with Fourier components of different $m$. Several interesting features can be noted. First of all, note that the typical radial wavelength of the instability increases as the disc mass increases. This can be easily understood even from the simple quadratic dispersion relation, Eq. (\ref{eq:wkb}), which predicts that the typical wavelength of the instability is proportional to the disc thickness $H$, Eq. (\ref{eq:wavelength}). Now, as we know, self-regulated discs have $H/R\approx M_{\rm disc}/M$ and therefore a more massive disc is also thicker and the resulting spiral structure tends to be more open. Also, note that while for the low mass case the different Fourier modes have similar amplitudes, as the disc mass increases, modes with lower $m$ tend to become more important. Also this behaviour can be understood in terms of the local dispersion relations described above. Indeed, when the disc mass is low, the parameter $J\approx mM_{\rm disc}/M$ is also small (except for very large $m$ which are not easily excited) and we can use the quadratic dispersion relation which predicts a similar behaviour for all $m$. On the other hand, for larger disc masses, $J$ can be of order unity already for small $m$ which then tend to be much more unstable, as confirmed in the simulations. If we increase the mass ratio further, going up to $M_{\rm disc}\sim M$ \cite{LR05} this trend continues and strong  $m=2$ disturbances become violently unstable. In this high mass case, self-regulation cannot be established and the disc shows a recurrent pattern of highly temporally variable instabilities, as shown in Fig. \ref{fig:massive}. During such episodes, the torques induced by self-gravity can be very large, and cause a relatively fast redistribution of gas within the disc.

The behaviour described above changes when the cooling time is decreased to smaller values \cite{gammie01}. In this case, the disc does not reach a quasi-steady self-regulated state but rather fragments into several bound objects. Figure \ref{fig:fragmentation} show the results of a simulation very similar to the one displayed in Fig. \ref{fig:simulation}, but where the cooling time is decreased to $\tau_{\rm cool}=3\Omega^{-1}$ \cite{RLA05}.
This result can be understood in the following way, by adopting a local approach to describe the instability.  In a gravitationally unstable disc, the typical growth timescale of unstable perturbations is of the order of the dynamical timescale $\Omega^{-1}$. The non-linear stabilization of the perturbation only works if the heat generated by compression and shocks is not removed efficiently from the disc through cooling. Since the perturbation grows on the dynamical timescale, if we want to avoid fragmentation, we require that cooling acts on a longer timescale. Note that the requirement that the cooling timescale be shorter than the dynamical timescale in order to result in fragmentation has been known for several years, even outside the context of disc instability \cite{rees76,silk77}. The exact value of the threshold for fragmentation does depend slightly on the specific numerical set-up and ranges from $\tau_{\rm cool}=3\Omega^{-1}$ to $\tau_{\rm cool}=6\Omega^{-1}$ \cite{gammie01,RLA05,CHL07}. An additional discussion on the dependence of the threshold on the ratio of specific heats and on its interpretation is provided in Section \ref{sec:fragmentation} below.  

More recent simulations have further improved the thermodynamical description of the disc, including a more realistic cooling prescription based on the opacity properties of the gas. In the simplest case, one can implement such `realistic' cooling in local simulations of infinitesimally thin discs, where the vertical radiative transfer is simply treated with a one-zone approximation \cite{johnson03}. More realistic, three-dimensional simulations including radiative transfer are being developed in the last few years \cite{whitehouse05} and start to provide their first results \cite{boley06,mayer07,stamatellos07}.

\section{TRANSPORT INDUCED BY GRAVITATIONAL INSTABILITIES}
\label{sec:transport}

The discussion in the last two sessions has shown that (a) accretion discs can be subject to gravitational instabilities whenever the parameter $Q$ (which describes linear stability in the tightly wound approximation) is of order unity and (b) under appropriate conditions  the non-linear evolution of the instability tends to self-regulate in such a way that the disc is kept very close to marginal stability ($Q\approx 1$) over a wide radial range. We have also seen that, in the limit of Keplerian rotation, the condition $Q\approx 1$ is roughly equivalent to $M_{\rm disc}/M\approx H/R$, so that if the thickness is small, even a relatively light disc can be gravitationally unstable. 

We now turn to the question of the transport of energy and of angular momentum induced by the instability. In a standard viscous accretion disc, both energy and angular momentum are transported through the disc by viscosity, and are therefore described by a single transport coefficient. As we will see, this might not be the case for self-gravitating accretion discs, and we will then have to determine if and under which conditions can such a self-gravitating disc be described in terms of a single `viscosity' coefficient. That a spiral density wave transports angular momentum was early recognized in the context of stellar dynamics \cite{shu70,lyndenbell72}. In the context of accretion discs, where most of the theoretical development has been achieved through the use of an {\it ad hoc} (cf. the $\alpha$-prescription) anomalous viscosity, the key question is understanding to what extent can this form of transport be described in terms of a `viscous' process. In other words, the $\alpha$-prescription for viscosity has always been used with the (often tacit) assumption that it actually corresponds to the transport induced by some kind of disc instability. It is therefore natural to ask whether the development of gravitational instability can indeed be this long sought cause of angular momentum transport. 

A possible way to proceed is to simply {\it assume} that the transport of angular momentum takes the same form of a viscous diffusion and to introduce a $Q$-dependent viscosity parameter $\alpha$ \cite{linpringle87,linpringle90}, such that
\begin{equation}
\alpha_{\rm sg}=\left\{
\begin{array}{cc}
\xi\left(\displaystyle\frac{\bar{Q}^2}{Q^2}-1\right)  & \hspace{2cm} Q<\bar{Q}  \\
\\
0                                                                                        & \hspace{2cm} Q>\bar{Q} 
\end{array}
\right.
\label{eq:linpringle}
\end{equation}
Here $\bar{Q}$ is the value of $Q$ at which the disc becomes unstable to non axisymmetric perturbations and $\xi$ is a parameter to measure the strength of the induced torques. The above formulation is useful in practical cases, for example, when one wants to incorporate in a simple way the self-regulation mechanism in simple time-dependent models of self-gravitating discs. However, 
it lacks one important feature elucidated from the numerical simulations described above. In this picture, $\alpha_{\rm sg}$ only depends on the local value of $Q$ and not on the cooling timescale $\tau_{\rm cool}$, that we have seen controls so efficiently the development of the instability. In particular, for self-regulated discs, we expect $Q\approx \bar{Q}$ and the formula above would then produce a negligibly small $\alpha_{\rm sg}$, while we know that a finite amplitude spiral structure is present in self-regulated discs and indeed it is this spiral structure that provides the heating to balance the imposed cooling rate. 

If we maintain the assumption that the transport induced by self-gravity can be described in terms of viscosity, a prescription for $\alpha_{\rm sg}$ consistent with the apparent behaviour of the numerical simulations described above can be obtained by the following argument, which is inspired from the self-regulation mechanism. We have seen that the behaviour of the disc is strongly dependent on the cooling timescale. In particular, for a given cooling timescale, the disc eventually reaches a steady thermal equilibrium state, where the amplitude of the spiral instability saturates at a value such that heating balances the imposed cooling. If we now decrease the cooling time, the disc will cool down and become more unstable, and the saturation amplitude of the instability will grow, thus providing an enhanced transport and dissipation, to match the new cooling configuration. A simple prescription can then be provided from the requirement of thermal equilibrium (cf. Eq. (\ref{eq:alphatcool}))
\begin{equation}
\alpha_{\rm sg}=\dln^{-2}\frac{1}{\gamma(\gamma-1)\Omega\tau_{\rm cool}}.
\label{eq:alphasg}
\end{equation}
Note that the above equation is simply a restatement of the energy balance equation, where the viscous heating term on the left-hand side is balanced by the cooling term on the right-hand side. A relation like the one described above necessarily holds whenever self-gravity provides the necessary heating to reach thermal equilibrium, and if transport can be described viscously. The practical inconvenient of Eq. (\ref{eq:alphasg}) is due to the fact that in general the cooling time in a given configuration is not known {\it a priori}, and it requires a careful evaluation of the cooling processes in order to determine the amount of stress in a specific case.

We still need to verify whether the condition for the validity of Eq. (\ref{eq:alphasg}) described above (i.e. that the transport induced by self-gravity is viscous) does hold. It is here that numerical simulations can provide the missing piece of information. Indeed, from the results of the simulation one is able to directly measure the relevant torques and stresses and compare the results with the models described above. The `viscosity' can be estimated from the simulations in two different ways. First, one can compare the secular evolution of the surface density as seen in the simulation with the one predicted for example from Eq. (\ref{eq:diffusion}) and fit the value of the viscosity to the observed evolution. This has been done \cite{laughlin94,laughlin96} for isothermal or polytropic equations of state and the evolution appears to be reproduced reasonably well with values of $\alpha_{\rm sg}$ in the range of 0.01-0.03. It should be noted that these simulations had relatively massive discs, with $M_{\rm disc}\approx M$. However, such estimates cannot be directly compared to what should be expected based on Eq. (\ref{eq:alphasg}), since these simulations, being isothermal, do not have a well defined cooling time. In some cases \cite{laughlin96}, the simulations require the adoption of a spatially varying $\alpha_{\rm sg}$, prompting the authors to speculate that this is a consequence of a `failing' of the $\alpha$-prescription due to the global nature of transport in their models. Now, while global transport might well operate in the massive ($M_{\rm disc}\approx M$) discs considered in those simulations, we note here that, as already remarked in Section \ref{sec:alpha} above, the simple fact that $\alpha$ is not constant does not call into question neither the use of the $\alpha$-prescription, nor the use of a local viscosity or a diffusion equation to describe the disc evolution. 

\begin{figure}
\centerline{\psfig{figure=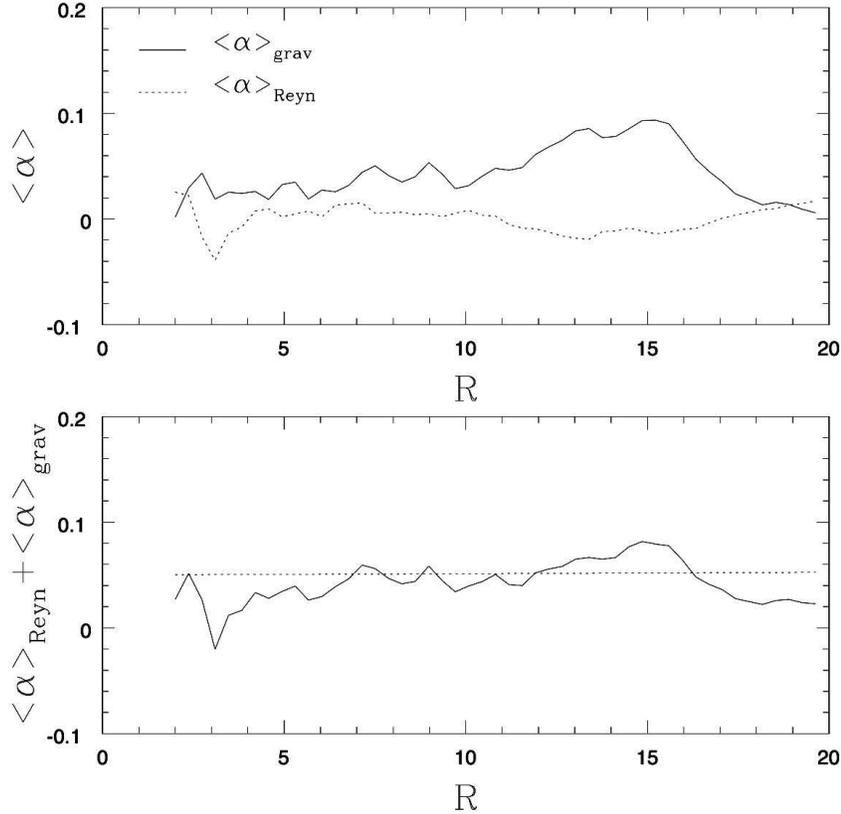,width=11.5cm}}
\caption{\small Upper panel: gravitational (solid line) and Reynolds (dotted line) contributions to the stress as a function of radius for the simulation shown in Fig. \ref{fig:simulation}. Lower panel: comparison between the total (Reynolds plus gravitational) stress (solid line) and the expected value from Eq. (\ref{eq:alphasg}).} 
\label{fig:alpha}
\end{figure}

A second possibility to estimate viscosity is to compute directly the torque exerted by the non axisymmetric disturbances developing in self-gravitating discs. As we will see, this will lead to more insight on the issue of whether transport can be described in terms of local quantities or rather global effects invalidate such a treatment. The starting point for this analysis is Euler equation for a dissipationless fluid, Eq. (\ref{eq:euler}). Consider the case where the fluid velocity is made up of two contributions, one from a (secularly varying) average component, that we call ${\bf v}$, and one from a rapidly varying fluctuating component, that we call ${\bf u}$. In a similar way, the self-gravitating potential can be divided in a quasi-steady component and a fluctuating one. It can be shown (see, for example, \cite{balbusreview,balbus03}) that in cylindrical coordinates, the vertically intergated Euler equation takes the following form:
\begin{equation}
\frac{\partial}{\partial t}\left(\Sigma Rv_{\phi}\right)+\frac{1}{R}\frac{\partial}{\partial R}(Rv_R \Sigma Rv_{\phi}) = 
-\sum_{i} \frac{1}{R}\frac{\partial}{\partial R}\left(R^2 \Sigma \langle u^{(i)}_{R}u^{(i)}_{\phi}\rangle \right).
\label{eq:fluctuations}
\end{equation}
In Eq. (\ref{eq:fluctuations}) the angle brackets indicate a vertical and azimuthal average and  the summation extends to the various fluctuating fields that contribute to the stress. In the case considered here, the two relevant fields are the velocity field and the gravitational field. The stress associated with the fluctuating velocity is usually called the ``Reynolds'' stress and is given by
\begin{equation}
T_{R\phi}^{\rm Re} = -\Sigma \langle u_{R}u_{\phi}\rangle,
\label{eq:reynoldsstress}
\end{equation}
while the term associated with self-gravity is given by 
\begin{equation}
T_{R\phi}^{\rm g} = -\Sigma \langle u^{\rm g}_{R} u^{\rm g}_{\phi}\rangle, 
\label{eq:gravitationalstress}
\end{equation}
where ${\mathbf u}^{\rm g}={\mathbf g}/\sqrt{4\pi G\rho}$, and ${\bf g}=-\nablav\Phi$ is the gravitational field \cite{lyndenbell72}. Equation (\ref{eq:fluctuations}) shows that in the presence of a non-zero correlation between the radial and azimuthal components of any of the fluctuating fields described above, angular momentum can be removed from the mean flow. Indeed, comparing Eq. (\ref{eq:fluctuations}) to Eq. (\ref{eq:angmom}), it is readily seen that such correlations play exactly the same role as a viscous stress tensor for what concerns angular momentum conservation. Of course, we can also use the $\alpha$-prescription, that is, we can measure the stress in units of the local vertical integrated pressure:
\begin{equation}
\alpha_{\rm sg}=\dln^{-1}\frac{ \langle u_{R}u_{\phi}\rangle+\langle u^{\rm g}_{R} u^{\rm g}_{\phi}\rangle}{c_{\rm s}^2}.
\label{eq:alphasim}
\end{equation}
The above formula is simply a measure of the torque provided by non-axisymmetric structures in units of the local pressure and obviously it can always be used to characterized the angular momentum transport due to self-gravity. The question of whether this can be interpreted as a `local' viscosity or not requires a consideration of two further issues: (a) Is the value of $\alpha_{\rm sg}$ determined locally or does it depend on the global disc configuration? (b) Does this process dissipate energy as a viscous term would do, that is according to Eq. (\ref{eq:power})?

Both questions above can be answered using the results of numerical simulations. Indeed, we can simply measure, based on the simulations, $\alpha_{\rm sg}$ from Eq. (\ref{eq:alphasim}). As we have seen above, in many cases the disc reaches a self-regulated state in which thermal equilibrium holds. We can then compare the torque computed from Eq. (\ref{eq:alphasim}) with the expectations based on Eq. (\ref{eq:alphasg}), that is derived under the assumption that the cooling term is balanced by local viscous heating. This exercise has been done in a number of papers \cite{gammie01,LR04,boley06} and the general results is that the two expressions agree relatively well over a large radial range. Note that the simulations reported in \cite{gammie01} were set up to be local (i.e. they only described a small patch in the disc) and there is therefore no surprise that possible global effects did not show up. On the other hand, the simulations of \cite{LR04} and those of \cite{boley06} are global models of the disc (which in these two cases is taken to be of relatively low mass $M_{\rm disc}/M\ll 1$) and global effects could in principle be present. Fig. \ref{fig:alpha} shows an example of such calculation, where the stress computed from Eq. (\ref{eq:alphasim}), (solid line in the lower panel) is compared to the expectation from Eq. (\ref{eq:alphasg}) (dotted line) for the simulation shown in Fig. \ref{fig:simulation}. An even clearer example is shown in Fig. 13 of \cite{boley06}, where the same comparison is made, again showing a remarkable agreement over a large portion of the disc. Note that the simulations of \cite{LR04} and those of \cite{boley06} employ significantly different cooling prescriptions, in that while in \cite{LR04} $\tau_{\rm cool}\Omega$ is set to be a constant, resulting in an approximately constant value of $\alpha_{\rm sg}$ (cf. Eq. (\ref{eq:alphasg})), in \cite{boley06} $\tau_{\rm cool}$ is either constant or self-consistently computed based on the solution of the radiative transfer equation, therefore resulting in a non constant value of $\alpha_{\rm sg}$. In all cases, however, energy appears to be dissipated according to the expectations from a viscous model. Clearly, such a comparison is only meaningful when the disc is in thermal equilibrium. As discussed above, for very massive discs \cite{LR05}, where  $M_{\rm disc}\lesssim M$, a quasi-steady thermal equilibrium is generally not reached and this kind of comparison cannot be made. 

It is also possible to estimate whether the value of $\alpha_{\rm sg}$ is determined locally or not. For example one can calculate the contribution to the gravitational stress at a given reference location coming from a region of finite size around it  \cite{LR04}. It can then be seen that the dominant contribution to the stress arise from a neighbouring region of size $\approx 10H$. A similar result has also been obtained using a different method for local simulations \cite{gammie01}. The local approximation thus breaks down if $10H\gtrsim R$, or $H/R\approx M_{\rm disc}/M\gtrsim 0.1$. Note once again the different behaviour of light and massive discs in this respect.  

We have thus seen that, at least in the case of low mass discs, the transport of energy and angular momentum induced by gravitational instabilities can be well reproduced in terms of a local, `viscous' model. Is this result to be expected? Low mass discs can support a relatively large spectrum of density waves, with different $m$ and with wavenumbers of the order of $1/H$. In the presence of effective cooling, many of such waves can be easily excited. One possibility is that such waves interact with one another, undergo shocks and dissipation, and therefore do not propagate over large radial distances, especially if their typical wavelength is small. In such a case, a local description of the transport induced by the waves would appear to be appropriate. Alternatively, another possible outcome is that a relatively small number of waves can propagate significantly and establish a pattern of global modes, which persist over a few orbits and extend over a large radial range. Such persistent, global modes have been observed in some hydrodynamical simulations of accretion discs, even in the case of low mass discs \cite{boley06}, but even in this case the calculated stress tensor agrees with the value required by energy balance in a viscous disc. 

The energy flux transported by self-gravitating density waves can also be calculated analytically \cite{shu70}. It can be shown \cite{balbus99} that in general the energy flux (contrary to what happens for a viscous process) is not proportional to the stress tensor, and that some other, `anomalous' terms are present, which would preclude a local treatment. These terms can be evaluated within a WKB analysis and it is shown \cite{balbus99} that they vanish at corotation (where the pattern speed of the wave $\Omega_{\rm P}=\Omega(R)$), their contribution increasing as we move away from corotation and becoming significant if $|\Omega_{\rm P}-\Omega(R)|\approx\Omega(R)$. Let us consider the case of an approximately Keplerian disc. In this case, for a wave generated close to corotation such global terms become important only if the wave has travelled a radial distance of the order of $\Delta R/R\approx 2^{2/3}\approx 1.58$.  Now, it is plausible to assume that the minimum distance travelled by a density wave is no less than its radial wavelength $\approx 2\pi H$. A necessary (but not sufficient!) condition for global effects to be negligible is thus $H/R\lesssim \Delta R/2\pi R\approx 0.25$. Once again, we see that thin and thick discs (or, equivalently, light and massive discs) behave in a significantly different way. While for thick and massive discs a local approach appears inadequate, in the case of lighter and thinner discs it is possible to have either a local or a global behaviour, depending on the coherence of the spiral modes involved. The numerical evidences described above appear to imply that a local approach is generally valid for thin discs, and indeed even in the cases where a global pattern is observed, its radial extent is not large enough to prevent the use of a local model \cite{boley06}.

Let us summarize what we have discussed so far. The behaviour of
self-gravitating discs depends on three dimensionless parameters: the
parameter $Q$ (which measures how hot the disc is), the parameter $M_{\rm
  disc}/M$ (which measures how massive the disc is) and the parameter
$\Omega\tau_{\rm cool}$ (which measures how fast does the disc cool). If $Q$
is significantly larger than unity the disc is gravitationally stable, while
if $Q$ approaches unity it becomes unstable. For an approximately Keplerian
rotation the condition $Q\approx 1$ is equivalent to requiring $M_{\rm
  disc}/M\approx H/R$. The behaviour of a gravitationally unstable disc
depends on $M_{\rm disc}/M$ and $\Omega\tau_{\rm cool}$. In particular, if
$\Omega\tau_{\rm cool}\lesssim 1$ the disc rapidly undergoes fragmentation and
produces a number of gravitationally bound clumps, while for larger values of
the cooling rate fragmentation is inhibited. In the latter case, if $M_{\rm
  disc}/M\lesssim 0.25$, the disc settles down in a quasi-steady
self-regulated state, the instability takes the form of a tightly wound spiral
structure and the transport of energy and angular momentum induced by it are
usually well described within a local `$\alpha$-like' approach. On the other
hand more massive discs, with $M_{\rm disc}/M\gtrsim 0.25$, cannot be
described within the local approximation and global transport is expected to
occur. In this case, a quasi-steady self-regulated state is not generally
observed in the simulations (see Fig. \ref{fig:massive}), which show instead a
series of recurrent strong episodes of instability, during which the
gravitational torques can be very large ($\alpha_{\rm sg}\approx 0.5$), and
determine a fast redistribution of the gas within the disc.

\section{THE RELATION BETWEEN FRAGMENTATION AND TRANSPORT}
\label{sec:fragmentation}

 \begin{figure}
\centerline{\psfig{figure=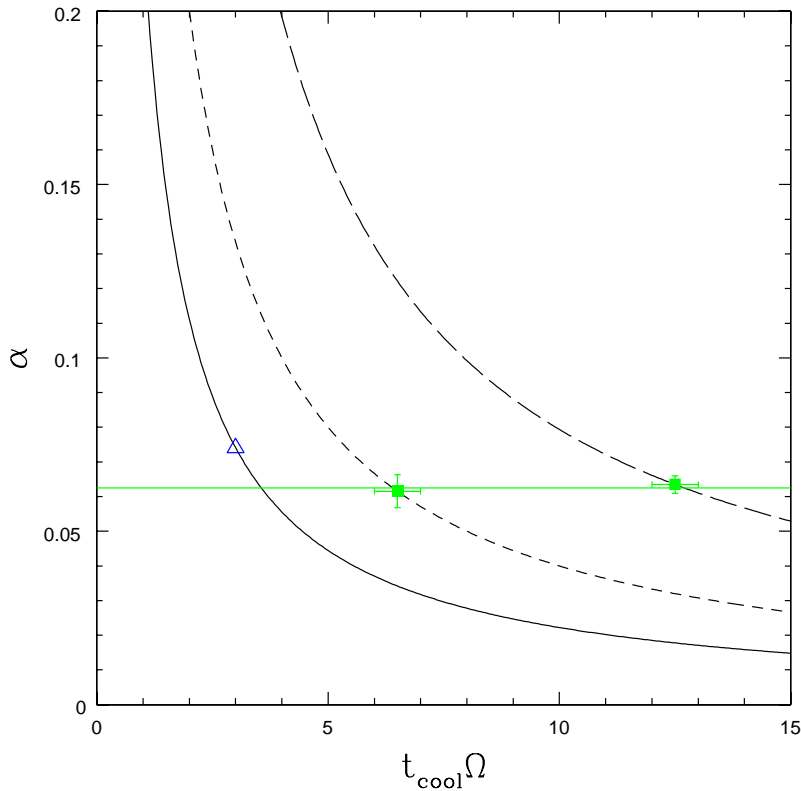,width=11cm}}
\caption{\small The relation between gravitationally induced viscosity $\alpha$ and the cooling time $\tau_{\rm cool}$ in thermal equilibrium, expressed in Eq. (\ref{eq:alphasg}), for three different values of $\gamma=2$ (solid line), $\gamma=5/3$ (short dashed line) and $\gamma=7/5$ (long-dashed line) (from \cite{RLA05}). The green squares indicate the value of $\tau_{\rm cool}$ at which fragmentation is observed in \cite{RLA05}, while the triangle refers to the 2D simulations presented in \cite{gammie01}. It appears that fragmentation occurs whenever the value of $\alpha$ needed to be provided by self-gravity in order to reach thermal balanceexceeds a maximum value of order $\alpha_{\rm max}\approx 0.06$.} 
\label{fig:alphamax}
\end{figure}

The above discussion on the transport induced by self-gravity in a light disc (i.e. for the case where a local description in terms of an $\alpha$ viscosity is valid) allows us to give a more detailed look at the issue of fragmentation. Let us then concentrate on light, quasi Keplerian discs. If these discs are self-gravitating, they settle down in a quasi-steady self-regulated state, with $Q\approx \bar{Q}$, in which they transport angular momentum with a torque strength that can be expressed in terms of $\alpha$ through Eq. (\ref{eq:alphasg}). However, from the discussion of Section \ref{sec:simulations}  we know that if the cooling rate is smaller than a given threshold $\tau_{\rm cool}\Omega<\beta_{\rm crit}\approx 3-6$, the disc undergoes fragmentation. Therefore, through Eq. (\ref{eq:alphasg}), a minimum value of $\Omega\tau_{\rm cool}$ to prevent fragmentation corresponds to a maximum value of $\alpha_{\rm sg}$. 
It is then possible to interpret the fragmentation threshold in a different way, in terms of how much angular momentum can be transported through self-gravity in a steady state. This interpretation is further reinforced by numerical simulations which employ different values of the ratio of specific heats $\gamma$ \cite{RLA05}. It appears that as $\gamma$ decreases, discs are more susceptible to fragmentation. For example, while for $\gamma=5/3$, the threshold for fragmentation is at $\beta_{\rm crit}\approx 6.5$, it increases to $\beta_{\rm crit}\approx 12.5$ for $\gamma=7/5$. The results of this analysis is shown in Fig. \ref{fig:alphamax}, where the three curves show the relation between gravitationally induced viscosity $\alpha$ and the cooling time $\tau_{\rm cool}$ in thermal equilibrium, expressed in Eq. (\ref{eq:alphasg}), for three different values of $\gamma=2$ (solid line), $\gamma=5/3$ (short dashed line) and $\gamma=7/5$ (long-dashed line). The green squares indicate the value of $\tau_{\rm cool}$ at which fragmentation is observed in \cite{RLA05}, while the triangle refers to the 2D simulations with $\gamma=2$ presented in \cite{gammie01}.  It is then interesting to note that fragmentation occurs whenever the value of $\alpha$ needed to be provided by self-gravity in order to reach thermal balance exceeds a maximum value of order $\alpha_{\rm max}\approx 0.06$. For relatively long cooling times, the strength of the torques induced by self-gravity does not need to be very large in order to provide the required heating to reach equilibrium. For shorter cooling times, the saturation amplitude of the spiral structure increases until it becomes too large to be sustained by the disc, which then undergoes fragmentation. The torque provided by the spiral structure at this maximum sustainable amplitude corresponds to $\alpha_{\rm max}\approx 0.06$. 

It is important to stress that the threshold value for $\alpha$ discussed above refers to a quasi-steady state, in which the disc stays in thermal equilibrium and the relevant physical quantities do not change
significantly on time scales shorter than the thermal timescale. This does not happen for massive discs (with masses comparable to that of the central object), that can generate transient strong spiral episodes, with correspondingly large values of the stress $\alpha$, which, however, do not last for longer than one dynamical timescale. 

If the fragmentation threshold is ultimately due to an inability of the disc to redistribute angular momentum on short timescales for a prolonged period of time, then it is possible to envisage a further possible route to fragmentation, in cases where the disc, even if the cooling timescale is long, is fed at a high accretion rate. This possibility, although it may play an important role in several contexts, has not yet been tested numerically. In particular, the maximum accretion rate sustainable at any given radius in the disc is only dependent on the local temperature and is given by $\dot{M}_{\rm max}=2\alpha_{\rm max}c_{\rm s}^3/G$.

\section{APPLICATION TO OBSERVED SYSTEMS: STAR AND PLANET FORMATION}
\label{sec:examples}

The discussion of the sections above has focussed on the basic physical processes that determine the evolution of self-gravitating accretion discs, without referring to any specific astrophysical system where such mechanisms can be at work. We have thus seen what are the basic parameters involved in the determination of the disc properties and evolution. We now turn to an examination of a few concrete cases and we will try to connect the theoretical picture described above to the observed properties of these systems. 

\subsection{Protostellar discs}
\label{sec:YSO}

The formation of protostellar discs is the natural outcome of the process of star formation. Stars are supposed to form from dense condensations within turbulent molecular clouds, called molecular cloud cores (see \cite{mckee07} for a recent review). The typical specific angular momentum of one such core is of the order of $10^{21}\mbox{cm}^2/\mbox{sec}$ and is therefore much larger than the specific angular momentum of a star (for comparison, the Sun has a specific angular momentum of the order of $10^{16}\mbox{cm}^2/\mbox{sec}$). The gas that is going to form a star therefore has to get rid of most of its angular momentum. The collapse of a molecular cloud core is thus bound to form initially a quite extended flattened object, a disc, with sizes of the order of tens to hundreds of AU. Angular momentum redistribution inside the disc allows accretion to proceed and to feed the forming protostar. Clearly, during this process the mass balance in the star-disc system moves from a disc-dominated initial stage, to a star-dominated evolved stage, after the main infall phase has finished. The effects of the disc self-gravity are then going to be more important for younger object than for evolved ones. This evolution has an observational counterpart in a frequently used classification \cite{lada84}, which identifies as Class 0 and Class I the youngest objects, which are still embedded into a dusty gaseous envelope. During this phase the disc is still actively accreting from the envelope and the disc mass is expected to be very large. Unfortunately, during this phase most of the activity which occurs in the disc is enshrouded by the envelope and is difficult to observe, with the exception of a few cases. Class II objects have cleared out most of the envelope and the protostar-disc system is now visible. Such objects are often called Classical T Tauri stars. Since the main infall phase has ended, disc masses in the T Tauri phase are expected to be smaller than in the Class 0 and Class I case. Finally, Class III objects are the most evolved class of protostars, where the gaseous disc has almost entirely dispersed, leaving a remnant `debris disc' mostly formed of rocks and pebbles (akin to the Zodiacal dust in our Solar System). 

The properties of protostellar discs are obviously a strong function of the stellar mass. Most of the data available at the moment refers to discs around solar type stars. However, in recent years a significant amount of data has started to become available for discs at the extremes of the stellar mass distribution, that is for brown dwarfs at the low mass end, and for massive O and B type stars at the upper end of the spectrum. In the latter case, in particular, there is strong evidence pointing to the presence of very massive discs, and these objects are therefore an interesting laboratory to test the models described in the previous sections. We will therefore discuss both the well known case of solar type protostars and the case of massive O-B protostars. 

\subsection{Solar mass young stars}

Disc masses in the T Tauri phase are generally estimated from sub-mm emission, since at these wavelengths dust emission is usually optically thin and we can thus easily convert the observed flux into a mass, if we know what is the dust opacity \cite{beckwith90}. Recent surveys of protostellar discs around solar mass stars in the Taurus-Auriga complex \cite{andrews05} indicate disc masses between $10^{-4}M_{\odot}$ and $10^{-1}M_{\odot}$, with a median value of $5~10^{-3}M_{\odot}$. In another case, \cite{andrews07} the results for a smaller sample (possibly biased to larger masses) indicate a median mass one order of magnitude larger and in a few cases discs as massive as $0.2M_{\odot}$ have been found. Similar values are also reported for the Orion region, where in a few cases masses as large as $0.4M_{\odot}$ have also been reported \cite{eisner06}. Since the stellar mass for these samples ranges from 0.1 to 1$M_{\odot}$, the mass ratios between the disc and the star is between $10^{-3}$ and a few times $10^{-1}$. The thickness of protostellar discs can be estimated from measurements of the temperature in the disc, and it turns out that in general $H/R\approx 0.1$. Bearing in mind that gravitational instabilities develop when $M_{\rm disc}/M_{\star}\sim H/R$, we see that objects in the upper end of the mass distribution for T Tauri stars are expected to be gravitationally unstable. 

It should be noted, however, that such measurements suffer from very large systematic errors, mostly due to uncertainties in the dust opacity. Indeed, in many cases there is evidence for relatively evolved dust grains, so that the dust size distribution can be significantly different than in the interstellar medium \cite{testi2001,natta04}. If dust grains have evolved to relatively large sizes, their opacity would be reduced and the inferred disc mass would correspondingly increase. Additionally, it is important to stress that such measurements are sensitive only to the dust content of the disc, which is expected to be a very small fraction of the total gas mass (in the interstellar medium, for example, the dust to gas ratio is of the order of 0.01). The dust mass measurements are then rescaled to obtain the gas mass, by adopting the standard interstellar medium scaling factor, the applicability of which is highly uncertain. In summary, it is likely that most of the above estimates are actually underestimating the real disc masses \cite{hartmann06}. We thus see that even in many T Tauri objects gravitational instabilities might contribute substantially to the angular momentum transport. The importance of self-gravity becomes even higher when one consider that T Tauri stars are expected to have already accreted much of their mass from the disc. If by this rather evolved stage self-gravity is still important, it must have dominated the dynamics of younger systems, such as Class I objects.

Accretion rates are generally measured from the strength of emission lines emitted as the gas reaches the star in the so called `magneto-spheric accretion', when the innermost parts of the disc are truncated by the stellar magnetic field, and the gas falls almost freely onto the star along magnetic field lines, or from the veiling of photospheric lines due to accretion shocks \cite{gullbring98,gullbring2000}. 
It should be noted here that these measurements refer to the accretion rate onto the star, which does not need to correspond to the accretion rate at larger distance, if significant sinks of matter are present, such as a disc wind (which is generally very small) or, more likely, as a massive planet, whose presence might act like a dam for the accretion flow \cite{lubow96}. In any case, typical values range between $10^{-9}-10^{-7}\msunyr$. With such relatively low values of the accretion rate, the expected disc luminosity is relatively small and the energy balance in the disc is determined mostly by irradiation from the central star \cite{chiang97}. While the detailed balance of course might vary from object to object, it is generally hard to look for accretion signatures in the broadband SED of T Tauri stars. 

What about the role of self-gravity? Theoretical models of the evolution of protostellar discs clearly indicate that in the early phases of star formation (in the Class I phase) the disc is massive enough to be self-gravitating and the transport should be dominated by self-gravity \cite{hartmann06,clarke07,kratter07}. Such effects are expected to become less important in the subsequent T Tauri phase. 

From the observational point of view, an intriguing possibility would be to directly observe a spiral morphology from high angular resolution observations. In the near future, the Atacama Large Millimeter Array (ALMA), might in principle detect such detailed structures in the dust continuum. We have seen in section \ref{sec:simulations} above that for low mass discs (such as those that might be expected around T Tauri stars), the spiral structure is tightly wound and characterized by high-$m$ and short wavelengths (see Fig. \ref{fig:modes}). This can be difficult to observe even at the high resolution achievable with ALMA. On the other hand, more open features, typical of more massive discs, might be detected more easily. In a couple of cases such extended spiral structure has been observed, in the case of AB Aur (an evolved intermediate mass protostar) \cite{lin06,pietu05} and in a case of a young Class 0 object \cite{rodriguez05}. The latter case is particularly interesting, since estimates of the disc mass indicate that the disc makes up more than 30\% of the system. On the other hand, in the case of the more evolved AB Aur, the disc mass is very low and estimates of the disc surface density and temperature indicate $Q\approx 10$, which is too high to argue in favour of a gravitational instability.

\subsection{FU Orionis objects}
A remarkable group of solar mass young stars are the so-called FU Orionis objects. FU Orionis objects are a small class of protostellar systems undergoing large outbursts, during which their luminosity increases  by as much as three orders of magnitude. It is currently believed that the origin of the outburst lies in a sudden enhancement of the accretion rate in the disc, that can rise to values of the order of $10^{-4}\msunyr$. Unfortunately only a very small number of such systems have been observed and in only three cases (the prototypical FU Ori, V1057 Cyg and V1515 Cyg) we have a detailed knowledge of the light curve over a long period of time \cite{CLMI}, which enables us to have some insight on their evolution and therefore on the dynamics of the outburst. 

FU Orionis objects are generally surrounded by dusty envelopes \cite{kenyon91}, and substantial infall from the envelope, at a rate of approximately $10^{-6}\msunyr$ is often required by detailed modeling of the outburst \cite{bellin94,LC04}. These characteristics tend to identify FU Orionis objects as transient episodes of enhanced activity in an otherwise quiescent Class I object.

The disc origin of the large luminosities observed in these systems has been confirmed recently by high-resolution interferometric observations \cite{millan06}, which indicate that emission comes from an extended (a few AU) source, like a disc. Additionally, it has been shown \cite{kenyon85,kenyon88} that the spectral energy distribution of FU Orionis objects is to first approximation consistent with the one expected from a standard steady state accretion disc (as sketched, for example, in Fig. \ref{fig:SED}). Indeed, this is essentially the only clear case where a protostellar disc displays the typical emission feature of an active disc. Detailed modeling \cite{kenyon88} shows that the typical accretion rates are of the order of $10^{-4}M_{\odot}/$yr. The agreement however only holds at relatively short wavelengths, in the optical and near infrared, while in the mid infrared a substantial excess emission with respect to the expected disc spectrum is observed. Given the youth of the system, we expect that the disc mass in these cases should be relatively large and it is therefore plausible to associate the excess emission at mid-infrared wavelengths with the effect of self-gravity. Indeed, self-regulated discs are hotter in their outer parts, due to the combined effect of the self-regulation process and of the flattening of the rotation curve when the disc mass is large. As a result, the SED tends to show some excess emission at long wavelengths (see Fig. \ref{fig:selfregulated}), consistent with observations of FU Orionis objects. Detailed modeling \cite{LB2001} has shown how it is possible to fit accurately the long-wavelength SED of the best studied FU Orionis objects, based on steady-state models of self-regulated discs (note that while clearly these outbursting objects are evolving, the evolution timescale is longer than the dynamical timescale, so that a steady-state model is reasonably appropriate). An example of such modeling is shown in Fig. \ref{fig:fuor} and refers to the prototypical case of FU Ori (full details can be found in \cite{LB2001}). These models do however require fairly massive discs, with $M_{\rm disc}\approx M_{\star}$. Such models predict deviations from Keplerian rotation, that might be detected from observations of the line profiles emitted by these systems \cite{LB03b}

\begin{figure}
\centerline{\psfig{figure=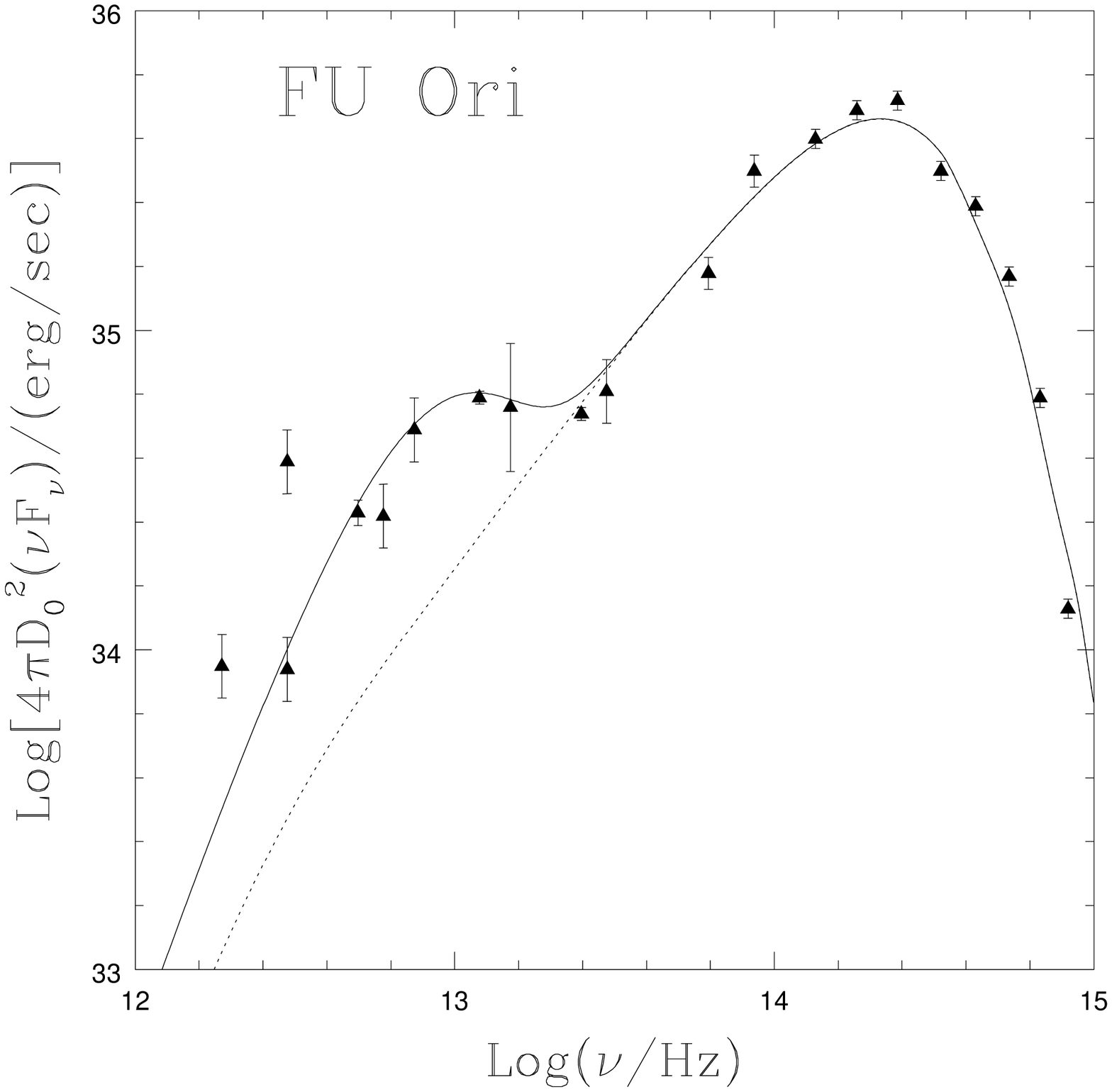,width=4.6cm}\psfig{figure=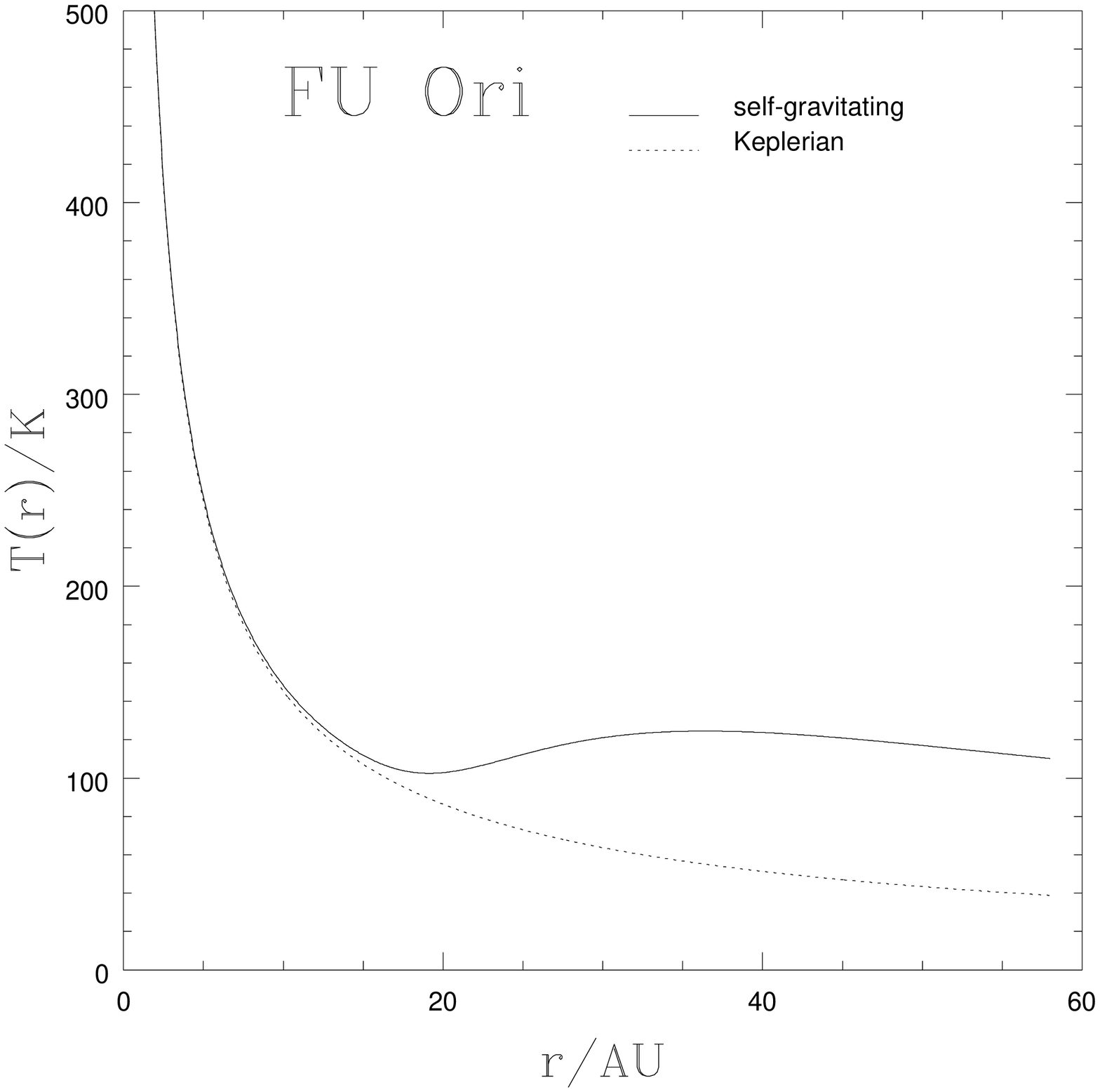,width=4.6cm}\psfig{figure=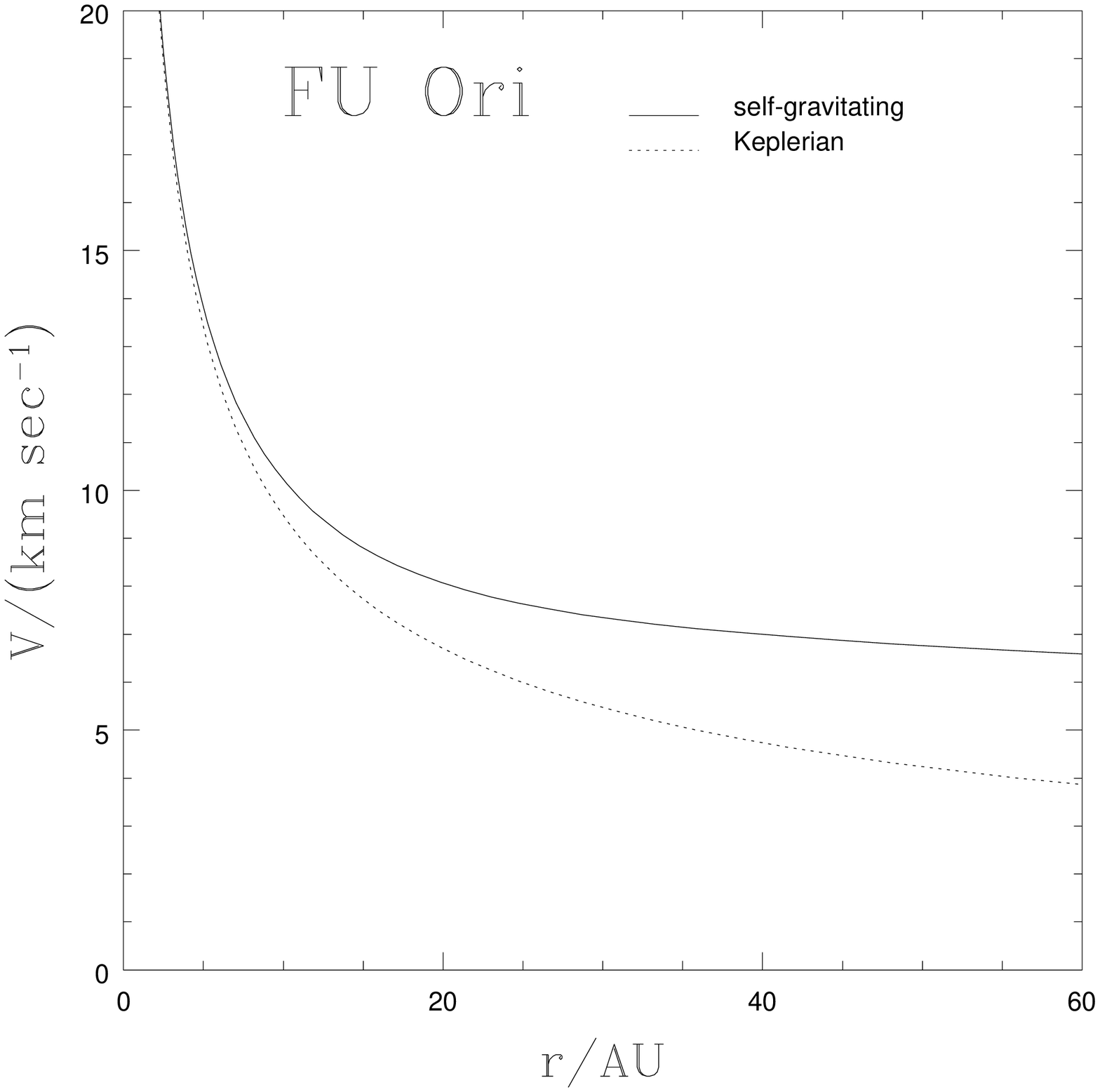,width=4.6cm}}
\caption{\small Modeling of FU Ori SED with a self-regulated self-gravitating disc. Left panel: observed (data points with error bars) and model SED (solid line). Middle panel: temperature profile implied by the self-reulated model (solid line) compared to a non-self-gravitating model with the same parameters (dashed line). Right panel: Rotation curve of the model (solid line) compared to a Keplerian curve (dashed line).} 
\label{fig:fuor}
\end{figure}

Traditional models to explain the cause of the outburst assume that this is due to a thermal instability in the disc, caused by the steep change in disc opacity as hydrogen becomes ionized \cite{bellin94,belletal95}. Such models have been extensively tested against observations and reproduce the broad features of the light curve. However, agreement is only obtained for rather extreme values of the parameters involved (for example, the $\alpha$ viscosity parameter is required to be as low as $10^{-4}-10^{-3}$) and by introducing some {\it ad hoc} triggering event \cite{LC04,clarkelin90}.

An alternative outburst model is that the increase of mass accretion rate is due to the development of a gravitational instability. As we have seen above, SED modeling is consistent with the presence of fairly massive discs. Now, when the disc mass is of the order of the central object mass, we are in the regime where we expect the development of large scale global gravitational instabilities (cf. Section \ref{sec:transport}). Numerical simulations of this regime \cite{LR05,vorobyov05} show that accretion can be highly variable, and indeed characterized by periods of quiescence followed by  recurrent episodes of enhanced accretion. A gravitational origin for FU Orionis outburst has been proposed \cite{vorobyov06} (or a combination of gravitational and MHD instabilities, \cite{armitage2001}), even though a detailed comparison of the model light curves with observations is still lacking.

\subsection{Massive protostars} 
\label{sec:highmass}
In the previous two subsections, we have concentrated on the properties of relatively low mass young stars, i.e. stars with a mass of the order of one solar mass. Stars of higher mass behave in a significantly different way. One of the main reasons behind this different behaviour lies in the fact that the typical contraction time of the forming protostar (that is roughly the time it takes for a forming star to ignite nuclear reactions in its center) is much shorter for high mass stars than for low mass stars. In particular, for stars of mass larger than $\approx 8M_{\odot}$, the star reaches the main sequence while still accreting matter from the surrounding protostellar core \cite{palla93}. Hence, while for low mass stars we can observe optically visible pre-main-sequence stars as T Tauri (Class II) stars after the main infall phase has terminated, which are therefore surrounded by a relatively low mass disc, this is not the case for stars of mass larger than $8M_{\odot}$. In the latter case, the protostar is always going to be found during the main infall phase, and we thus expect that, if discs are present also here, they should be characterized by much larger masses than in T Tauri stars, and their properties should reflect more those of younger protostars (Class 0 or Class I).

Clearly, observations of high mass protostars and of their circumstellar
environment (including the possible presence of discs) are difficult, due to
the fact that they are still embedded within their parent core. Nevertheless,
observations in the last few years have been able to reveal a significant
number of them, and there is now a substantial evidence that protostellar
discs are present also for high mass stars (see \cite{cesaroni07} for a recent
review). It should be mentioned however that the evidence for the presence of
discs is mostly circumstantial. In most cases, the `disc' is found as a
structure that lies perpendicular to a large scale outflow, which is common
around young stars, and is easy to detect. Such structures are either seen in
infrared continuum observations as dark silhouette discs against a bright
background \cite{chini04}, or through observations of thermal
\cite{cesaroni94,cesaroni05} and non-thermal (usually maser) emission lines
\cite{torrelles98,greenhill04,patel05,edris05}, which have the advantage of
providing information about the velocity structure of the circumstellar
environment, so that a disc can be associated with a velocity gradient
perpendicular to the outflow axis, usually interpreted as a sign of rotation.

The general result is consistent with the expectations outlined above. The
discs are indeed found to be very massive, with masses comparable to, and in
some cases even larger than, that of their central star. Although the data are
still very uncertain, there is also some evidence for non-Keplerian rotation,
as expected when the disc mass is high \cite{cesaroni05}. Observed structures
can be broadly divided in two groups: some are very extended (with sizes of
$\approx 10^{4}$AU), very massive (with masses much larger than their central
stars) and the luminosity of the central object is generally consistent with
an O-type protostar (mass larger than $20M_{\odot}$). The second group of
objects is characterized by smaller sizes ($\sim 10^{3}$ AU), smaller mass
($M_{\rm disc}\lesssim M$), and are generally associated with B-type stars
(with masses between 8 and 20 $M_{\odot}$). Objects in the first group are so
large that their typical orbital timescale is of the order of the lifetime of
the structure. Such objects are therefore unlikely to be in centrifugal
equilibrium and is therefore difficult to identify them as accretion discs
(even though a smaller size disc can in principle be hidden within these
structure). On the other hand, objects in the second group may be regarded as
proper discs.

Let us now consider what accretion rates are associated with the discs in the
second group of objects. Here again, observations are very difficult. A well
constrained parameter is the outflow rate, which is produced in the inner
parts of the disc (at a distance $\lesssim 1$ AU) from the star, and it is of
the order of $10^{-4}-10^{-3}\msunyr$. Clearly, the accretion rate through the
disc should be at least of this order of magnitude to feed the outflow. Can
the disc sustain such large accretion rates? Let us first consider the case
where transport can be simply described in terms of a local, viscous
phenomenon. As we have seen above, this is the expectation when the transport
is induced by a gravitational instability in low mass discs. We have also seen
that in this case a quasi-steady, self-regulated spiral structure does not
provide a torque larger than $\alpha\approx 0.06$ (when measured in terms of
the $\alpha$-prescription) without undergoing fragmentation. It is then easy
to estimate the accretion rate expected under these conditions, from
$\dot{M}\approx 2\alpha c_{\rm s}^3/G$, where $c_{\rm s}$ is the sound speed.
Typical temperatures in the outer disc for these objects are of the order of
170K (this is the case of IRAS 20126+4104, the best studied among these
objects \cite{cesaroni05}), which correspond to a maximum accretion rate of
$\dot{M}\approx 2~10^{-5}\msunyr$, well below the rate needed to steadily feed
the outflow from the inner disc. On the other hand, we have to keep in mind
that the picture where gravitational instabilities provide a local, viscous
source of angular momentum transport is only appropriate for relatively low
mass discs. Here, the disc masses are much higher and hence the situation is
more similar to the case shown in Fig. \ref{fig:massive}, than to the one
shown in Fig. \ref{fig:modes}. We should thus expect the development of a
series of recurrent global spiral structures, which can temporarily (for a
period of the order of the outer dynamical timescale, which in this case is of
the order of $10^4$ years) transport angular momentum much more efficiently
than a local process would do. As discussed above, during such episodes the
stress tensor induced by the spiral structure can be up to a factor ten larger
than in a quasi-steady self-regulated case, and we can thus reconcile in this
way the observationally implied accretion rates with the rates expected from
theoretical arguments.

\subsection{Planet formation}
\label{sec:planet}
The process of planet formation is intimately linked to the process of star
formation. The interest in this subject has grown significantly in the last
ten years, and in particular after the discovery of the first planetary system
around a solar type star \cite{mayor95}, outside our own Solar System. To
date, hundreds of extra-solar planets have been discovered, and the list is
ever-growing. Dedicated space missions, such as Darwin and the Terrestrial
Planet Finder, are going to be launched in the near future. This large set of
data allow us to compare theoretical models of planet formation with a
statistical sample, rather than to the only one example (the Solar system)
that was available until very recently. Our theories about planet formation
are going to be significantly improved and modified as new data become
available.

The properties of extra-solar planets appear to be very different from the
ones of the planets in our Solar System, although in many cases they reflect
the limitations intrinsic to the detection methods used. In fact, most
extra-solar planets are discovered through the `radial velocity' technique,
based on the measurement of the radial velocity induced by the presence of the
planet on the motion of the host star. This method naturally favours the
detection of planets of high mass and large eccentricities, which would induce
the largest radial velocity. Also, planets at large separations, which have
very long periods, cannot be detected this way. An additional observational
limitation is given by the fact that the radial velocity provides information
about $M_{\rm p}\sin i$, where $M_{\rm p}$ is the planetary mass and $i$ is
the inclination of the planetary orbit. If (as it often happens) the system is
not eclipsing, we do not have information on the inclination and thus the
planetary mass is only known to within this potentially large unknown factor.
In any case, the inferred planetary masses range from 5 Earth masses, in the
case of the `super-Earths' recently discovered \cite{udry07}, up to 20 Jupiter
masses, much larger than the most massive planets in the Solar System. It
should be noted that many of these massive planets are found at very small
separations from their host stars, smaller than 0.1 AU, where the locally
available mass is too small to form the planet, thus suggesting that during
the course of the planet formation substantial radial migration can happen.
Additionally, while in the Solar System the orbital eccentricity is generally
low, below 0.1 (except for Mercury and Pluto), extra-solar planets essentially
span all the possible eccentricities, reaching values as high as 0.9.

Parallel to the investigation of extra-solar planets, we have also improved
our knowledge of the structure of planets within the Solar system.
Particularly relevant to the discussion on theories of planet formation are
the most recent evidence for the presence of a solid core in the interior of
Jupiter and Saturn \cite{guillot99}. In fact, the estimates of the mass of
this solid core have been revised and they now appear to be smaller than
previously thought. In particular, while for Saturn the core mass is of the
order of $20M_{\oplus}$, in the case of Jupiter the upper limit is
$14M_{\oplus}$ and models with no core at all are also not excluded.

Currently, there are two models for the formation of planets. The first (and
the more widely accepted) is the so called `core accretion' model, while the
second (which was usually disregarded, but has recently gained progressively
more attention) is the so called `gravitational instability' model. Clearly,
in the context of the present paper, the second model appears to be more
interesting. However, as we shall see, gravitational instabilities may play an
important role even in the core accretion model. While a thorough review of
planet formation theories is clearly beyond the scope of this paper, I here
briefly summarize the main features of the two models.

\subsubsection{\bf The core accretion model}
In the core accretion model \cite{pollack96} (see also \cite{lissauer_PPV} and
\cite{armitagereview} for two recent reviews), planets are supposed to form
sequentially out of the solid component present in protostellar discs. The
first step in planet formation is the growth of interstellar dust through
inelastic collisions up to a size of tens of centimeters. This growth is seen
in observations of protostellar discs, whose opacities often indicates the
presence of large dust grains \cite{testi2001}, and where high-resolution
spectroscopy in the infrared indicate the presence of structured silicate
grains \cite{vanboekel05,apai05} (incidentally, the large uncertainties in
protostellar disc masses mentioned in section \ref{sec:YSO}s are essentially
due to the unknown level of dust growth). For sizes larger than one meter, the
collisional agglomeration of dust is not effective. The growth in this mass
range, up to the formation of kilometer-sized planetesimals, is still largely
unknown (but see below). Once km-sized planetesimals have formed they
essentially decouple from the gas and keep growing through their mutual
interaction, enhanced by gravitational focussing. In this way a few planetary
cores emerge from the background, and if the core reaches a size of the order
of $5-10M_{\oplus}$ it is able to further accrete a large gaseous envelope,
hence becoming a giant planet, like Jupiter. An important attractive of this
model is that it is able to produce both small, Earth-like, rocky planets, and
the large Jupiter-like gaseous planets. Note that within this model, Jupiter
is bound to have a massive rocky core, the evidence in favour of which, as
mentioned above, is not robust. Apart from this, two other problems affect
this model for planet formation. First of all, it is a relatively slow
process. Detailed models \cite{pollack96} require 10 million years in order to
form Jupiter, which is just within the estimates of the lifetime of
protostellar discs. A second important problem is related to the formation of
planetesimals. It is in this respect that the disc self-gravity might play a
fundamental role in planet formation and I thus discuss it here in some
detail.

\begin{figure}
\centerline{\psfig{figure=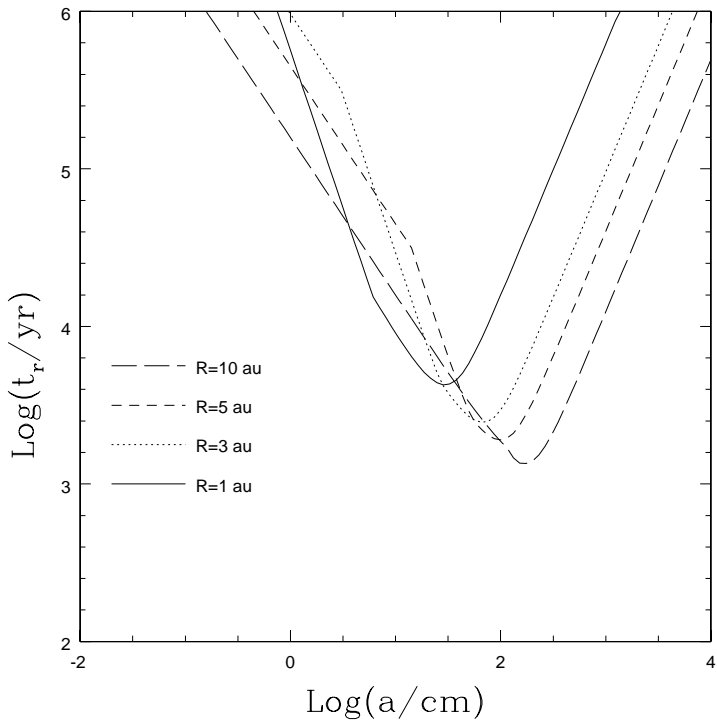,width=10cm}}
\caption{\small Solid particles migration timescale as a function of particle
  size for a typical protostellar disc configuration. The disc is here assumed
  to have a smooth pressure distribution, uniformly decreasing outward as a
  power law. The four lines refer to four different radial locations in the
  disc. From \cite{rice04}.}
\label{fig:migration}
\end{figure}

The issue arises when one considers the motion of meter sized solid particles
within a gaseous disc. Even in the case where the disc contribution to the
radial gravitational field is negligible, the gas rotation can be
non-Keplerian, due to the effects of pressure gradients. As discussed in
section \ref{sec:radial}, pressure gradients provide additional support to the
gas against gravity, so that the rotation curve can be non-Keplerian to some
extent (Eqs. (\ref{eq:gasdrift1}) and (\ref{eq:gasdrift2})). The degree of
non-Keplerianity in this case is determined by $(c_{\rm s}/v_{\phi})^2\approx
(H/R)^2$.  The sign of this contribution, depends on the pressure gradient, so
that if the pressure decreases with increasing radius, the rotation is
sub-Keplerian. Solid particles, on the other hand, do not feel the effect of
pressure and therefore their rotation curve to first approximation is
Keplerian. This thus determines a velocity difference between the solids and
the gas (note that this effect is analogous to the so called `asymmetric
drift' which occurs in galaxy dynamics, cf. \cite{cava06}). If the solids and
the gas are coupled via a drag force (for example, the usual Stokes or Epstein
drag, which are usually dependent on the solid particle size) this may have
some sizable effect on the solid's motion. Let us consider a smooth disc with
pressure decreasing outwards \cite{weiden77}, in this case the solids would
move faster than the gas and the drag force therefore acts in such a way to
remove angular momentum from them and thus determines an inward radial
migration. The magnitude of such radial motion and the timescale for radial
migration depends on the particle size. In particular, for very small sizes
the solids are so strongly coupled to the gas that they essentially follow the
gas motion and their velocity difference with respect to the gas is
negligible. For very large sizes, on the other hand, the coupling becomes very
weak, and the solids move in essentially Keplerian orbits, with a very small
inward velocity. The most interesting behaviour occurs for intermediate sizes,
for which the solids are coupled to the gas, but not so strongly to completely
follow the gas streamlines. The migration timescale as a function of particle
size is shown in Fig. \ref{fig:migration} for a typical smooth protostellar
disc configuration at different radial distances from the star. It can be seen
that for sufficiently small and for sufficiently large particle sizes the
migration timescale is very long. For meter-sized objects however, the effect
becomes stronger and the migration time can be as small as a few thousand
years, a very short timescale with respect to the typical solid growth time at
these size. This then poses a severe problem for the core accretion model, in
that as soon as the dust grows in size and reaches the meter-size region, it
suffers from this very fast radial migration and would essentially disappear
into the innermost regions of the disc before being able to grow further and
form planetesimals.

\begin{figure}
\centerline{\psfig{figure=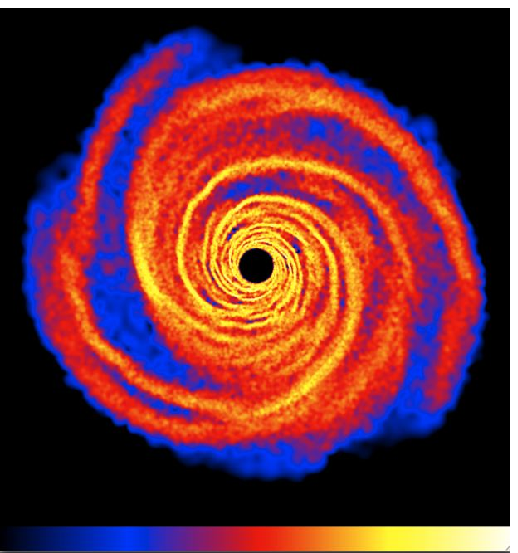,width=6.6cm},\psfig{figure=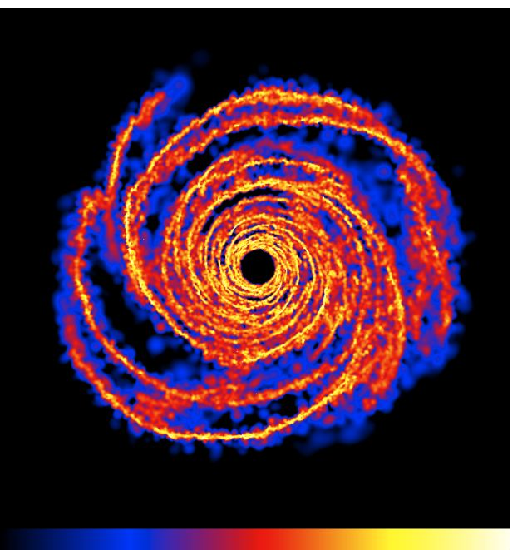,width=6.6cm}}
\caption{\small Surface density of the solid component in a self-gravitating
  gaseous discs, where the gas and the solids are coupled through drag. Left
  panel: solid size equal to 10 meters. Right panel: solid size equal to 50
  cm. While in the 10 meter case the solid density is very similar to the
  corresponding gas density, in the 50 cm case the solid concentrate very
  efficiently into filamentary structures at the peaks of the spiral arms.
  From \cite{rice04}.}
\label{fig:solids}
\end{figure}

The above behaviour occurs for a smooth disc, in which the pressure is a
monotonic function of radius. If the disc is characterized by a long-lasting
spiral structure, in which the density (and hence the pressure) presents local
maxima, the situation can be significantly different
\cite{haghighipour03a,haghighipour03b}. Indeed, if the pressure increases with
radius, the gas motion is locally super-Keplerian and the direction of the
drag force is reverted, leading to a fast outward migration. In turn, in the
presence of a local pressure maximum, the solid tend to migrate and
concentrate to the maximum. The timescale for this process can be even faster
than the migration timescale. First of all, recall that the typical size of a
spiral perturbation due to the disc self-gravity is of the order of $H$. This
means that the pressure gradient is enhanced with respect to a smooth disc by
a factor $R/H$. Secondly, in order to migrate to the peak of the spiral
structure, the solids have to move a distance $H/R$ shorter than in a smooth
disc. We thus expect this migration to the peak to occur on a timescale
$(H/R)^2$ shorter than the timescale shown in Fig. \ref{fig:migration}.
Simulations of a two component self-gravitating disc, where both a gaseous and
a solid component are present have been run \cite{rice04,rice06} and
essentially confirm the scenario outlined above. Fig. \ref{fig:solids} shows a
comparison of the density structure of solids of different sizes, i.e. 10
meters (left) and 50 cm (right) \cite{rice04}. Note that in these simulations
the parameter $\Omega \tau_{\rm cool}$ was taken to be much larger than unity
in order to prevent the gaseous disc to fragment. In the 10 meter case, the
solid surface density essentially follows that of the gas, while in the 50 cm
case, the solid concentrate in a filamentary structure at the peak of the
spiral arm. This strong concentration can have important effects on the growth
rate of solids in this size range. First of all, the enhanced density favours
and speeds up collisional growth. Additionally (and possibly more
importantly), the solid component might become gravitationally unstable, both
because of the high density and because the velocity dispersion of solid
particles is strongly reduced as the particles get `trapped' within the
spiral. Simulations of the same setup, and including the self-gravity of the
solid component, have shown that meter sized solids in the filaments easily
collapse and form gravitationally bound larger objects \cite{rice06}.
Unfortunately, the mass resolution of such simulations is too small to
determine whether the resulting solid fragments would be of the right mass
range for planetesimals.

\subsubsection{\bf The gravitational instability model} In the gravitational
instability model gaseous giant planets are simply formed from the
fragmentation of a self-gravitating gaseous disc
\cite{boss00,boss02,mayer02,PPV_GI}. As we have seen in section
\ref{sec:simulations} above, this naturally follows from the development of
gravitational instabilities, {\it if the cooling rate in the disc is
  sufficiently large}, such that $\Omega \tau_{\rm cool}<1$. If this happens,
then gravitationally bound objects are able to form rapidly, within a few
orbital timescales. With respect to the core accretion model, the
gravitational instability model has the advantage of being a fast process,
hence obviating to one of the difficulties associated with core accretion. A
downside of it is that it can only account for the gas giant planets and would
not explain the formation of terrestrial ones. Additionally, in this model gas
giants would not have a solid core. Now, while for Jupiter this might also be
the case, a solid core is thought to be present in Saturn. In a few cases,
extra-solar planets have also been detected through eclipses of their parent
stars \cite{charbonneau07}, which allows to put constraints on their size and
hence on their inner density, which can then give hints on the presence of a
solid core. In one extreme case, HD 149026b, which has a total mass of
$0.36M_{\rm Jupiter}$, it is estimated that the core would be as massive as
$70M_{\oplus}$ \cite{fortney06}, which would then exclude the gravitational
instability model (but would also be very difficult for the core accretion
one!).

What would be the typical mass of a planet formed by gravitational
instability? A simple estimate can be obtained by considering that, in a thin
disc, the most unstable wavelength is $\lambda=2\pi H$. The typical mass of a
fragment would then be of the order of $M_{\rm frag}\approx \Sigma\lambda^2$.
This can be simply rewritten in terms of the mass of the star $M$ and the
aspect ratio $H/R$, as
\begin{equation}
M_{\rm frag}\approx \frac{4\pi}{Q}\left(\frac{H}{R}\right)^3M.
\end{equation} 
Now, a gravitationally unstable disc has $Q\approx 1$, the typical aspect
ratio of protostellar discs is $H/R\approx 0.1$, and if we take the central
star mass $M=M_{\odot}$, we get $M_{\rm frag}\approx 12 M_{\rm Jupiter}$. It
then appears that this model would only form the most massive observed
planets.

However, most of the debate over the gravitational instability model for
planet formation concentrates on whether the actual cooling time in
protostellar discs can be short enough to induce fragmentation. This has been
estimated analytically \cite{rafikov05} and the results is that the cooling
time is short enough to induce fragmentation only at very large distances from
the central star, of the order of 50-100 AU. On the other hand, it has also
been argued that if the disc becomes convectively unstable, then cooling might
become more effective, producing the correct conditions to induce
fragmentation (\cite{boss04}, but see \cite{rafikov07} for a rebuttal of this
hypothesis). Recent simulations that employ sophisticated (but still
uncertain) radiative cooling prescriptions \cite{boley06,mayer07} indicate
that indeed fragmentation of protoplanetary discs only occurs at very large
distances from the central star.

Even if the cooling time is long, it has been argued that the possible
interaction with a companion or the close passage of a nearby star might
trigger disc fragmentation by tidally increasing the local density. Early
numerical isothermal simulations did indeed show such a behaviour
\cite{boffin98,watkins98a,watkins98b}. However, it should be noticed that, as
mentioned above, the response of an isothermal disc to a gravitational
instability is bound to be relatively violent, since the stabilizing feedback
loop associated with self-regulation cannot happen under isothermal
conditions. More recently, the same issue has also been investigated with a
non-isothermal equation of state and the results appear to be different. It
has been found that both in the case of a disc in a binary system
\cite{mayer05b} and in the case of a stellar fly-by \cite{LMCR07}, the effect
of tidal heating is much more effective in stabilizing the disc than the
de-stabilizing effect of the associated density enhancement, so that even in
this case the ultimate criterion for disc fragmentation is that the cooling
time be shorter than the dynamical time (see \cite{boss06} for an alternative
view).

However, not all hope is lost for the gravitational instability model. As we
have seen, this mechanism is indeed able to form large gaseous planets (or
small brown dwarfs) at large radial distances from the host star, of the order
of 100 AU. Such a new planetary population is still not excluded from
observations. Note that the radial velocity technique requires tracking the
stellar velocity for a period of the order of the planetary orbital period in
order to detect a planet. Now, planets at very large distances have very long
periods and might therefore require more time in order to be detected.
Additionally, in the future such a different planet population might become
detectable by other techniques, such as direct imaging.

In this respect, a very interesting case is that of the system 2MASS 1207. It
consists of a brown dwarf primary, with a mass of $25M_{\rm Jupiter}$, and a
planetary mass companion, with a mass of $5M_{\rm Jupiter}$, orbiting at a
distance of 55 AU from the central brown dwarf \cite{chauvin04} (recent
revision of the distance to the system has provided slightly different orbital
parameters \cite{mamajek05}, with revised masses of 21 and 3 $M_{\rm Jupiter}$
for the primary and the secondary and a semi-major axis of 41 AU, but the
following arguments remain essentially unchanged). The companion, 2MASS 1207b,
is the first planetary mass object to be ever imaged. While clearly the
absolute mass of the companion is in the planetary regime, it is questionable
whether this system should be really considered as a planetary system around a
brown dwarf or rather as an extremely low mass binary, since the mass ratio
and the separation are perfectly consistent with the statistical properties of
binaries of larger mass \cite{LDC05}. In any event, what is relevant here is
to understand whether such a planetary mass object formed through the core
accretion scenario or rather through gravitational instability. It has been
shown \cite{LDC05} (see also \cite{payne07}) that given the large separation
of the system and the very low mass of the primary, it is extremely unlikely
to form this object via core accretion. On the other hand, this is a good
example where the gravitational instability scenario is expected to work,
given the large mass ratio and the large separation. Whether such an object is
a rare case, and especially whether cases like this also occur around larger
mass primaries, is still unknown but future observations might provide
interesting constraints.

\section{APPLICATION TO OBSERVED SYSTEMS: AGN}
\label{sec:AGN}

After describing the impact of the disc self-gravity on the dynamics of
accretion discs at the smallest scales, that is on the AU scale of
protostellar and planet forming discs, we now turn to explore how it may
affect the structure and dynamics of accretion discs at the largest scales,
that is at the parsec scales of discs feeding supermassive black holes in
Active Galactic Nuclei.

\subsection{Active Galactic Nuclei}

\subsubsection{\bf Feeding the hole}
It is well known that accretion discs feed the growth of supermassive black
holes in the nuclei of galaxies. Here the typical scale of the system is
obviously much larger than in the case of protostellar discs. The central
supermassive black hole has a mass that ranges from $10^6$ to $10^9M_{\odot}$.
The radial extent of the disc is more uncertain. While the high energy
emission from the AGN comes from the hotter innermost parts of the disc (from
a few to a few tens Schwarzschild radii), there is also substantial evidence
that the disc can extend much further out. One important piece of information
in this context comes from the observation of water maser emission from AGN
\cite{miyoshi95,greenhill97,kondratko06,greenhill96}, which reveals the
presence of a thin disc of gas in almost Keplerian rotation out to a distance
of the order of 1 pc, which corresponds to $10^5R_{\bullet}$ for a
$10^8M_{\odot}$ black hole, where $R_{\bullet}$ is the Schwarzschild radius.
In some cases the kinematics of these maser spots presents some interesting
features that will be discussed in more detail below. A very important
difference between AGN discs and protostellar discs is that AGN discs are
generally much thinner. Detailed models of the structure of the
non-self-gravitating inner disc \cite{collin90} indicate that the aspect ratio
is of the order of $H/R\approx 10^{-3}$, although in the self-gravitating
regime it can be larger, of the order of a few times $10^{-2}$ \cite{sirko03}.
This has important consequences for the issue of the role of the disc
self-gravity. As we have often mentioned above, a simple way to check the
importance of the disc self-gravity is to compare the mass ratio $M_{\rm
  disc}/M$ with the aspect ratio $H/R$. Now, given the small aspect ratio of
AGN discs, a very light disc, with $M_{\rm disc}/M\approx 10^{-3}$ can already
be subject to gravitational instabilities. It is then relatively easy to
calculate the distance at which we would expect the cumulative disc mass to be
high enough to make the disc unstable \cite{LB03a,goodman03}. It tuns out that
this distance is of the order of $10^3R_{\bullet}$, or $10^{-2}$ pc for a
$10^8M_{\odot}$ black hole.

There is still some debate about the radial extent of the disc feeding the
central black hole. Some models \cite{KingPringle2007} postulate that the disc
outer radius corresponds to the distance where it becomes self-gravitating,
i.e. $\approx 0.01-0.1$ pc, while some others allow the disc to extend out to
$10-100$ pc \cite{thompson05}. Probably reality falls somewhere in between. In
fact, on the one hand observations of water maser emission clearly indicate
the presence of a rotating gaseous disc at radii of the order of 1 pc, i.e.
well within the self-gravitating portion of the disc. On the other hand it
might be difficult to reconcile with observations discs which extend much
further beyond this point. Let us see this in more details.

The main difference between between AGN discs and protostellar discs with
respect to their behaviour in the self-gravitating regime lies in the
different cooling rate, which for AGN is typically very short. This is usually
seen by noting that the heating rate needed to keep the disc marginally stable
at $Q\approx 1$ is much larger than what can be provided by a viscous
accretion disc, with reasonable values of $\alpha$ \cite{goodman03,BL2001}.
Put it in other terms, this essentially means that the cooling timescale in
AGN is much shorter than the dynamical timescale. If we now consider the
results of the simulations described above in Section \ref{sec:simulations},
we would thus conclude that the disc would fragment and form stars
\cite{shlosman89,collin99}. This would be disastrous in terms of AGN feeding
because star formation would essentially remove most of the fuel for accretion
and turn it into stars. Indeed, such arguments are the main arguments in
favour of AGN feeding from a small scale disc, but leave us with the problem
of explaining the presence of large maser emitting discs.

One possible way out of this apparent paradox can be found if energy and
angular momentum are transported by global spiral waves in the disc, rather
than by a local $\alpha$-like process. As we have seen above, this usually
happens if the disc mass is a sizable fraction of the central object mass. In
this case, as already discussed, the energy balance should include some extra
``global'' terms \cite{balbus99}, arising from wave transport of energy, that
might provide the required energy to prevent fragmentation in the outer disc.
In this picture, a density wave might remove free rotational energy from the
inner disc, but rather than dissipating it locally (as would a standard
viscous process do), it might carry it a long way out along the wave, and
release it at large radii, where the wave is dissipated \footnote{Note that
  even a standard viscous process does transport energy between different
  regions of the disc, but this does not affect the arguments above, where we
  assume that in a wave-dominated transport the relative contributions of
  dissipation and transport of energy are significantly different from a
  viscous one.}. As seen above, this requires the presence of quite massive
discs, and the evolution in this case is generally highly variable, with
episodes of strong accretion and black hole feeding followed by more quiescent
periods where the accretion rate is small. Such a time variable accretion
model has also been sometimes proposed \cite{collin07}. In some way, this
solution is similar to the proposed solution for the high mass accretion rates
observed around high mass young stars above (Section \ref{sec:highmass}).

Alternatively, even if fragmentation does occur, it is still unclear whether
accretion would be inhibited altogether, or whether we might still have
accretion with only a relatively small amount of mass ending up in stars.
Numerical simulations \cite{nayak07} show that while it is true that for small
cooling times the disc undergoes fragmentation, if a small amount of energy is
released from the forming stars (either from accretion luminosity from the
stars or from the ignition of nuclear burning), this is sufficient to prevent
further fragmentation (see also \cite{nayak06}). Depending on the parameters
(and in particular if the cooling time is close to the fragmentation
threshold) the lifetime of the gaseous disc can be very long. In this case,
the missing energy to keep the disc self-regulated would be provided by star
formation feedback rather than by global energy transport, as in the model
above.

In both models described above, the disc is then allowed to extend outside
$\sim 10^3R_{\bullet}$, that is the radius where it becomes self-gravitating,
provided that some extra-heating source is able to maintain it close to
marginal stability in a self-regulated way. We have already noticed that
self-regulated discs are thus characterized by a secondary bump in the
broadband Spectral Energy Distribution, which can be prominent at infrared
wavelengths. As shown in Fig. \ref{fig:selfregulated}, the luminosity in this
long wavelength bump is an increasing function of the location of the outer
radius of the disc \cite{LB2001}. This fact can then be used to put
constraints on the disc outer edge \cite{sirko03}. It turns out that if the
disc extends beyond $\approx 1$ pc, (and assuming that the self-regulation
condition is determined by thermal pressure, rather than by turbolent motions
of gas clumps and clouds within the disc) then the infrared bump would become
too luminous, exceeding the typical observed SED of Active Galactic Nuclei. We
thus see that, while an approximately self-regulated accretion disc can be
present well beyond 0.01 pc, thus potentially explaining the presence of dense
gaseous discs to produce the water maser emission, it is unlikely to extend
far beyond 1 pc.

\subsubsection{\bf Star formation in the Galactic Center}
In the context delineated above, an important role is played by the evidence
that has been gathered in the last few years, which points to the presence of
a large number of young stars very close to the supermassive black hole at the
center of our own Milky Way \cite{genzel03,ghez05}. In particular, most of
these stars appear to belong to two distinct stellar discs orbiting at roughly
the same distance to the black hole, i.e. at a distance of 0.05-0.5 pc
\cite{levin03,paumard06}. The most likely explanation for the origin of these
stars in that they formed {\it in situ} and in particular from the
fragmentation of a self-gravitating accretion disc \cite{levin03,nayak05b}.
Such observations thus fit naturally in the context described above, since we
know that at parsec distances an AGN accretion disc would be self-gravitating
and its cooling time is expected to be short enough to induce fragmentation.
The conditions in the Galactic Center might be typical of other galaxies,
where a nuclear starburst can be a result of the very same mechanism
\cite{nayak06,levin07}.

\subsubsection{\bf Kinematics of water masers} 
In the discussion above, we have often referred to the detection of water
maser emission from the inner parts of AGN. Indeed, such observations are a
valuable tool to study the disc at the radial distances where self-gravity is
expected to be important. In the section above, we used water maser
observations simply to indicate the mere presence of discs at radii of the
order of 0.1-1 pc. On the other hand they also offer a way to probe the disc
geometry (for example, the presence of a warp
\cite{herrnstein96,papaloizou98,greenhill03}) and, most importantly, a way to
probe its kinematic. In fact, water masers are usually observed to arise from
a series of distinct ``maser spots'', at different radial distances from the
nucleus, and from the Doppler shift of the maser emission line it is then
possible to reconstruct the rotation curve. In some cases, for example for NGC
4258, the resulting rotation curve is very close to Keplerian
\cite{miyoshi95}, and it thus allows a very precise determination of the mass
of the central BH, which for NGC 4258 is $3.6~10^7M_{\odot}$.

\begin{figure}
\centerline{\psfig{figure=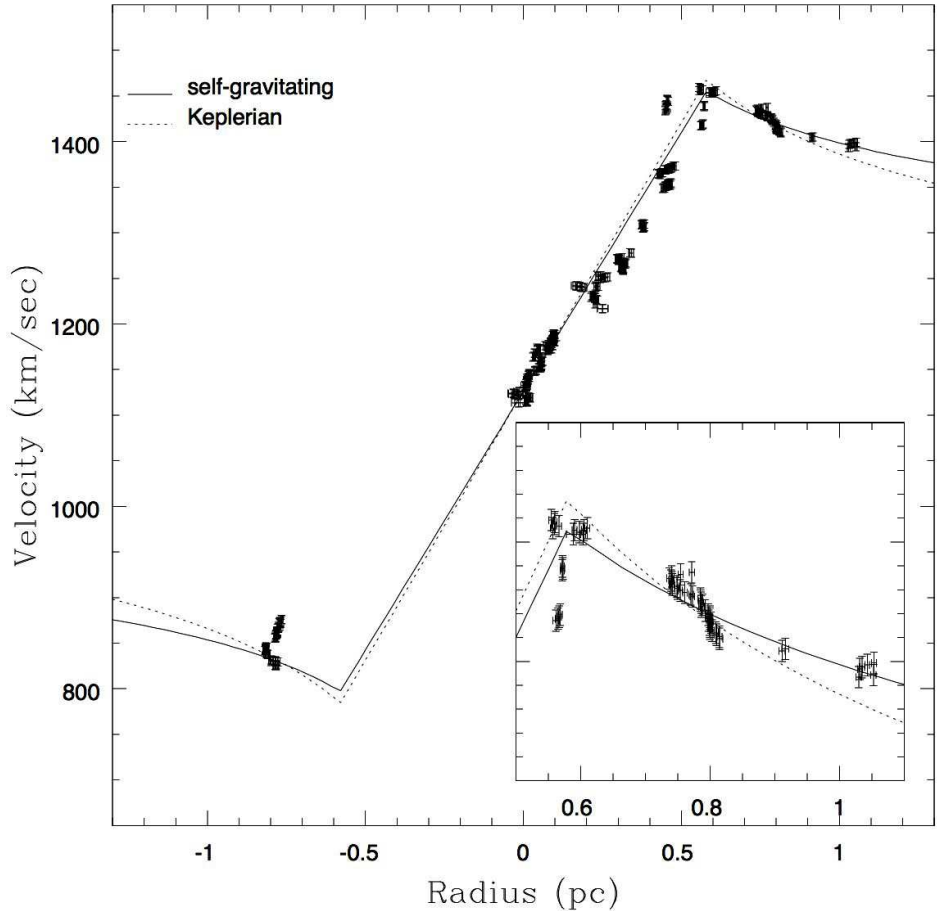,width=13cm}}
\caption{\small The data points show the observed rotation curve measured from
  water maser emission in the AGN NGC 1068. The solid line is the best -fit
  (non-Keplerian) model to the data, obtained from a self-regulated,
  self-gravitating disc model (see Section \ref{sec:selfregulation}). In this
  model both the disc and the black hole contribute to the radial
  gravitational field, and we have $M_{\rm disc}\approx M= (8.0\pm
  0.3)~10^6M_{\odot}$. The dashed line is a fit fit obtained from a Keplerian
  model, where the mass of the black hole is $M\approx
  (1.50\pm0.02~10^7M_{\odot}$.}  \label{fig:ngc1068} \end{figure} However, in
many other cases the rotation curve, while still displaying a smooth declining
profile, as would be expected for a rotating disc, does not follow exactly
Kepler's law. This is for example the case of NGC 1068
\cite{greenhill96,greenhill97}, of the Circinus galaxy \cite{greenhill03} and
of NGC 3079 \cite{kondratko05}. In particular, for the case of NGC 1068 the
maser data are consistent with a circular velocity $v_{\phi}\propto r^{-0.31}$
\cite{greenhill96}. Given the discussion above, which shows that at a scale of
a fraction of a parsec, where the maser spots are detected, the disc can be
self-gravitating, it is then tempting to attribute such (often small)
deviation from Keplerian rotation to the contribution of the disc
self-gravity.

A detailed fit to the circular velocity traced by water masers in NGC 1068
with a model which incorporates both the gravitational field of the black hole
and that of the disc has been performed \cite{LB03a} (see also \cite{hure02}),
by using the self-regulated disc models described in Section
\ref{sec:selfregulation}. The results of such modeling are shown in Fig.
\ref{fig:ngc1068}, which illustrates the observed rotation curve and the fit
with a self-regulated model \cite{LB03a}. As seen in Fig. \ref{fig:ngc1068},
the observed rotation curve of the maser spots can be divided in two different
regions: 1) at small impact parameter the masers show an apparently rising
rotation curve; 2) starting from a radial distance $\approx 0.6$ pc from the
center the rotation curve declines, following a sub-Keplerian profile.

The natural interpretation of the declining part of the rotation curve is that
it arises from material that moves parallel to our line of sight (i.e. that
lies on a disc diameter perpendicular to the line of sight).  The best
argument in favor of this interpretation is that maser amplification is
largest for material that lies close to the line of the nodes. On the other
hand, the rising part of the observed ``rotation curve'' is thought to
originate from one quarter of the disc at the inner maser disc radius, so that
the rising curve is an effect of velocity projection along the line of sight
(see also \cite{miyoshi95}). According to this interpretation, the inner
radius of the maser disc is located at $\approx 0.6$ pc and the outer disc
radius is at $\approx 1$ pc.

The best fitting self-gravitating model is shown in Fig. \ref{fig:ngc1068}
with a solid line. There are essentially two parameters that determine the
model, which can be traced back to the black hole mass $M$ and the disc
surface density $\Sigma$, which in turn (see Section \ref{sec:selfregulation})
gives us information on $\dot{M}/\alpha$. The resulting black hole mass is $M=
(8.0\pm 0.3)~10^6M_{\odot}$ and the disc mass is approximately equal to the
black hole mass. From the required disc surface density it is then possible to
obtain $\dot{M}=(28.1\pm 0.2)~\alpha M_{\odot}/$yr. The mass accretion rate
$\dot{M}$ can be estimated for example from the bolometric luminosity as
$\dot{M}\approx 0.23 \msunyr$, and we thus obtain $\alpha\approx 8.3~10^{-3}$,
which is of the right order of magnitude as would be expected from the
transport induced by gravitational instabilities.

The best-fit curve resulting from the self-gravitating disc model is not a power-law. However, as mentioned above the data are consistent with a rotation curve of the form $V\propto r^{-0.31}$ \cite{greenhill96}. Indeed, computing the quantity $\mbox{d ln}v_{\phi}/\mbox{d ln} R$ for the best fit model, it is easily seen that it ranges from $-0.35$ at the inner disc edge of the disc to $-0.30$ at the outer disc edge. 

The dashed line in Fig. \ref{fig:ngc1068} shows the result of a fit done with a Keplerian disc model. The fit is significantly worse than the non-Keplerian one (a detailed statistical analysis can be found in \cite{LB03a}). The resulting best-fit value of the black hole mass in the Keplerian fit is $M\approx (1.50\pm0.02)
~10^7M_{\odot}$. Note that the total mass (disc + black hole) of the self-gravitating best fit model is roughly the same as the black hole mass of the Keplerian fit. Therefore, a non-self-gravitating fit,
which attributes all the mass to the central object, gives the correct value for the total mass, but fails to provide the correct slope of the rotation curve.

\subsection{Supermassive black hole formation}
\label{sec:SMBH}

In the section above we have concentrated on supermassive black holes in the local Universe, i.e. in the Milky Way and in nearby AGN. We now turn our attention to very high redshifts, where we expect the emergence of the seeds of the first supermassive black holes. We will then see that even at such early epochs, phenomena related to the development of gravitational instabilities might play a key role in determining the properties of the black hole population.

Observations of AGN at high redshifts, up to $z\sim 6$ \cite{fan04,fan06}, indicate that supermassive black holes, with masses up to $10^9M_{\odot}$ were already in place when the Universe was only $10^9$ years old. This clearly requires that the black hole growth occurred at very high rates, with an average of $1\msunyr$. Such a rapid early growth poses serious challenges to models of their formation. Some models  \cite{haiman98,volonteri05,wyithe05}  assume that the seeds of supermassive black holes are the remnants of the zero-metallicity first stars (the so-called Population III stars), that are expected to be relatively massive \cite{abel00,bromm02} and thus produce black holes with a mass of up to 100 $M_{\odot}$. However, unless the efficiency of conversion of matter into energy through the accretion process is very low, it is impossible to grow the seeds to the required masses by $z\sim 6$ through Eddington-limited accretion \cite{volonteri05b}. The problem here is that when the accretion rate is large, the radiation pressure produced by the accretion luminosity can exceed the gravitational force of the black hole and hence lead to an expulsion instead of accretion of matter. The limiting luminosity at which this happens is called the Eddington limit. Now, if the accretion efficiency $\epsilon=L/\dot{M}c^2$exceeds $\approx 0.1$ (where $L$ is the accretion luminosity and $c$ is the speed of light), the Eddington limit does not allow the large accretion rates needed to grow the seeds fast enough to become bright AGN by $z\sim 6$ \cite{volonteri05b}. Note also that the Eddington limit is linearly proportional to the black hole mass, so that the problem of accreting at very high rates is particularly important in the earliest phases of the growth, when the black hole mass is small. Once the black hole mass has exceeded $10^4-10^5M_{\odot}$, further growth at $\approx 1\msunyr$ can proceed at sub-Eddington rates. 

The efficiency is in turn dependent on the spin of the black hole, with high
spin producing very large efficiencies $\epsilon\sim 0.5$. Accretion of matter
naturally tends to spin up the hole \cite{volonteri05} and hence to increase
the efficiency, thus exceeding the Eddington limit for relatively low
$\dot{M}$ and preventing a fast growth of the hole. While recent calculations \cite{KLOP,LP06} show that it is possible to keep the hole spin low if the growth occurs through several small randomly oriented accretion episodes \cite{king06}, we still have to face the issue of how to produce the high infall rates required. 

Alternative models propose the direct formation of more massive seeds with masses of about $10^5\,M_{\odot}$ directly out of the collapse of dense gas \cite{haehnelt93,umemura93,loeb94,eisenstein95,bromm03,koushiappas04,begelman06}. The key limiting factor for these models is the disposal of the angular momentum. Recently, it has been proposed \cite{koushiappas04,begelman06,LN06,LN07} that large scale gravitational instabilities developing during the growth of pre-galactic discs is the missing ingredient, able to funnel the required amount of gas into the center of the galaxy.

According to such models, the formation of the seeds of supermassive black holes occurs at a redshift $z\sim10-15$, when the intergalactic medium had not been yet enriched by metals forming in the first stars. As a consequence, the chemical composition of the gas at this early epoch is essentially primordial, i.e. the gas is mostly hydrogen and helium. The cooling properties of this gas are therefore relatively simple. In particular, in the absence of molecular hydrogen, the main coolant is provided by atomic hydrogen, for which the cooling timescale becomes extremely long for temperatures smaller than $\sim 10^4$ K, and we thus expect the gas to reach thermal equilibrium at a temperature $T_{\rm gas}$ of the order of $10^4$K. 

Now, consider a dark matter halo (modeled, for simplicity, as a truncated singular isothermal sphere) of mass $M_{\rm halo}$ and circular velocity $V_{\rm h}$, extending out to $r_{\rm h}=GM_{\rm halo}/V_{\rm h}^2$. We also assume that the halo contains a gas mass $M_{\rm gas}=m_{\rm d}M_{\rm halo}$, where $m_{\rm d}$ is of the order of the universal baryonic fraction, $\approx 0.1$, whose angular momentum is $J_{\rm gas}=j_{\rm d}J$, where $j_{\rm d}\sim m_{\rm d}$. The angular momentum of the dark matter halo $J$ is expressed in terms of its spin parameter $\lambda=J|E|^{1/2}/GM_{\rm halo}^{5/2}$, where $E$ is its total energy.  The probability distribution of the spin parameter of dark matter halos can be obtained from cosmological N-body simulations \cite{warren92} and is well described by a log-normal distribution peaking at $\lambda=0.05$. 

If the virial temperature of the halo $T_{\rm vir}\propto V_{\rm h}^2$ is larger than the gas temperature $T_{\rm gas}$, the gas collapses and forms a rotationally supported disc, with circular velocity $V_{\rm h}$, determined by the gravitational field of the halo. For low values of the spin parameter $\lambda$ the resulting disc can be compact and dense. In this case, during the infall of gas onto the disc, its density rises until the stability parameter $Q$ becomes of the order of unity. At this point, the disc starts developing a gravitational instability, which as we have seen above is able to efficiently redistribute angular momentum and allow accretion. Further infall of gas does not cause the density to rise much further, but rather it promotes an increasingly high accretion rate into the center. This process goes on until infall is over and the disc has attained a surface density low enough to be marginally gravitationally stable, i.e. with $Q=\bar{Q}$. It is then possible to calculate what fraction of the infalling mass needs to be transported into the center to make the disc marginally stable, as a function of the main parameters involved. In this way, we get \cite{LN06,LN07}:
\begin{equation}
M_{\rm BH}=m_{\rm d}M_{\rm halo}\left[1-\sqrt{\frac{8\lambda}{m_{\rm d}\bar{Q}}
\left(\frac{j_{\rm d}}{m_{\rm d}}\right)\left(\frac{T_{\rm gas}}{T_{\rm
vir}}\right)^{1/2}}\right],
\label{eq:mbh}
\end{equation}
where we have suggestively called $M_{\rm BH}$ the accreted mass, since this mass is the total mass available for the formation of the black hole seed in the center. 

\begin{figure}
\centerline{\psfig{figure=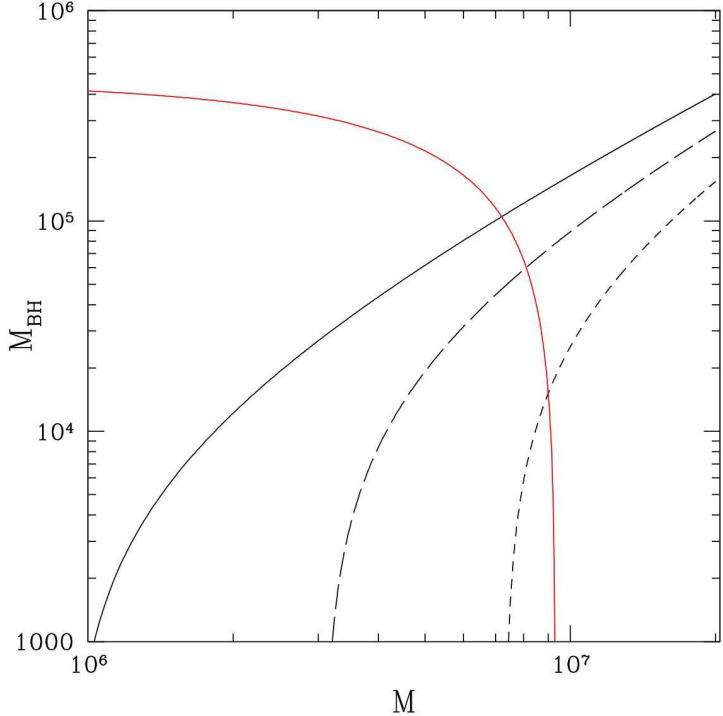,width=10cm}}
\caption{\small Mass available for the formation of the seed of a supermassive black hole in the center of pre-galactic discs as a function of the mass of the parent dark matter halo. The plots refer to the following choice of parameters: $\bar{Q}=2$, $T_{\rm gas}=4000$K, $m_{\rm d}=j_{\rm d}=0.05$, $\lambda = 0.01$ (solid line), $\lambda=0.015$ (long-dashed line), $\lambda=0.02$ (short-dashed line). The red curve shows the threshold for fragmentation from Equation (\ref{eq:frag}), with $\alpha_{\rm c}=0.06$. Halos on the right of the red line give rise to fragmenting discs.}
\label{fig:seed}
\end{figure}

However, for large halo mass, the internal torques needed to redistribute the excess baryonic mass become too large to be sustained by the disc, which might then undergo fragmentation. We have seen in the previous sections that the maximum torque that can be delivered by a quasi-steady self-regulated disc is of the order of $\alpha_{\rm c}\approx 0.06$. Since the infall rate of gas from the halo is proportional to $T_{\rm vir}^{3/2}$, we expect fragmentation when the virial temperature exceeds a critical value $T_{\rm max}$, given by (see \cite{LN06} for details):
\begin{equation}
\frac{T_{\rm max}}{T_{\rm gas}}>\left(\frac{4\alpha_{\rm c}}{m_{\rm
d}}\frac{1}{1+M_{\rm BH}/m_{\rm d}M_{\rm halo}}\right)^{2/3},
\label{eq:frag}
\end{equation}
Although it is possible, as often mentioned in this paper, that accretion proceeds even for larger values of $\alpha$ in a highly time-variable way when the disc mass is large, and it is also possible that accretion proceeds even in a fragmenting disc, we make here the conservative assumption that all halos that violate Eq, (\ref{eq:frag}) do fragment and do not accrete.  Fig. \ref{fig:seed} illustrates the relationship between halo mass and black hole mass based on Equation (\ref{eq:mbh}) for three different values of the spin parameter $\lambda$. The red line in Fig. \ref{fig:seed} correspond to Eq. (\ref{eq:frag}), so that halos on the right of the red line are expected to fragment. We can thus see that the typical mass fed into the center of such pre-galactic disc is of the order of $10^3M_{\odot}$ up to $10^5M_{\odot}$. The typical accretion rates during this early epochs is of the order of $10^{-2}\msunyr$ \cite{LN06}.  If such high masses are assembled as seeds of supermassive black holes at redshift $10-15$, it is then easy to grow through Eddington-limited accretion to $10^9M_{\odot}$ by $z=6$, as required by observations. 

\begin{figure}
\centerline{\psfig{figure=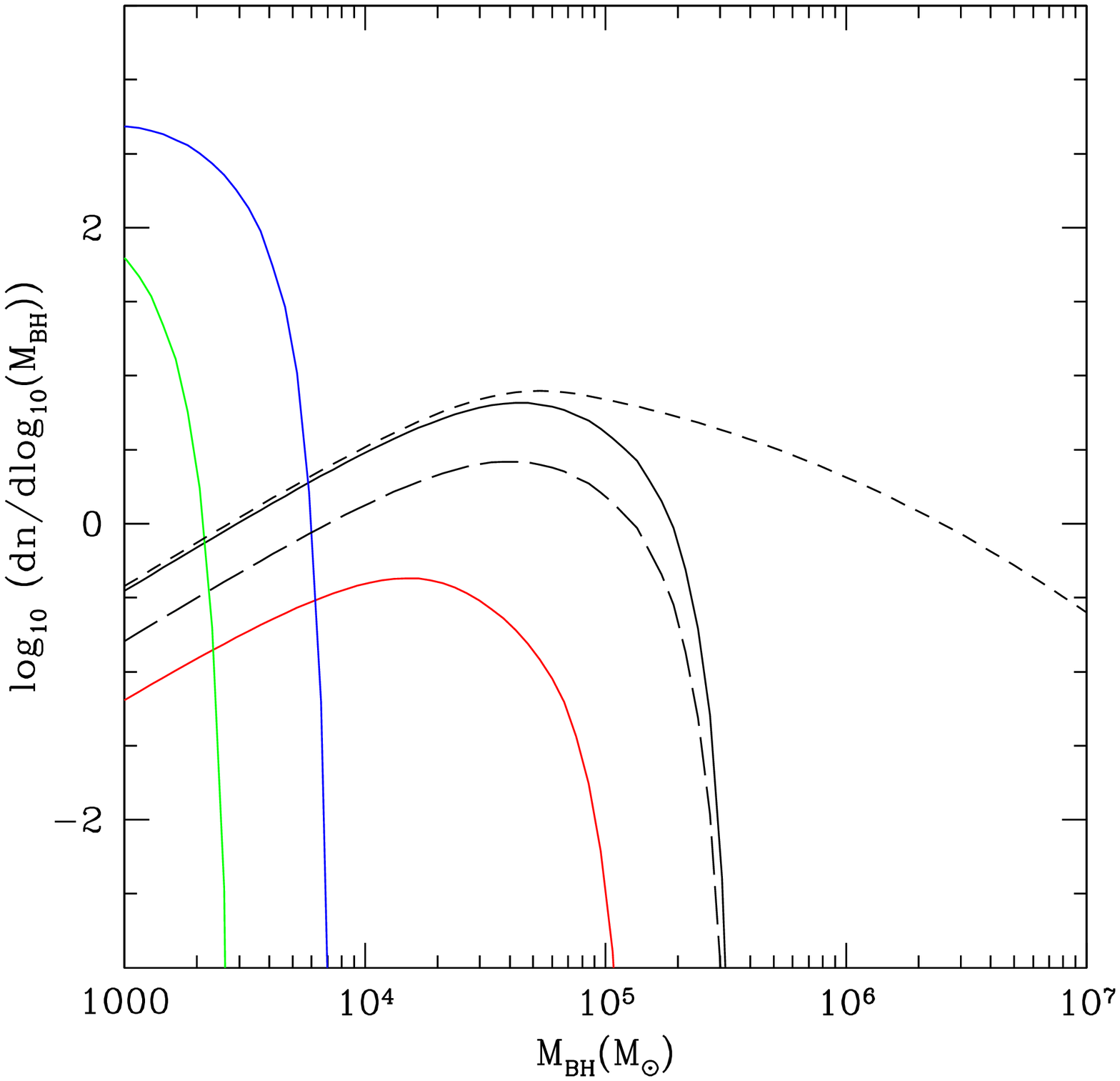,width=7cm}\psfig{figure=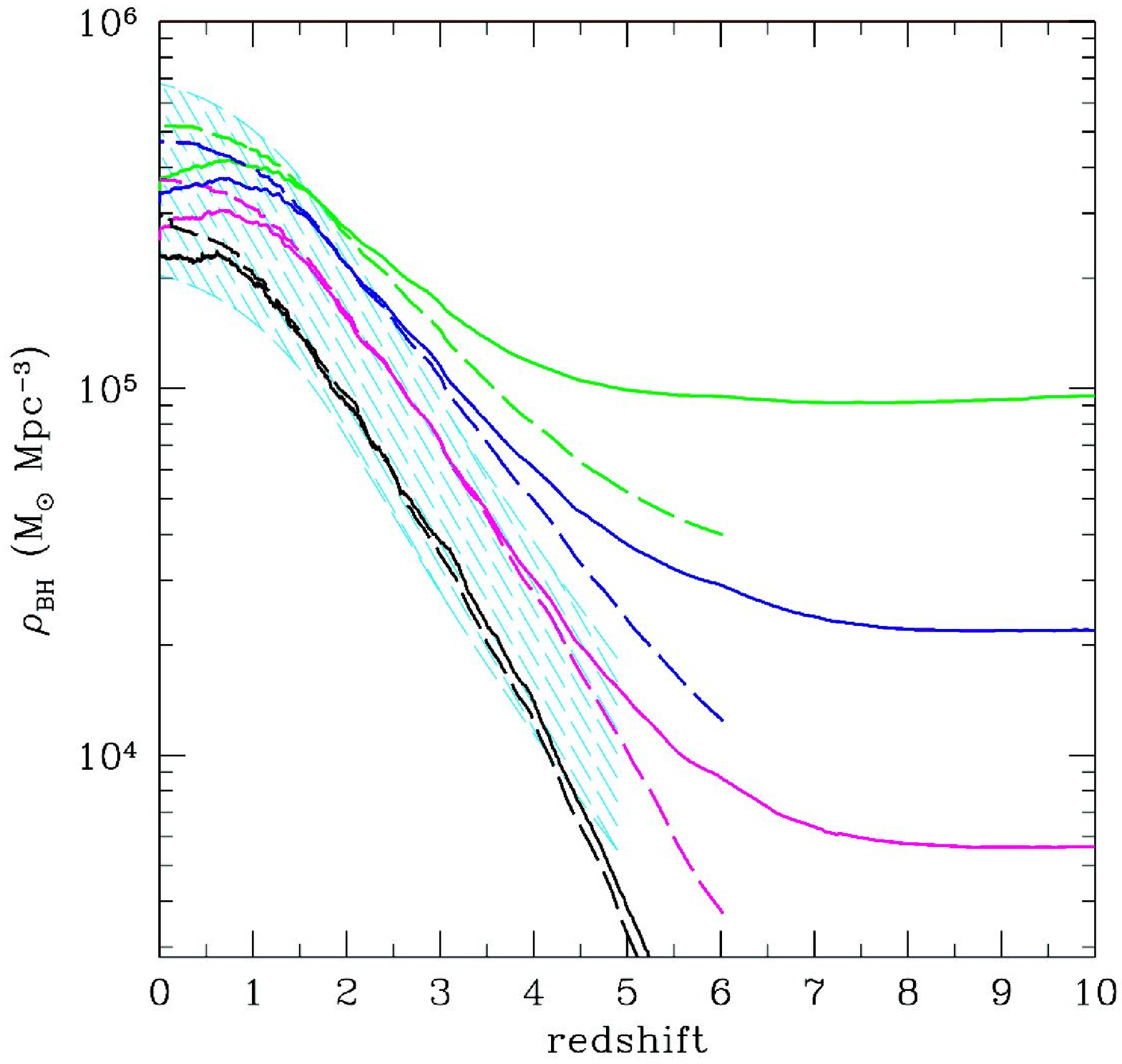,width=7.3cm}}
\caption{\small Left panel: Mass function of seed black holes predicted by the model based on Eqs. (\ref{eq:mbh}) and (\ref{eq:frag}). The black solid line refers to $z=10$, while the red line refers to $z=20$. The long-dashed line shows the effect of reducing $\bar{Q}$ from 2 to 1.5. The short dashed line shows the effect of not including the possibility of fragmentation (more details can be found in \cite{LN07}). Right panel: the integrated density of black holes predicted from a merger tree evolution of the black hole seed population. The three solid line refer to three different choices of the parameters (see \cite{VLN07} for details). The dashed area indicate the observationally permitted region from estimates of the black hole density at low redshifts. At high redshift ($z\gtrsim 6$) the density reflects the one attained after the seed formation phase, while the rise at lower redshifts indicates the growth through AGN activity.}
\label{fig:BH_pop}
\end{figure}

Equation (\ref{eq:mbh}) provides a powerful link between the properties of dark matter haloes and the mass of massive seed black holes that can grow within them. As shown, the amount of mass that will
be concentrated in the central regions of these pre-galactic discs depends only on halo properties (such as the spin parameter $\lambda$ and the fraction of baryonic mass that collapses to the disc $m_{\rm
d}$), on the ratio between gas temperature and halo virial temperature, and on the threshold value of $Q$, which has a very small range of variation around $\bar{Q}\approx 1$. This simple model has been used to calculate several properties of the black hole population at high redshift. In particular, from the distribution of halo masses and angular momentum it is straightforward to derive the mass function of the supermassive black hole seeds \cite{LN07}, which turns out to be strongly peaked at around $10^5M_{\odot}$, as shown in the left panel of Fig. \ref{fig:BH_pop}. Furthermore, it is also possible to include such a simple prescription within evolutionary models that track the properties of the black hole population along cosmic time, such as merger tree models \cite{VLN07} (such models are usually based on the so called hierarchical model for the growth of structures, where the structures observed in the present day Universe are the result of the subsequent merger of smaller structures at early times). It is then interesting to see that the evolution of such a primordial seed population can naturally account for the current estimates of the density of black holes at low redshift (Fig. \ref{fig:BH_pop}, right panel). In addition, an important and testable prediction of such models is that dwarf galaxies, that did not have any progenitor massive enough to seed a black hole, should not host a supermassive black hole. In particular if the velocity dispersion of the galaxy is below $\sim 50$km/sec, the probability of hosting a black hole turns out to be negligibly small \cite{VLN07}.

\section{CONCLUSIONS AND CHALLENGES FOR THE FUTURE}
\label{sec:conclusions}

As we have seen, self-gravitating accretion discs are an important aspect of the modeling of several systems, and thus play a key role in modern astrophysics. The influence of the disc self-gravity extends from the very small scales, associated with the formation of planetary systems, up to the large scales, associated with the formation and the feeding of supermassive black holes in the nuclei of galaxies. While clearly the specific properties of systems associated with such hugely different scales are going to differ substantially, there is yet a unifying framework related to the scale-free nature of the gravitational force.  

Because of self-gravity, accretion discs can develop gravitational instabilities. These are going to deeply affect the structure and the evolution of the accretion disc. Firstly, we have seen that the instability usually takes the form of a regular spiral structure, that can efficiently redistribute the disc angular momentum and thus allow the accretion process to take place. In this respect the role of the disc self-gravity is complementary to that of other instabilities, such as magneto-hydrodynamic (MHD) instabilities \cite{balbusreview,balbus03}. Indeed, while MHD instabilities are likely to be the most important source of angular momentum transport in the inner and hotter parts of the disc, it is not clear whether they are able to provide the necessary torques in colder environments, where the ionization level of the gas is low and the coupling to the magnetic field is weaker. This is, for example, the case of the outer parts of protostellar discs \cite{fromang02}. On the other hand, such cold parts of the discs are naturally subject to gravitational instabilities. It is then possible to envisage a situation where gravitational instabilities are responsible for bringing the accreting material from large distances (which might be of the order of tens of AU for protostellar discs, and a fraction of a parsec for AGN discs) into the inner disc, where MHD instabilities take over and are responsible for the ultimate deposition of the matter onto the central object. In this respect, it would be important to develop models which include both the effects of MHD induced turbulence and that of gravitational instabilities. Semi-analytical models along these lines have been discussed \cite{kratter07}, but it is still unclear what would be the interplay between magnetic and gravitational instabilities in the general case \cite{fromang04a,fromang04b}.

A second aspect related to gravitational instabilities is the natural tendency associated with self-gravity to collapse and produce gravitationally bound objects. Disc fragmentation is a possible outcome of the onset of gravitational instabilities. As discussed above, it is now well established that the conditions leading to fragmentation are essentially related to the disc thermodynamics and to the cooling rate. From the modeling point of view, fragmentation has a twofold aspect. On the one hand, if one is interested in producing smaller objects within the disc, it offers a natural way to accomplish the task. On the other hand, there is the danger that most of the disc mass would end up in the fragments, with very little being accreted onto the central object. In this respect, fragmentation is a dangerous outcome, which might inhibit accretion.

We have seen that the biggest difference between protostellar discs and AGN discs with respect to self-gravity lies indeed in their thermal properties, and in particular in the fact that AGN discs are expected to be cold (in the sense that $H/R\ll 1$) and with a short cooling timescale (in the sense that $\tau_{\rm cool}\Omega\ll 1$), while protostellar discs are hot ($H/R\approx 0.1$) and with a long cooling timescale ($\tau_{\rm cool}\Omega\gg 1$, at least within 100 AU). The behaviour of these two kind of systems in this respect is then significantly different. Most protostellar discs are not expected to fragment, except perhaps in their outermost regions. This is the main problem affecting models of planet formation which rely on direct disc fragmentation. At the other end of the scales that we have considered, we see that AGN discs instead are likely to fragment. This is good news in view of forming young massive stars in the vicinity of the central supermassive black hole, as are observed, for example, close to the black hole in our Galactic Center. On the other hand, we have discussed above the problems that might arise from disc fragmentation in this context, related to the feeding of the black hole. 

The last 10-15 years have witnessed a significant improvement of our understanding of the evolution of self-gravitating accretion discs. This has been made possible through the interplay between simple analytical models and the results of detailed numerical simulations of the disc hydrodynamics. The big advances in computing power in the last decade have made it possible to run such numerical simulations with unprecedented resolution, and progressively including the effects of several physical processes in more detail. In this paper, emphasis has been put on the important role of the gas cooling rate in determining the outcome of the instability. This has been often introduced in a simplified way in the simulations, which has led to a valuable investigation of the different regimes, in a somewhat academic way. The theoretical challenge in this respect is to introduce in the numerical codes a more realistic cooling function, in order to understand the behaviour of actual systems. For example, most of the current debate on the applicability of the fragmentation scenario for planet formation is ultimately due to our ignorance of the actual thermodynamics at work in protostellar discs. The difficulty here is to treat numerically the radiative transfer within very optically thick media. Only very recently have numerical codes implemented radiation physics and, although still preliminary, some results are starting to emerge \cite{boley06,mayer07}. Codes which couple radiative transfer with the disc hydrodynamics are also desired to include the possibly important effect of irradiation from the central object, which has mostly been neglected in the simulations run so far (with some exceptions \cite{cai06}). 

If we look at the problem of star formation within AGN accretion discs, here the issue is ultimately whether accretion can proceed at significant rates even if the disc fragments. In other words, it is still unclear how much gas needs to be turned into stars before fragmentation is quenched. Here, it would be important to include in the models the energy feedback arising from the forming stars. Even in this context then, the key ingredient to be added into numerical codes is a proper treatment of radiative transfer, coupled with hydrodynamics.

\begin{acknowledgments}

ACKNOLEDGMENTS. I would like to dedicate this paper to the memory of Eduardo Delgado-Donate, who contributed to the development of some of the ideas presented here. I acknowledge several interesting discussions with Giuseppe Bertin, Cathie Clarke, Andrew King, Sergei Nayakshin and Jim Pringle. I would also like to thank the various collaborators with whom I have worked on these issues along the years, and in particular Phil Armitage, Priya Natarajan and Ken Rice. 

\end{acknowledgments}

\bibliographystyle{varenna}
\bibliography{lodato}

\begin{thebibliography}{100}
\expandafter\ifx\csname url\endcsname\relax\def\url#1{\texttt{#1}}\fi
\expandafter\ifx\csname urlprefix\endcsname\relax\def\urlprefix{URL }\fi

\bibitem{bertinbook}
\NAME{Bertin G. \atque Lin C.~C.}, \TITLE{Spiral Structure in Galaxies: a
  Density Wave Theory} (MIT Press, Cambridge) 1996.

\bibitem{miyoshi95}
\NAME{{Miyoshi} M. \etal}, \IN{Nature }{373}{1995}{127}.

\bibitem{greenhill97}
\NAME{{Greenhill} L.~J. \atque {Gwinn} C.~R.}, \IN{Astrophysics \& Space
  Science }{248}{1997}{261}.

\bibitem{kondratko06}
\NAME{{Kondratko} P.~T., {Greenhill} L.~J. \atque {Moran} J.~M.}, \IN{ApJ
  }{652}{2006}{136}.

\bibitem{franck}
\NAME{Frank J., King A. \atque Raine D.}, \TITLE{Accretion Power in
  Astrophysics} (Cambridge University Press, Cambridge) 2002.

\bibitem{PPV}
\NAME{{Reipurth} B., {Jewitt} D. \atque {Keil} K.} (Eds.), \TITLE{{Protostars
  and Planets V}} (Arizona University Press) 2007.

\bibitem{pacinski78}
\NAME{Paczy\'nski B.}, \IN{Acta Astronomica }{28}{1978}{91}.

\bibitem{kolikalov79}
\NAME{Kolykhalov P.~I. \atque Sunyaev R.~A.}, \IN{Soviet Astronomy Letters
  }{5}{1979}{180}.

\bibitem{linpringle87}
\NAME{Lin D. N.~C. \atque Pringle J.~E.}, \IN{MNRAS }{225}{1987}{607}.

\bibitem{linpringle90}
\NAME{Lin D. N.~C. \atque Pringle J.~E.}, \IN{ApJ }{358}{1990}{515}.

\bibitem{pringle81}
\NAME{Pringle J.~E.}, \IN{ARA\&A }{19}{1981}{137}.

\bibitem{binney}
\NAME{{Binney} J. \atque {Tremaine} S.}, \TITLE{{Galactic dynamics}}
  (Princeton, NJ, Princeton University Press, 1987, 747 p.) 1987.

\bibitem{BL99}
\NAME{Bertin G. \atque Lodato G.}, \IN{A\&A }{350}{1999}{694}.

\bibitem{mestel63}
\NAME{{Mestel} L.}, \IN{MNRAS }{126}{1963}{553}.

\bibitem{bertin97}
\NAME{Bertin G.}, \IN{ApJ }{478}{1997}{L71}.

\bibitem{beckwith90}
\NAME{{Beckwith} S.~V.~W., {Sargent} A.~I., {Chini} R.~S. \atque {Guesten} R.},
  \IN{AJ }{99}{1990}{924}.

\bibitem{cesaroni94}
\NAME{Cesaroni R. \etal}, \IN{ApJ }{435}{1994}{137}.

\bibitem{cesaroni05}
\NAME{{Cesaroni} R., {Neri} R., {Olmi} L., {Testi} L., {Walmsley} C.~M. \atque
  {Hofner} P.}, \IN{A\&A }{434}{2005}{1039}.

\bibitem{LB03a}
\NAME{Lodato G. \atque Bertin G.}, \IN{A\&A }{398}{2003}{517}.

\bibitem{weiden77}
\NAME{Weidenschilling S.}, \IN{MNRAS }{180}{1977}{57}.

\bibitem{rice04}
\NAME{Rice W. K.~M., Lodato G., Pringle J.~E., Armitage P.~J. \atque Bonnell
  I.~A.}, \IN{MNRAS }{355}{2004}{543}.

\bibitem{rice06}
\NAME{Rice W. K.~M., Lodato G., Pringle J.~E., Armitage P.~J. \atque Bonnell
  I.~A.}, \IN{MNRAS }{372}{2006}{L9}.

\bibitem{spitzer42}
\NAME{Spitzer L.}, \IN{ApJ }{95}{1942}{329}.

\bibitem{toomre64}
\NAME{Toomre A.}, \IN{ApJ }{139}{1964}{1217}.

\bibitem{bertinbook2}
\NAME{Bertin G.}, \TITLE{Dynamics of Galaxies} (Cambridge University Press,
  Cambridge) 2000.

\bibitem{shakura73}
\NAME{Shakura N.~I. \atque Sunyaev R.~A.}, \IN{A\&A }{24}{1973}{337}.

\bibitem{hartmann98}
\NAME{{Hartmann} L., {Calvet} N., {Gullbring} E. \atque {D'Alessio} P.},
  \IN{ApJ }{495}{1998}{385}.

\bibitem{lasota01}
\NAME{Lasota J.~P.}, \IN{New Astronomy Reviews }{45}{2001}{449}.

\bibitem{bellin94}
\NAME{Bell K.~R. \atque Lin D. N.~C.}, \IN{ApJ }{427}{1994}{987}.

\bibitem{LC04}
\NAME{{Lodato} G. \atque {Clarke} C.~J.}, \IN{MNRAS }{353}{2004}{841}.

\bibitem{balbusreview}
\NAME{Balbus S.~A. \atque Hawley J.~F.}, \IN{Reviews of Modern Physics
  }{70}{1998}{1}.

\bibitem{hartmann06}
\NAME{{Hartmann} L., {D'Alessio} P., {Calvet} N. \atque {Muzerolle} J.},
  \IN{ApJ }{648}{2006}{484}.

\bibitem{clarke07}
\NAME{{Clarke} C.~J.}, \IN{MNRAS }{376}{2007}{1350}.

\bibitem{kratter07}
\NAME{{Kratter} K.~M., {Matzner} C.~D. \atque {Krumholz} M.~R.}, \IN{ArXiv
  e-prints }{0709.4252}{2007}{}.

\bibitem{lyndenbell74}
\NAME{Lynden-Bell D. \atque Pringle J.~E.}, \IN{MNRAS }{168}{1974}{603}.

\bibitem{hartmann}
\NAME{Hartmann L.}, \TITLE{Accretion Processes in Star Formation} (Cambridge
  University Press, Cambridge) 1998.

\bibitem{lyndenbell69}
\NAME{Lynden-Bell D.}, \IN{Nature }{223}{1969}{690}.

\bibitem{adams88}
\NAME{Adams F.~C., Lada C.~J. \atque Shu F.~H.}, \IN{ApJ }{326}{1988}{865}.

\bibitem{chiang97}
\NAME{Chiang E.~P. \atque Goldreich P.}, \IN{ApJ }{490}{1997}{368}.

\bibitem{dullemondPPV}
\NAME{{Dullemond} C.~P., {Hollenbach} D., {Kamp} I. \atque {D'Alessio} P.},
  \IN{Protostars and Planets V }{}{2007}{555}.

\bibitem{hohl71}
\NAME{Hohl F.}, \IN{ApJ }{168}{1971}{343}.

\bibitem{ostriker73}
\NAME{{Ostriker} J.~P. \atque {Peebles} P.~J.~E.}, \IN{ApJ }{186}{1973}{467}.

\bibitem{laubertin78}
\NAME{Lau Y.~Y. \atque Bertin G.}, \IN{ApJ }{226}{1978}{508}.

\bibitem{bertin89}
\NAME{{Bertin} G., {Lin} C.~C., {Lowe} S.~A. \atque {Thurstans} R.~P.}, \IN{ApJ
  }{338}{1989}{104}.

\bibitem{LR04}
\NAME{Lodato G. \atque Rice W. K.~M.}, \IN{MNRAS }{351}{2004}{630}.

\bibitem{LR05}
\NAME{Lodato G. \atque Rice W. K.~M.}, \IN{MNRAS }{358}{2005}{1489}.

\bibitem{vandervoort70}
\NAME{{Vandervoort} P.~O.}, \IN{ApJ }{161}{1970}{87}.

\bibitem{bertin88}
\NAME{Bertin G. \atque Romeo A.~B.}, \IN{A\&A }{195}{1988}{105}.

\bibitem{LB2001}
\NAME{Lodato G. \atque Bertin G.}, \IN{A\&A }{375}{2001}{455}.

\bibitem{vorobyov07}
\NAME{{Vorobyov} E.~I. \atque {Basu} S.}, \IN{MNRAS }{}{2007}{855}.

\bibitem{boley06}
\NAME{{Boley} A.~C., {Mej{\'{\i}}a} A.~C., {Durisen} R.~H., {Cai} K., {Pickett}
  M.~K. \atque {D'Alessio} P.}, \IN{ApJ }{651}{2006}{517}.

\bibitem{sirko03}
\NAME{Sirko E. \atque Goodman J.}, \IN{MNRAS }{341}{2003}{501}.

\bibitem{miller70}
\NAME{{Miller} R.~H., {Prendergast} K.~H. \atque {Quirk} W.~J.}, \IN{ApJ
  }{161}{1970}{903}.

\bibitem{hohl73}
\NAME{Hohl F.}, \IN{ApJ }{184}{1973}{353}.

\bibitem{anthony83}
\NAME{{Anthony} D.~M. \atque {Carlberg} R.~G.}, \IN{ApJ }{332}{1988}{637}.

\bibitem{laughlin94}
\NAME{Laughlin G. \atque Bodenheimer P.}, \IN{ApJ }{436}{1994}{335}.

\bibitem{laughlin96}
\NAME{Laughlin G. \atque R\`o\.zyczka M.}, \IN{ApJ }{456}{1996}{279}.

\bibitem{laughlin98}
\NAME{{Laughlin} G., {Korchagin} V. \atque {Adams} F.~C.}, \IN{ApJ
  }{504}{1998}{945}.

\bibitem{pickett98}
\NAME{Pickett B.~K. \etal}, \IN{ApJ }{504}{1998}{468}.

\bibitem{pickett2000}
\NAME{Pickett B.~K. \etal}, \IN{ApJ }{529}{2000}{1034}.

\bibitem{gammie01}
\NAME{Gammie C.~F.}, \IN{ApJ }{553}{2001}{174}.

\bibitem{rice03a}
\NAME{Rice W. K.~M., Armitage P.~J., Bate M.~R. \atque Bonnell I.~A.},
  \IN{MNRAS }{338}{2003}{227}.

\bibitem{rice03b}
\NAME{Rice W. K.~M., Armitage P.~J., Bate M.~R., Bonnell I.~A., Jeffers S.~V.
  \atque Vine S.~G.}, \IN{MNRAS }{346}{2003}{L36}.

\bibitem{rice03c}
\NAME{Rice W. K.~M., Armitage P.~J., Bate M.~R. \atque Bonnell I.~A.},
  \IN{MNRAS }{339}{2003}{1025}.

\bibitem{RLA05}
\NAME{{Rice} W.~K.~M., {Lodato} G. \atque {Armitage} P.~J.}, \IN{MNRAS
  }{364}{2005}{L56}.

\bibitem{pickett03}
\NAME{{Pickett} B.~K., {Mej{\'{\i}}a} A.~C., {Durisen} R.~H., {Cassen} P.~M.,
  {Berry} D.~K. \atque {Link} R.~P.}, \IN{ApJ }{590}{2003}{1060}.

\bibitem{mejia05}
\NAME{Mejia A.~C., Durisen R.~H., Pickett M.~K. \atque Cai K.}, \IN{ApJ
  }{619}{2005}{1098}.

\bibitem{mayer04}
\NAME{{Mayer} L., {Quinn} T., {Wadsley} J. \atque {Stadel} J.}, \IN{ApJ
  }{609}{2004}{1045}.

\bibitem{rees76}
\NAME{{Rees} M.~J.}, \IN{MNRAS }{176}{1976}{483}.

\bibitem{silk77}
\NAME{{Silk} J.}, \IN{ApJ }{214}{1977}{152}.

\bibitem{CHL07}
\NAME{{Clarke} C., {Harper-Clark} E. \atque {Lodato} G.}, \IN{MNRAS
  }{381}{2007}{1543}.

\bibitem{johnson03}
\NAME{Johnson B.~M. \atque Gammie C.~F.}, \IN{ApJ }{597}{2003}{131}.

\bibitem{whitehouse05}
\NAME{{Whitehouse} S.~C., {Bate} M.~R. \atque {Monaghan} J.~J.}, \IN{MNRAS
  }{364}{2005}{1367}.

\bibitem{mayer07}
\NAME{{Mayer} L., {Lufkin} G., {Quinn} T. \atque {Wadsley} J.}, \IN{ApJ
  }{661}{2007}{L77}.

\bibitem{stamatellos07}
\NAME{{Stamatellos} D., {Whitworth} A.~P., {Bisbas} T. \atque {Goodwin} S.},
  \IN{A\&A }{475}{2007}{37}.

\bibitem{shu70}
\NAME{{Shu} F.~H.}, \IN{ApJ }{160}{1970}{99}.

\bibitem{lyndenbell72}
\NAME{Lynden-Bell D. \atque Kalnajs A.~J.}, \IN{MNRAS }{157}{1972}{1}.

\bibitem{balbus03}
\NAME{{Balbus} S.~A.}, \IN{ARA\&A }{41}{2003}{555}.

\bibitem{balbus99}
\NAME{Balbus S.~A. \atque Papaloizou J. C.~B.}, \IN{ApJ }{521}{1999}{650}.

\bibitem{mckee07}
\NAME{{McKee} C.~F. \atque {Ostriker} E.~C.}, \IN{ARA\&A }{45}{2007}{565}.

\bibitem{lada84}
\NAME{{Lada} C.~J. \atque {Wilking} B.~A.}, \IN{ApJ }{287}{1984}{610}.

\bibitem{andrews05}
\NAME{{Andrews} S.~M. \atque {Williams} J.~P.}, \IN{ApJ }{631}{2005}{1134}.

\bibitem{andrews07}
\NAME{{Andrews} S.~M. \atque {Williams} J.~P.}, \IN{ApJ }{659}{2007}{705}.

\bibitem{eisner06}
\NAME{{Eisner} J.~A. \atque {Carpenter} J.~M.}, \IN{ApJ }{641}{2006}{1162}.

\bibitem{testi2001}
\NAME{Testi L. \etal}, \IN{ApJ }{554}{2001}{1087}.

\bibitem{natta04}
\NAME{Natta A., Testi L., Neri R., Shepherd D.~S. \atque Wilner D.~J.},
  \IN{A\&A }{416}{2004}{179}.

\bibitem{gullbring98}
\NAME{Gullbring E., Hartmann L., Brice\~{n}o C. \atque Calvet N.}, \IN{ApJ
  }{492}{1998}{323}.

\bibitem{gullbring2000}
\NAME{Gullbring E., Calvet N., Muzerolle J. \atque Hartmann L.}, \IN{ApJ
  }{544}{2000}{927}.

\bibitem{lubow96}
\NAME{Artymowicz P. \atque Lubow S.~H.}, \IN{ApJ }{467}{1996}{77L}.

\bibitem{lin06}
\NAME{{Lin} S.-Y. \etal}, \IN{ApJ }{645}{2006}{1297}.

\bibitem{pietu05}
\NAME{{Pi{\'e}tu} V., {Guilloteau} S. \atque {Dutrey} A.}, \IN{A\&A
  }{443}{2005}{945}.

\bibitem{rodriguez05}
\NAME{Rodriguez L.~F., Loinard L., D'Alessio P., Wilner D. \atque P.T.P. H.},
  \IN{ApJ }{621}{2005}{L133}.

\bibitem{CLMI}
\NAME{{Clarke} C., {Lodato} G., {Melnikov} S.~Y. \atque {Ibrahimov} M.~A.},
  \IN{MNRAS }{361}{2005}{942}.

\bibitem{kenyon91}
\NAME{Kenyon S.~J. \atque Hartmann L.}, \IN{ApJ }{383}{1991}{664}.

\bibitem{millan06}
\NAME{{Millan-Gabet} R. \etal}, \IN{ApJ }{641}{2006}{547}.

\bibitem{kenyon85}
\NAME{{Hartmann} L. \atque {Kenyon} S.~J.}, \IN{ApJ }{299}{1985}{462}.

\bibitem{kenyon88}
\NAME{{Kenyon} S.~J., {Hartmann} L. \atque {Hewett} R.}, \IN{ApJ
  }{325}{1988}{231}.

\bibitem{LB03b}
\NAME{Lodato G. \atque Bertin G.}, \IN{A\&A }{408}{2003}{1015}.

\bibitem{belletal95}
\NAME{{Bell} K.~R., {Lin} D.~N.~C., {Hartmann} L.~W. \atque {Kenyon} S.~J.},
  \IN{ApJ }{444}{1995}{376}.

\bibitem{clarkelin90}
\NAME{Clarke C.~J., Lin D. N.~C. \atque Pringle J.~E.}, \IN{MNRAS
  }{242}{1990}{439}.

\bibitem{vorobyov05}
\NAME{{Vorobyov} E.~I. \atque {Basu} S.}, \IN{ApJ }{633}{2005}{L137}.

\bibitem{vorobyov06}
\NAME{{Vorobyov} E.~I. \atque {Basu} S.}, \IN{ApJ }{650}{2006}{956}.

\bibitem{armitage2001}
\NAME{Armitage P.~J., Livio M. \atque Pringle J.~E.}, \IN{MNRAS
  }{324}{2001}{705}.

\bibitem{palla93}
\NAME{Palla F. \atque Stahler S.~W.}, \IN{ApJ }{375}{1993}{288}.

\bibitem{cesaroni07}
\NAME{{Cesaroni} R., {Galli} D., {Lodato} G., {Walmsley} C.~M. \atque {Zhang}
  Q.}, in proc. of \TITLE{Protostars and Planets V}, edited by \NAME{{Reipurth}
  B., {Jewitt} D. \atque {Keil} K.} 2007, p. 197.

\bibitem{chini04}
\NAME{Chini R. \etal}, \IN{Nature }{429}{2004}{155}.

\bibitem{torrelles98}
\NAME{{Torrelles} J.~M., {G{\'o}mez} J.~F., {Garay} G., {Rodr{\'{\i}}guez}
  L.~F., {Curiel} S., {Cohen} R.~J. \atque {Ho} P.~T.~P.}, \IN{ApJ
  }{509}{1998}{262}.

\bibitem{greenhill04}
\NAME{{Greenhill} L.~J., {Gezari} D.~Y., {Danchi} W.~C., {Najita} J., {Monnier}
  J.~D. \atque {Tuthill} P.~G.}, \IN{ApJ }{605}{2004}{L57}.

\bibitem{patel05}
\NAME{{Patel} N.~A., {Curiel} S., {Sridharan} T.~K., {Zhang} Q., {Hunter}
  T.~R., {Ho} P.~T.~P., {Torrelles} J.~M., {Moran} J.~M., {G{\'o}mez} J.~F.
  \atque {Anglada} G.}, \IN{Nature }{437}{2005}{109}.

\bibitem{edris05}
\NAME{{Edris} K.~A., {Fuller} G.~A., {Cohen} R.~J. \atque {Etoka} S.}, \IN{A\&A
  }{434}{2005}{213}.

\bibitem{mayor95}
\NAME{{Mayor} M. \atque {Queloz} D.}, \IN{Nature }{378}{1995}{355}.

\bibitem{udry07}
\NAME{{Udry} S., {Bonfils} X., {Delfosse} X., {Forveille} T., {Mayor} M.,
  {Perrier} C., {Bouchy} F., {Lovis} C., {Pepe} F., {Queloz} D. \atque
  {Bertaux} J.-L.}, \IN{A\&A }{469}{2007}{L43}.

\bibitem{guillot99}
\NAME{Guillot T.}, \IN{Planetary and Space Sciences }{47}{1999}{10}.

\bibitem{pollack96}
\NAME{Pollack J.~B. \etal}, \IN{Icarus }{124}{1996}{62}.

\bibitem{lissauer_PPV}
\NAME{{Lissauer} J.~J. \atque {Stevenson} D.~J.}, in proc. of \TITLE{Protostars
  and Planets V}, edited by \NAME{{Reipurth} B., {Jewitt} D. \atque {Keil} K.}
  2007, p. 591.

\bibitem{armitagereview}
\NAME{{Armitage} P.~J.}, \IN{ArXiv Astrophysics e-prints }{0701485}{2007}{}.

\bibitem{vanboekel05}
\NAME{{van Boekel} R. \etal}, \IN{A\&A }{437}{2005}{189}.

\bibitem{apai05}
\NAME{{Apai} D., {Pascucci} I., {Bouwman} J., {Natta} A., {Henning} T. \atque
  {Dullemond} C.~P.}, \IN{Science }{310}{2005}{834}.

\bibitem{cava06}
\NAME{{Bertin} G. \atque {Cava} A.}, \IN{A\&A }{459}{2006}{333}.

\bibitem{haghighipour03a}
\NAME{Haghighipour N. \atque Boss A.~P.}, \IN{ApJ }{583}{2003}{996}.

\bibitem{haghighipour03b}
\NAME{Haghighipour N. \atque Boss A.~P.}, \IN{ApJ }{598}{2003}{1301}.

\bibitem{boss00}
\NAME{Boss A.~P.}, \IN{ApJ }{536}{2000}{L101}.

\bibitem{boss02}
\NAME{Boss A.~P.}, \IN{ApJ }{576}{2002}{462}.

\bibitem{mayer02}
\NAME{{Mayer} L., {Quinn} T., {Wadsley} J. \atque {Stadel} J.}, \IN{Science
  }{298}{2002}{1756}.

\bibitem{PPV_GI}
\NAME{{Durisen} R.~H., {Boss} A.~P., {Mayer} L., {Nelson} A.~F., {Quinn} T.
  \atque {Rice} W.~K.~M.}, in proc. of \TITLE{Protostars and Planets V}, edited
  by \NAME{{Reipurth} B., {Jewitt} D. \atque {Keil} K.} 2007, p. 607.

\bibitem{charbonneau07}
\NAME{{Charbonneau} D., {Brown} T.~M., {Burrows} A. \atque {Laughlin} G.}, in
  proc. of \TITLE{Protostars and Planets V}, edited by \NAME{{Reipurth} B.,
  {Jewitt} D. \atque {Keil} K.} 2007, p. 701.

\bibitem{fortney06}
\NAME{{Fortney} J.~J., {Saumon} D., {Marley} M.~S., {Lodders} K. \atque
  {Freedman} R.~S.}, \IN{ApJ }{642}{2006}{495}.

\bibitem{rafikov05}
\NAME{Rafikov R.}, \IN{ApJ }{621}{2005}{69}.

\bibitem{boss04}
\NAME{{Boss} A.~P.}, \IN{ApJ }{610}{2004}{456}.

\bibitem{rafikov07}
\NAME{{Rafikov} R.~R.}, \IN{ApJ }{662}{2007}{642}.

\bibitem{boffin98}
\NAME{{Boffin} H.~M.~J., {Watkins} S.~J., {Bhattal} A.~S., {Francis} N. \atque
  {Whitworth} A.~P.}, \IN{MNRAS }{300}{1998}{1189}.

\bibitem{watkins98a}
\NAME{{Watkins} S.~J., {Bhattal} A.~S., {Boffin} H.~M.~J., {Francis} N. \atque
  {Whitworth} A.~P.}, \IN{MNRAS }{300}{1998}{1205}.

\bibitem{watkins98b}
\NAME{{Watkins} S.~J., {Bhattal} A.~S., {Boffin} H.~M.~J., {Francis} N. \atque
  {Whitworth} A.~P.}, \IN{MNRAS }{300}{1998}{1214}.

\bibitem{mayer05b}
\NAME{{Mayer} L., {Wadsley} J., {Quinn} T. \atque {Stadel} J.}, \IN{MNRAS
  }{363}{2005}{641}.

\bibitem{LMCR07}
\NAME{{Lodato} G., {Meru} F., {Clarke} C.~J. \atque {Rice} W.~K.~M.}, \IN{MNRAS
  }{374}{2007}{590}.

\bibitem{boss06}
\NAME{{Boss} A.~P.}, \IN{ApJ }{641}{2006}{1148}.

\bibitem{chauvin04}
\NAME{{Chauvin} G., {Lagrange} A.-M., {Dumas} C., {Zuckerman} B., {Mouillet}
  D., {Song} I., {Beuzit} J.-L. \atque {Lowrance} P.}, \IN{A\&A
  }{425}{2004}{L29}.

\bibitem{mamajek05}
\NAME{{Mamajek} E.~E.}, \IN{ApJ }{634}{2005}{138}.

\bibitem{LDC05}
\NAME{{Lodato} G., {Delgado-Donate} E. \atque {Clarke} C.~J.}, \IN{MNRAS
  }{364}{2005}{L91}.

\bibitem{payne07}
\NAME{{Payne} M.~J. \atque {Lodato} G.}, \IN{MNRAS }{381}{2007}{1597}.

\bibitem{greenhill96}
\NAME{{Greenhill} L.~J., {Gwinn} C.~R., {Antonucci} R. \atque {Barvainis} R.},
  \IN{ApJ }{472}{1996}{L21}.

\bibitem{collin90}
\NAME{{Collin-Souffrin} S. \atque {Dumont} A.~M.}, \IN{A\&A }{229}{1990}{292}.

\bibitem{goodman03}
\NAME{Goodman J.}, \IN{MNRAS }{339}{2003}{937}.

\bibitem{KingPringle2007}
\NAME{{King} A.~R. \atque {Pringle} J.~E.}, \IN{MNRAS }{377}{2007}{L25}.

\bibitem{thompson05}
\NAME{{Thompson} T.~A., {Quataert} E. \atque {Murray} N.}, \IN{ApJ
  }{630}{2005}{167}.

\bibitem{BL2001}
\NAME{Bertin G. \atque Lodato G.}, \IN{A\&A }{370}{2001}{342}.

\bibitem{shlosman89}
\NAME{{Shlosman} I., {Frank} J. \atque {Begelman} M.~C.}, \IN{Nature
  }{338}{1989}{45}.

\bibitem{collin99}
\NAME{Collin S. \atque Zahn J.~P.}, \IN{A\&A }{344}{1999}{433}.

\bibitem{collin07}
\NAME{{Collin} S. \atque {Zahn} J.-P.}, \IN{ArXiv e-prints
  }{0709.3772}{2007}{}.

\bibitem{nayak07}
\NAME{{Nayakshin} S., {Cuadra} J. \atque {Springel} V.}, \IN{MNRAS
  }{379}{2007}{21}.

\bibitem{nayak06}
\NAME{{Nayakshin} S.}, \IN{MNRAS }{372}{2006}{143}.

\bibitem{genzel03}
\NAME{{Genzel} R. \etal}, \IN{ApJ }{594}{2003}{812}.

\bibitem{ghez05}
\NAME{{Ghez} A.~M. \etal}, \IN{ApJ }{620}{2005}{744}.

\bibitem{levin03}
\NAME{{Levin} Y. \atque {Beloborodov} A.~M.}, \IN{ApJ }{590}{2003}{L33}.

\bibitem{paumard06}
\NAME{{Paumard} T., {Genzel} R., {Martins} F., {Nayakshin} S., {Beloborodov}
  A.~M., {Levin} Y., {Trippe} S., {Eisenhauer} F., {Ott} T., {Gillessen} S.,
  {Abuter} R., {Cuadra} J., {Alexander} T. \atque {Sternberg} A.}, \IN{ApJ
  }{643}{2006}{1011}.

\bibitem{nayak05b}
\NAME{{Nayakshin} S. \atque {Sunyaev} R.}, \IN{MNRAS }{364}{2005}{L23}.

\bibitem{levin07}
\NAME{{Levin} Y.}, \IN{MNRAS }{374}{2007}{515}.

\bibitem{herrnstein96}
\NAME{{Herrnstein} J.~R., {Greenhill} L.~J. \atque {Moran} J.~M.}, \IN{ApJ
  }{468}{1996}{L17}.

\bibitem{papaloizou98}
\NAME{{Papaloizou} J.~C.~B., {Terquem} C. \atque {Lin} D.~N.~C.}, \IN{ApJ
  }{497}{1998}{212}.

\bibitem{greenhill03}
\NAME{{Greenhill} L.~J., {Kondratko} P.~T., {Lovell} J.~E.~J., {Kuiper}
  T.~B.~H., {Moran} J.~M., {Jauncey} D.~L. \atque {Baines} G.~P.}, \IN{ApJ
  }{582}{2003}{L11}.

\bibitem{kondratko05}
\NAME{{Kondratko} P.~T., {Greenhill} L.~J. \atque {Moran} J.~M.}, \IN{ApJ
  }{618}{2005}{618}.

\bibitem{hure02}
\NAME{{Hur{\'e}} J.-M.}, \IN{A\&A }{395}{2002}{L21}.

\bibitem{fan04}
\NAME{Fan X. \etal}, \IN{AJ }{128}{2004}{515}.

\bibitem{fan06}
\NAME{{Fan} X. \etal}, \IN{AJ }{131}{2006}{1203}.

\bibitem{haiman98}
\NAME{{Haiman} Z. \atque {Loeb} A.}, \IN{ApJ }{503}{1998}{505}.

\bibitem{volonteri05}
\NAME{{Volonteri} M., {Madau} P., {Quataert} E. \atque {Rees} M.~J.}, \IN{ApJ
  }{620}{2005}{69}.

\bibitem{wyithe05}
\NAME{{Wyithe} J.~S.~B. \atque {Loeb} A.}, \IN{ApJ }{634}{2005}{910}.

\bibitem{abel00}
\NAME{{Abel} T., {Bryan} G.~L. \atque {Norman} M.~L.}, \IN{ApJ
  }{540}{2000}{39}.

\bibitem{bromm02}
\NAME{{Bromm} V., {Coppi} P.~S. \atque {Larson} R.~B.}, \IN{ApJ
  }{564}{2002}{23}.

\bibitem{volonteri05b}
\NAME{{Volonteri} M. \atque {Rees} M.~J.}, \IN{ApJ }{633}{2005}{624}.

\bibitem{KLOP}
\NAME{{King} A.~R., {Lubow} S.~H., {Ogilvie} G.~I. \atque {Pringle} J.~E.},
  \IN{MNRAS }{363}{2005}{49}.

\bibitem{LP06}
\NAME{{Lodato} G. \atque {Pringle} J.~E.}, \IN{MNRAS }{368}{2006}{1196}.

\bibitem{king06}
\NAME{{King} A.~R. \atque {Pringle} J.~E.}, \IN{MNRAS }{373}{2006}{L90}.

\bibitem{haehnelt93}
\NAME{{Haehnelt} M.~G. \atque {Rees} M.~J.}, \IN{MNRAS }{263}{1993}{168}.

\bibitem{umemura93}
\NAME{{Umemura} M., {Loeb} A. \atque {Turner} E.~L.}, \IN{ApJ
  }{419}{1993}{459}.

\bibitem{loeb94}
\NAME{{Loeb} A. \atque {Rasio} F.~A.}, \IN{ApJ }{432}{1994}{52}.

\bibitem{eisenstein95}
\NAME{{Eisenstein} D.~J. \atque {Loeb} A.}, \IN{ApJ }{443}{1995}{11}.

\bibitem{bromm03}
\NAME{{Bromm} V. \atque {Loeb} A.}, \IN{ApJ }{596}{2003}{34}.

\bibitem{koushiappas04}
\NAME{{Koushiappas} S.~M., {Bullock} J.~S. \atque {Dekel} A.}, \IN{MNRAS
  }{354}{2004}{292}.

\bibitem{begelman06}
\NAME{{Begelman} M.~C., {Volonteri} M. \atque {Rees} M.~J.}, \IN{MNRAS
  }{370}{2006}{289}.

\bibitem{LN06}
\NAME{{Lodato} G. \atque {Natarajan} P.}, \IN{MNRAS }{371}{2006}{1813}.

\bibitem{LN07}
\NAME{{Lodato} G. \atque {Natarajan} P.}, \IN{MNRAS }{377}{2007}{L64}.

\bibitem{warren92}
\NAME{{Warren} M.~S., {Quinn} P.~J., {Salmon} J.~K. \atque {Zurek} W.~H.},
  \IN{ApJ }{399}{1992}{405}.

\bibitem{VLN07}
\NAME{{Volonteri} M., {Lodato} G. \atque {Natarajan} P.}, \IN{ArXiv e-prints
  }{0709.0529}{2007}{}.

\bibitem{fromang02}
\NAME{{Fromang} S., {Terquem} C. \atque {Balbus} S.~A.}, \IN{MNRAS
  }{329}{2002}{18}.

\bibitem{fromang04a}
\NAME{{Fromang} S., {Balbus} S.~A. \atque {De Villiers} J.-P.}, \IN{ApJ
  }{616}{2004}{357}.

\bibitem{fromang04b}
\NAME{{Fromang} S., {Balbus} S.~A., {Terquem} C. \atque {De Villiers} J.-P.},
  \IN{ApJ }{616}{2004}{364}.

\bibitem{cai06}
\NAME{{Cai} K., {Durisen} R.~H., {Michael} S., {Boley} A.~C., {Mej{\'{\i}}a}
  A.~C., {Pickett} M.~K. \atque {D'Alessio} P.}, \IN{ApJ }{636}{2006}{L149}.

\end{thebibliography}

\end{document}